Title: Development of a Thermal Control System with Mechanically Pumped
CO2 Two-Phase Loops for the AMS-02 Tracker on the ISS

Article Type: Research Paper

Section/Category: Special Applications

Corresponding Author: Prof. Zhenhui He,

Corresponding Author's Institution:

First Author: Zhenhui He

Order of Authors: Zhenhui He; ZHAN ZHANG

Abstract: To provide a stable thermal environment for the AMS-Tracker, a
thermal control system based on mechanically pumped CO2 two-phase loops
was developed. It has been operating reliably in space since May 19,
2011. In this article, we summarize the design, construction, tests, and
performance of the AMS-Tracker thermal control system (AMS-TTCS).

1  Development of a Thermal Control System with Mechanically Pumped $CO_2$

2  Two-Phase Loops for the AMS-02 Tracker on the ISS

3  G.Alberti[d], A.Alvino[d], G. Ambrosi[d], M. Bardet[c], R.Battiston[d], S. Borsini[d], J.F. Cao[h], Y. Chen[a,b], J. van

4  Es[c], C.Gargiulo[f], K.H.Guo[b], L.Guo[h], Z.H.He[✉,a, b], Z.C.Huang[b], V. Koutsenko[e], E. Laudi[d], A. Lebedev[g],

5  S.C. Lee[k], T.X.Li[b], Y.L. Lin[i], S.S.Lv[b], M. Menichelli[d], J.Y. Miao[h], D.C.Mo[b], J.Q.Ni[b], A. Pauw[c],

6  X.M.Qi[b], G.M. Shue[j], D.J. Sun[j] , X.H.Sun[b], C.P.Tang[b], B. Verlaat[c], Z.X.Wang[b], Z.L.Weng[b], W.J.

7  Xiao[b], N.S.Xu[a,b], F.K. Yang[i], C.C. Yeh[j], Z.Zhang[b], T. Zwartbol[c]

9   a.  State Key Laboratory of Optoelectronic Materials and Technologies, School of Physics and

10      Engineering, Sun Yat-sen University, 510275 Guangzhou, People's Republic of China

11   b.  Center for Space Technology, Sun Yat-sen University, 519082 Zhuhai, People's Republic of

12      China

13   c.  National Aerospace Laboratory NLR, Voorsterweg 31, 8316 PR Marknesse, the Netherlands

14   d.  Università and Sezione INFN di Perugia, Laboratorio  per lo Studio degli Effetti delle Radiazioni

15      Ionizzanti (SERMS), Via Pentima Bassa 21, 05100 Terni, Italy

16   e.  National Institute for Subatomic Physics Mechanical Engineering, Science Park 105, 1098 XG

17      Amsterdam, the Netherlands

18   f.  Istituto Nazionale Fisica Nucleare sez.Roma, Italy

19   g.  Massachusetts Institute of Technology, Cambridge, MA 02139, USA

20   h.  System Engineering Department, Chinese Academy of Space and Technology, 100094 Beijing,

21      People's Republic of China

22   i.  Electronic Systems Research Division, Chung Shan Institute of Science and Technology, Lung-

23      Tan, Taoyuan, Taiwan

24    j.    Aerospace Industrial Development Corporation, Taichung city 40760, Taiwan

25    k.    Institute of Physics, Academia Sinica, Taipei 11529, Taiwan

26    Key words: space thermal control system, two-phase loop, design, construction, performance

27    ✉: stshzh@mail.sysu.edu.cn  Fax: +86 20 84113398

28    Abstract

29        To provide a stable thermal environment for the AMS-Tracker, a thermal control

30    system based on mechanically pumped $CO_2$ two-phase loops was developed. It has been

31    operating reliably in space since May 19, 2011. In this article, we summarize the design,

32    construction, tests, and performance of the AMS-Tracker thermal control system (AMS-

33    TTCS).

## 1   Introduction

Alpha Magnetic Spectrometer AMS-02([1] [2] [3] [4]) is an astroparticle physics experiment that has been running on the International Space Station (ISS, see Fig.1) since May 19, 2011. The AMS Silicon Tracker is a sub-detector of the AMS. It is inside a Magnet, to detect charge particles, which can distinguish antiparticle from particle.

To achieve stability with a superconducting magnet providing a magnetic field of 0.87 Tesla, temperature stability of 3 $^{o}$C per orbital period for the Tracker silicon wafers and their front-end electronics, with a temperature difference less than 10 $^{o}$C between any of the silicon wafers. The operation temperature of the Tracker is between −10 $^{o}$C to 25 $^{o}$C, and the survival temperature is between −20 $^{o}$C to 40 $^{o}$C. Total dissipated heat to be carried away was about 156W, including 144 W (±10%) out of the Tracker's 192 front-end electronics hybrids, 2W from the silicon wafer, and 10W from the Star Tracker. The life time was required to be 3 to 5 years. To meet such requirement, a mechanically pumped $CO_2$ two-phase-loop thermal control system (called TTCS) had been developed [5], to provide the required thermal boundaries to the Tracker to be working in a superconducting magnet. The design was verified [6].

After NASA announced to extend the life of the ISS to 2020, AMS-02 was upgraded [7] by replacing the superconducting magnet with the permanent magnet that was developed for the AMS-01, in order to extend the life. To maintain the Tracker's resolution in the lower magnetic field of the permanent magnet, the Tracker's configuration as well as the TTCS evaporator was rearranged [7].

In this paper, we present the design, the construction, the tests, and the performance of the TTCS.

## 2   Tracker thermal control system concept design

### 2.1   Analysis of the basic requirement

Two-phase technologies, such as heat pipes and loop heat pipes, are widely used in space thermal control. For the AMS Tracker, however, there are 166 (192 in the original design) heat-dissipating

58    front-end hybrids distributed around the Tracker. To collect all the heat from the hybrids, the total

59    heat path is tens of meters long. Neither heat pipes, which are limited for high heat fluxes and long

60    distance cooling applications, nor loop heat pipes, which are not suitable for distributed heat sources

61    cooling applications, is applicable for the requirement. Moreover, complicate thermal environment on

62    the ISS requires an active control system. A mechanically pumped two-phase loop system was thus

63    proposed to meet the unique set of requirements [5].

64    There are strict budget mass and power for space applications. The Tracker thermal control

65    system has 72kg and 120W for the mass and power budget respective. In addition, the evaporators,

66    the heat collectors locate close to the detector, must be small enough to minimize secondary particles.

67    (Such particles would mislead the detector to take them as cosmic particles.) For a certain fluid in a

68    given flow rate, smaller tube leads to much higher frictional pressure drop ( $h_f = 8fLQ^2/g\pi^2D^5$, where $h_f$

69    is the frictional head loss; L is the length of the pipe; D is the hydraulic diameter of the pipe; Q is the

70    volumetric flow rate; g is the local acceleration due to gravity; $f$ is a dimensionless coefficient called

71    the Darcy friction factor, respectively). Such pressure drop must be controlled within a certain range,

72    so that the two-phase saturated temperature in the evaporators, which is a function of the pressure, is

73    less than 3 $^oC$ in order to provide a uniform temperature boundary. Among the common used

74    refrigerants applicable in the required operation temperature range, carbon dioxide ($CO_2$) has the

75    lowest viscosity. Calculation showed that, $CO_2$ provides lower frictional pressure-drop as compared

76    with ammonia and propylene, which have even wider working window, and are commonly used in

77    space. In addition, the strict safety requirement (no release of harmful materials) for instruments on

78    the manned space station also makes the $CO_2$ the most suitable working fluid for the two-phase loop

79    application.

## 2.2   Configuration and working principle

81    The Tracker Thermal Control System (TTCS) has two mutual redundant fluid-loops. Each loop

82    contains a component box (TTCB)[8], two evaporators (top and bottom) in parallel, and two

83    condensers (RAM and WAKE) in parallel, which are thermally contacted to the two radiators. (see

84   Fig.2, with the radiators hidden for clearness). Assembled in one TTCB are one accumulator, one heat

85   exchanger with redundant start-up heater, two redundant pumps, four redundant pre-heaters, two

86   redundant cold-orbit heater, and pressure sensors. The basic functions of the components of the

87   thermal control system are summarized in Table 1. The $CO_2$ driven by the pump cycles around the

88   loop, collecting the waste heat from the evaporators by partial evaporation (converting into latent heat

89   of evaporation), and releasing the heat in the condensers by condensing the vapor $CO_2$ into sub-cooled

90   liquid $CO_2$. About 5 K sub-cooling is required at the pump to prevent cavitation. The condensation

91   heat is transferred to the radiators, and then to the space. The accumulator is a fluid reservoir to

92   compensate for the required liquid charge in the system at different heat loads as well as during

93   transients of the thermal environments. The other function of the accumulator is to control the loop

94   temperature. The $CO_2$ in the accumulator is in vapor-liquid two-phase state. The accumulator is

95   connected to the loop that has low pressure drop, the pressure in the accumulator is almost the same as

96   that in the evaporators. Therefore, controlling the saturated pressure of the accumulator is equivalent

97   to controlling the saturated temperature of the evaporators, provided that the fluid in the evaporators is

98   in two-phase. To insure that the fluid flowing into the evaporators is in two-phase, preheaters are used

99   to heat the sub-cool liquid to the saturated temperature. To save heating power of the preheaters, a

100  heat exchanger is employed between the fluid paths, in which $CO_2$ flowing in and out of the

101  evaporators. The heat exchanger uses the two-phase $CO_2$ flown out of the evaporators to warm up the

102  sub-cooled liquid flowing in the evaporators. It also uses the sub-cooled liquid to reduce the vapor

103  quality of the out-flowing $CO_2$ to reduce the flow resistance in the transportation tubes (see the

104  schematic in Fig.3).

## 2.3   Design challenges and solutions

106  There are design challenges in extremely cold conditions, such as to prevent the $CO_2$ from

107  freezing inside the condensers, and the Tracker from being cooled below $-20\,^{\circ}$C.

108  Keep the Tracker above $-20\,^{\circ}$C as required. The loop must be running before the Tracker is

109  switched on. In an extremely cold condition, cold fluid below $-40\,^{\circ}$C from the condensers would flow

110  into the evaporator, and the low power of the preheaters could not heat the in-flowing liquid above

111  $-20\,^{\circ}C$. To prevent this from happening, a start-up heater was designed to heat the sub-cooled liquid

112  during start-up when the Tracker is not switched on. It can also help to ensure the $CO_2$ flowing in the

113  evaporators is in the saturated state when in an extremely cold condition during the TTCS operation.

114      The power of the pre-heaters and start-up heater is high enough to keep the Tracker electronics in

115  the required temperature range, but not enough to keep $CO_2$ inside the condensers above its triple

116  point $(-56\,^{\circ}C)$, below which $CO_2$ freezes. Increase the power of pre-heater and start-up heater could

117  prevent this from happening, but increase the pressure drop along the evaporators by increasing the

118  vapor quality of $CO_2$ in the evaporators. Putting the heat just before the inlet of the condensers is

119  much more effective for the anti-freezing. A heater with this function is implemented in the design

120  and called cold orbit heater.

121      Another driving requirement is the high pressure of $CO_2$. The maximum storage temperature is

122  $+65\,^{\circ}C$ resulting in a maximum designed pressure of 16 MPa. In addition, to keep the mechanical

123  pumps in a health condition, MIL-STD-1246 C class 100 is applied as the cleanness requirement for

124  the loop. Strict leak tightness is required lower than $1\times10^{-7}$ Pa$\cdot$m$^3$/s for $CO_2$ at 16MPa to guarantee

125  enough fluid in the required life time.

126      Special design that ensures the condensers to withstand possible high pressure in uncontrollable

127  melting of frozen $CO_2$ is given in section 4.2

128  # 3  Simulation and Detailed Design

129      To verify the feasibility of the designed loop, and to obtain the specifications of the TTCS

130  components, both static and dynamic simulations are required. A combined model with thermal sub

131  model and fluid models based on the SINDA/FLUINT was built up.

## 3.1   Space Model description

132

133   The TTCS space model consists of one fluid model and 18 thermal sub-models. The fluid

134   model is to simulate the mass and heat transfer along the loop. The thermal sub-models are built for

135   the Tracker, the Tracker radiators, and the TTCB, to simulate their thermal status. Seventeen heat-

136   exchange macros and four tie links are used to calculate the heat exchange between the thermal

137   models and the fluid model (see Fig.4). Besides, eight line macros are used to simulate the long pipe

138   without heat exchange, such as the evaporator or condenser feed line and the return line.

139   Homogeneous assumption was chosen for lumps and paths calculation. The correlation of

140   Lockhart-Martinelli was chosen for pressure drop calculation. The heat transfer correlation of Dittus-

141   Boelter was selected for single-phase flow; while that of Rohsenow for condensing, and that of Chen

142   for boiling of two-phase flows.

143   The TTCS/AMS-02 thermal environment is influenced by the following parameters:

144   ● Impinging solar, albedo and Earth radiative fluxes

145   ● Radiation toward deep space

146   ● Radiative heat exchange with other ISS surfaces

147   ● Temperature of the conductive interfaces

148   In addition to these factors, AMS-02 internal dissipation, geometry, surface coatings, and

149   materials have their own influences on the Tracker and TTCS temperatures. Deeply influenced by the

150   beta angle of the ISS and the Euler angles (that determine the ISS attitude) as well as by the resulting

151   motion of surfaces (e.g. solar arrays and radiators), the impinging radiative fluxes and the heat

152   exchange keep changing in the orbit. See Table 2 for the ranges of the beta angle and the Euler angles

153   of the ISS.

154   Then a series of temperature boundaries and heat sources is generated with the overall AMS

155   model according to the hottest and the coldest cases. There are 258 thermal boundaries in the TTCS

156   model. They are distributed in the TTCB, the Tracker, and the Tracker radiators. All the thermal

157   boundaries are orbit dependent.

158        The pressure of the TTCS loop is controlled by an accumulator, the temperature of which is
159        controlled by the accumulator heater and Peltier elements with PI algorithm. It was modelled as a tank
160        with heated shell, the heat leak to the TTCB base plate. The pump was assumed as an ideal one which
161        provided the loop with a constant mass flow rate. In the model, the mass flow rate along the main loop
162        was set constant. However the flow rate distribution between the branches in evaporators and
163        condensers depended on the geometry of the tubes and the fluid state, which would be calculated by
164        the model.

## 3.2  Simulation results

### 3.2.1 System performance

167        Simulation results show that the designed TTCS performs well in both the cold case and the hot
168        case. Shown in Fig.5 are two cases: (a) in a cold case with a mass flow rate of 2g/s, and (b) in a hot
169        case hot case with a mass flow rate of 3g/s. Because the fluid of the hot inlet comes from the
170        evaporators, while the cold inlet affected by the cold $CO_2$ is from the condensers, the TL symbols
171        related to TL labeled reveal the temperatures' variation from the evaporators and the condensers. The
172        Rad temperature is the result of the balance between the fluid, the orbital heat flux, and the radiation
173        to the environment. It implies the overall impact of the radiator boundaries.

174        More simulation results showed that the design parameters of the TTCS meet the requirement.

### 3.2.2 Sizing the radiators and the cold-orbit heater

176        To define the size of the radiators, a rough model without fluid loop, in which the integrated ISS
177        was taken into account, was built to calculate the balance temperature of the radiator in the orbital
178        environment by CGS. Two radiators, each with area of 1.22m$^2$, are big enough to radiate the waste
179        heat to the space in the extremely hot case, and not too big to minimize the power budget of the cold
180        orbit heater.

181      In a hot case (+75,-15,-20,-15), the dynamic simulation result (see Fig.6) shows that, the mass

182    flow rate of 4g/s can provide required sub cooling of $CO_2$ liquid at the pump inlet, (though that of 2g/s

183    cannot). This means the radiator size is just enough with high mass flow rate in the hot case.

184      In extremely cold cases, the $CO_2$ in the condenser will get frozen if no extra heat is applied (see

185    Fig.7). It means that the radiator is too large in the extremely cold cases. In order to meet requirement

186    in both hot and cold cases, radiators are optimized with proper size and with extra loop heating

187    capacity by employing the cold-orbit heater.

188      In combination of the special design of the condenser (see 4.2 for the condenser structure), where

189    thermal contact between the inlet tubes of the condenser and the second heat pipe of the radiator in the

190    entrance could provide the lowest anti-freezing power [9], By choosing the right flow direction of the

191    condenser, the power of the cold-orbit heater can be optimized to an acceptable budget of 60W [10].

192    Based on the analysis of the simulation, the condenser's configuration is also finalized.

193      Similarly, the powers of the preheaters, the start-up heater, and the accumulator heater were also

194    determined, within the power budget.

195      The final design keeps the Tracker temperatures within a band of less than 1℃ over the complete

196    operation temperature window (see Fig.5 for the system performance).

197    ## 3.3  Safety and structure analysis

198      The TTCB assembly and its components such as the accumulator and the heat exchanger must be

199    safe in terms of displacement, stresses, natural frequency, reactions, and constraint locations at

200    interfaces under critical load conditions in launching, landing, and in-orbit operation according to the

201    NASA's requirements. We analyzed the TTCB assembly and the main components using finite

202    element analysis to verify that the Yield and the Ultimate Margins of Safety are positive:

203       ● Static structural analysis;

204       ● Connection safety analysis such as bolt-insert-washer, bolt-nut-washer, with or without

205          thermal washers;

206       ● Fail-Safe structural analysis

207 ● Fail-safe connection safety analysis;

208 In addition, Modal analysis was performed to ensure that the first significant structure natural

209 frequency is above 50 Hz.

210 The TTCB is an overall assembly that holds the TTCS components except for the evaporators

211 and the condensers. The accumulator is an important component mount in the TTCB with high inner

212 pressure (maximum design pressure of 16 MPa) and extreme temperature range, thus the safety issue

213 is very important. We only present the analyses of the TTCB assembly and the accumulator; without

214 including those of other components

## 3.3.1 The Accumulator

216 The structure of the accumulator is described in 4.4.2. The accumulator was modeled as an

217 assembly. Most of its components were meshed as 3D solid elements and some were modeled as

218 beam element if they are long and thin. Those masses that were not embodied in the FEM mesh model,

219 such as that of the fluid, were added to the model as allotted mass. Connections between different

220 parts were modeled as common nodes, connecting elements, coupled Degrees of Freedom, or bolt

221 connections, depending on the corresponding characteristics, or if the reaction forces at those

222 locations were needed.

223 Two essential load cases were analyzed for the structural verification of the accumulator.

224 In the Launching and Landing load case, only the pressure load and the acceleration load were

225 applied in the calculation, because separate analysis showed that temperature load is ignorable. The

226 acceleration load of ±40 g was applied in one direction with ±10 g simultaneously applied in the other

227 two directions. The pressure loads inside the accumulator and the accumulator heat pipe (AHP) were

228 16 MPa and 5.7 MPa, respectively. Altogether, six acceleration load cases and one pressure load with

229 two different pressures in two components, and their combination were considered in the calculation.

230 In the In-orbit load case, there are two kinds of loads: the pressure load and the thermal load

231 related to heating. The pressure load is 16 MPa inside the accumulator (at 65$^\circ$C) and 5.7 Mpa inside

232 the AHP, respectively. The pressure loads and the temperature gradient in three hot cases were combined. The Factor of Safety for yield was 1.5 and that for ultimate was 2.5 or 4.0 depending on the tube diameter. The minimum Margins of Safety of all the components and bolt connections in all the load cases are summarized in Table 3.

A Fail-Safe analysis was performed with the highest loaded fastener removed, and all the Margins of Safety were recalculated (see Table 4). The Factor of Safety used for fail-safe analysis was 1.0 for both yield and ultimate.

From the calculation results, we found that the accumulator had smallest margin of safety, but was still acceptable for the NASA's safety standards. All the other components and bolt connections had even larger margins of safety. In the modal analysis, the first natural frequency was 399.8 HZ, much higher than the required 50 Hz (see Fig.8), indicating no further vibration test was needed.

A pressure test was performed, which qualified all the box components, and the integrated box assembly, respectively. The vibration tests also verified the simulation.

Thermal analysis was carried out for the accumulator for the safety verification. The results showed that even in the case of double failures of the thermostats (TSs), the accumulator's temperature, and thus its saturated pressure, is under the maximum design temperature (65ºC), and pressure (16 MPa).

### 3.3.2    The TTCB

The TTCB was modeled as an assembly. It consists of different mesh types such as 3D solid elements, 2D shells with certain thickness and beam elements, depending on the mechanical characteristics and dimensions of different components.

Bolt connections between different parts were modeled by coupling the nodes Degrees of Freedom (DoF) of different parts at their respective screws locations. Other components that were not modeled in the FEM model were considered as allotted masses. The connections of the TTCB with the USS were modeled as constrained translations and rotations at the corresponding locations.

The total estimated mass was 20.19 kg for one TTCB, therefore ±31 g acceleration load should be applied in one direction with ±7.75 g simultaneously applied in the other two directions. Different

259  load cases were considered by sweeping the direction of the acceleration vector. For bolts, 60%

260  preload was used for connection safety analysis. For structural and connection analysis, the Factor of

261  Safety for yield was 1.25, and that for ultimate was 2.0.

262      A Fail-Safe analysis was also performed with the highest loaded fastener removed and all the

263  Margins of Safety are recalculated. Factor of Safety is 1 in the fail-safe analysis.

264      The calculated first mode is at 53.9 Hz (see Table 5), higher than the required 50Kz. Analysis

265  showed that the designed TTCB met the safety requirement.

# 4  Construction and Verification

266

## 4.1  The Radiators

267

268      Due to the strict temperature requirement of the Tracker, dual radiators were designed for the

269  TTCS (called Tracker radiators), which are linked in parallel, and facing to different directions: the

270  ram side and the wake side of the ISS (see Fig.1). Such orientation design can avoid suffering from

271  high temperature in the extremely hot cases; because, when the sun shines directly on one of the

272  radiators for small beta angles of the ISS, it shines on the other at large incident angles, which suffer

273  less from the environmental high heat flux, and can still radiate most of the heat from the loop to the

274  space. To meet the requirement of radiation capability in the hottest case, the Tracker radiators are

275  trapezoid in shape with radiation area of 1.16 $m^2$ each (see Fig.9 (a)). The thickness of the radiator is

276  14.2 mm, and that of the top and bottom face sheets are 0.5 mm, respectively.

277      Seven heat pipes (see Fig.9(b) for its cross section) are embedded in the ROHACELL foam, and

278  they are sealed in an aluminum shell along the length direction (see Fig.10 for the photo). Condensers

279  are mounted on the back side of the radiators; in a way that each of them thermally connects with all

280  the seven heat pipes, so that the condensation heat from any of the condensers can spread out to the

281  whole radiator. The surfaces of the radiators were painted with the white paint (SG 121FG) to

282  optimize the heat rejection.

283     The radiators were manufactured in AIDC Taiwan, according to qualified composites
284     manufacture procedures.

285     On-ground tests were performed for the manufactured radiators in both horizontal and vertical
286     orientation. IR camera was employed to check the functioning of the heat pipe. In the horizontal
287     orientation (to simulate the microgravity condition), all of the seven heat pipes functions normally. In
288     the vertical orientation, simulation showed that the Tracker radiator could function normally even two
289     out of the seven heat pipes fail to start up, depending on the heat load. The thermal vacuum test (TVT)
290     showed that the radiators, which were in vertical orientation, could start up and had the ability of heat
291     radiation as required. Together with the simulation results, such test were further confirmed in the
292     integrated verification in the TVT ( see  [6] ).

293     ## 4.2   The Condensers

294     The two condensers of the primary loop were mounted onto the RAM and WAKE radiators on
295     the port side; and those of the secondary loop onto the starboard side, respectively.

296     The high triple-point of the $CO_2$ ($-56.6$ °C) makes the $CO_2$ freezing highly possible in the TTCS
297     condenser, for example, in the case of a full AMS power shutdown or in the phase of transferring
298     AMS from the shuttle to the ISS, where $CO_2$ in the condenser might drop down to $-120$ $^o$C. The
299     expansion of $CO_2$ during the melting creates a high inner pressure. Therefore, if the de-freezing
300     heating of the condenser by environmental heating is out of control, the in-homogenous heating on the
301     solid $CO_2$ could induce a local high pressure up to 300 MPa for enclosed liquid $CO_2$ between the still
302     frozen parts [9]. This is quite a challenge to the condenser design. To withstand such a high pressure,
303     Inconel mini-tubes (outer diameter D_out = 3.15 ±0.05 mm and wall thickness t=1.1 ± 0.1 mm) were
304     used, which can survive up to 1200 MPa by test (see Fig.11 for the structure of the condenser) [11].

305     To construct the condenser, seven small and smooth Inconel tubes were meanderingly imbedded
306     between two aluminum plates, glued with thermal conductive adhesive (Master Bond Polymer System
307     EP21TDC-2LO) to transfer heat from tube to bottom plate and to release possible stress between the

308    tubes and the condenser plates because of the different thermal expansion coefficients between
309    inconel alloy and aluminum, together with the large magnitude changes of the condenser temperature.
310    The inlet and outlet of the seven tubes were brazed into an inlet and an outlet manifold, respectively.
311    The tube length embedded in the condenser plate was 2.49 m and that from the manifold to the base
312    plate inlet and the outlet were 0.45 m, respectively. The length and the width of the plate were 460
313    mm and 340 mm, respectively the effective area was about $0.07 \text{m}^2$ (see Fig.11 (a) and (b) for the
314    configuration of the condenser). The thermal conductance of a single condenser two-phase flow state
315    is about 85 W/K at the working temperature of $-5°C$, which is from the measured heat transfer
316    coefficients [12].

317    Based on the condenser EM test results, the contact thermal conductance between the condensers
318    and the cold plates were about 30 to 60 W/K, depending on the cold plate temperature and the
319    mounting procedure.

320    The manufactured QM condenser (see Fig.12 for the photo) was mounted on a cold plate that
321    simulated the Tracker radiator. It passed the $CO_2$ freezing and de-freezing tests (thermal cycling test)
322    and performance test performed at SYSU.

### 4.3   The Evaporators

324    There are two evaporators (the top and the bottom) in parallel connected to the loop. The two
325    evaporators are stainless steel tubes with outer diameter of 3 mm and thickness of 0.2 mm. For the
326    first design, both evaporators have two rings in series, along which, and thermally connected with are
327    the front-end electronics of the AMS Tracker. The inner ring with many bends connects with 24
328    pieces of copper brands that thermally connect to the front-end electronics through carbon fiber bars;
329    the outer ring to 30 through copper braids. (see Fig.13 and Fig.14)

330    For the upgraded Tracker, the original layer 8 was removed. Part of it was put on the top of the
331    transition radiation detector (named as new layer 1). This layer is no longer thermally controlled by
332    the TTCS but by passive radiation cooling and an additional heater circuit. The rest of layer 8 formed
333    a new bottom Tracker layer between RICH and the electromagnetic calorimeter, (named layer 9, see

334 Fig. 3 in [4] for more details). Correspondingly, the TTCS bottom evaporator was also rearranged to

335 collect waste from the front-end electronics of the layer 9 (see Fig.15). The outer ring of the bottom

336 evaporator was replaced by a rectangular ring, with total length of the tube unchanged. About 19.5 W

337 of waste heat shifted correspondingly with the new layer 1 to the top of the AMS, indicating that the

338 bottom evaporator has 19.5W less heat load. Compared to the original design, the upgraded TTCS

339 now has asymmetric evaporators in terms of both heat load and tube routing geometry. It faces more

340 challenges from the cold environment because the radiators become over-designed now; and the

341 asymmetric evaporators may result in unequal flow distribution between the two evaporators, and thus

342 risk of dry-out for the evaporator at low flow-rate. The in-orbit results below showed that the concept

343 is robust for the late design upgrade.

344 ## 4.4 The Component boxes

345 The component box (TTCB) is an assembly that integrates the core components of the TTCS,

346 except for the evaporators and the condensers. They are two pumps, one accumulator, one heat

347 exchanger, two absolute pressure sensors, two differential pressure sensors, four preheaters, two start-

348 up heaters, and two cold orbit heaters. The mechanical concept box was designed by INFN Italy and

349 the detailed by NLR, the Netherlands. TTCBs were manufactured in AIDC Taiwan.

350 The two pumps are mounted on a start-up radiator, aluminum (7075) plate painted with white

351 paint, and facing the outer space as much as possible. This keeps the pumps the coldest components in

352 the TTCB before any of the pumps is switched on. This could prevent the pump from cavitation. The

353 accumulator, the heat exchanger, and the pressure sensors are mounted on aluminum (7075) plate that

354 is the base of the TTCB, and also the mechanical interface of the box to be mounted on the USS. In

355 the front edge of the TTCB base plate, a cold orbit heater is mounted (see Fig.16, the rectangular

356 copper plate, into which armored heater-wires were soldered). Armored heater-wires were soldered

357 onto the shell of the heat exchanger, to operate as the start-up heaters. Armored heater-wires were

358 soldered onto the tubes connected to the inlets of the evaporators.

359      Performance tests were carried out for the integrated TTCB QM (see Fig.17) at SYSU, China, by

360      making use of the experimental setup for the TTCS EM tests. Thermal vacuum test and EMI/EMC

361      test were performed at SERMS, Italy, for the TTCB FMs.

### 4.4.1 The Pumps

363      The pumps to cycle the working fluid are the hearts of the two-phase loops. The rotation speed of

364      the centrifugal pump can be adjusted independently to change the flow rate of the cycling $CO_2$.

365      Therefore, it has the advantage of decoupling the driving force of the pump from the heat load, as

366      compared to those capillary force pumps, and thus provides much higher systematic stability and

367      reliability. Because the pumps contain moving parts that always limit its life time. Centrifugal pumps

368      were chosen because they are good at long life-time among mechanical pumps.

369      Pacific Design Technology (PDT) developed small and light centrifugal pumps for the path-

370      finder to Mars, which could be used to guarantee the life-time design requirement. The continuous test

371      time of 12 months indicated that the life time of such a pump could be three to five years. PDT

372      modified the pumps to adapt to the $CO_2$ for the TTCS application (see Fig.18). Two mutual redundant

373      pumps are employed in one loop, i.g., four for two redundant loops. The possible total life time

374      including redundancy is 12 years.

375      The rotation speed of the pumps can be adjusted to provide from a normal flow rate of 4 ml/sec

376      at 460 mbar pressure head, to a maximum flow rate of 7 ml/sec at 1400 mbar pressure head. Since the

377      life time of the pumps is very much related to their rotation speed in operation, the rotation speed is

378      always kept as low as possible, yet just enough to support a stable and normal operation of the TTCS.

379      The pump rotation speed is set and changed by ground command.

### 4.4.2 The Accumulator

381      The accumulator is made of stainless steel to withstand an inner pressure of 24 MPa, the

382      designed yield pressure of the accumulator. Wick materials with special design structure were applied

383      in the inner accumulator to manage the $CO_2$ liquid in microgravity. A heat pipe (AHP) extending from the inside to the outside of the accumulator was welded to and through the accumulator shell. On the outer part of the heat pipe, armored heater wires (THERMOCOAX) were soldered to heat the accumulator more uniformly. On the heater wires, six thermostats (Comepa) were mounted to physically switch off the heaters at a set-point temperature of 55ºC, to meet the safety requirement for pressure vessels. To cool the accumulator, Peltier elements were mounted between the accumulator wall and the loop tube in which sub-cooled liquid is flowing through (see Fig.3 for the schematic, and Fig.19 for the 3D drawing of the accumulator, and Fig.20 for the outcome product of a FM). Three thermostats (with set temperature 45 ºC) were mounted on the Peltier saddle to meet the safety requirement.

The Pt-1000 sensors that used to measure the accumulator temperature for the temperature control were glued on the outer surface of the accumulator shell. Inside the corresponding position of the accumulator, the $CO_2$ is in where two-phase state (see Fig.16 for the Pt-1000s' position).

### 4.4.3 The Heat Exchanger

The TTCS heat exchanger is a cross-flow plate heat exchanger. In the normal operation condition, two-phase $CO_2$ flows in the hot side; while the sub-cooled $CO_2$ liquid in the cold side. To withstand the high pressure, the heat exchanger is made of Inconel alloy.

Two armored wire heaters were soldered onto the shell of the heat exchanger, which are used as the mutual-redundant start-up heaters (50W each). Fig.21 shows a heat exchanger flight model. Six thermostats (3 in series per heater, with a set-point temperature of +80 $^o$C) were mounted on the heat exchanger shell.

### 4.4.4 Integration

The whole TTCS loops were integrated onto the AMS. The evaporators were first mounted on the Tracker, with good thermal contacts to the Tracker front-end electronics. The condensers were then mounted onto the Tracker radiators respectively. The component boxes were mounted on the USSs correspondingly; and the TTCE crate was mounted on the main radiator. Finally, transportation

409     tubes and cables were connected to the corresponding components, respectively. After integration,
410     carbon dioxide was filled into the loops.

411        ### 4.4.5 Verification

412     Except verifications at component level and box level, the two TTCS loops were verified at
413     system level, together with the AMS. They pass all the tests, including Helium leak tests, proof
414     pressure tests, functional test. The TTCS loops were also thoroughly tested during the AMS02
415     thermal vacuum test and EMC test at ESA ESTEC in the Netherlands. Please refer to [6] for details of
416     the thermal vacuum test of the TTCS.

417 ## 5   Control electronics and control scenarios

418        ### 5.1   Control electronics

419     TTCS thermal control electronics (TTCE) is a combined unit of engineering data acquisition and
420     thermal control. Orientated to the redundant TTCS loops, TTCE adopts cross redundant design: each
421     loop is mounted with two sets of sensors and actuators (A and B), which, together with the TTCE
422     board A and B, are correspondingly composed of the redundant TTCE control loops (see Fig.22).

423     Dual CAN buses (CAN A and CAN B) connect the TTCE with the four JMDCs. The (dual
424     redundant) TTCE acts as four (2×2) slaves. Dual redundant power (+28 VDC / 10A) will be offered
425     by the PDS (power distribution system) to the Tracker thermal electronic power board (TTEP) inside
426     the TTCE crate (see Fig.22).

427     Inside the TTCE crate, there are three kinds of electronic boards; the TTEC, the TTEP, and the
428     TTPP (see Fig.23). TTEC board is a core electronic control board. It communicates with JMDC
429     (mission computer) through dual redundant CAN buses; collects temperature signals from three-fold
430     Pt1000s at each sensing point, and DS1820 sensors; interpreters control commands; runs control
431     algorithm and implements low-level control by outputting control signals. TTEP board is a power
432     supply for TTCE. More precisely, it provides +5V and +15V DC power to TTEC, and electrically isolated control output powers for: 1) PWM (pulse width modulation) power control for the accumulator Peltier elements; 2) PWM power control for the accumulator heaters; 3) on-off power output for the preheaters, cold-orbit heaters, and the start-up heaters, respectively. TTPP is a pump control board, and at the same time, it collects pressure information from the absolute pressure sensors (APSs) and differential pressure sensors (DPSs) of the loops.

The design, the manufacture, and the tests of the TTEC boards were given in more details in [13].

## 5.2   Control scenarios

A TTCS control loop controls temperature at a given set-point with the temperature-control sensors' values as inputs. Three sensors are used to implement a voting mechanism for reliable determination of the true value. Once on power, the TTCE reads the state parameters, such as temperatures, pressures, and sends those values to JMDC (red arrows in Fig.24), which are then sent to ground through the ISS. Those temperature-control sensors' values are used for the TTCS operation control at TTCE (see Fig.24). Control parameters, such as set-point temperatures, PI parameters, pump rotation speed, and control modes and ground commands are sent through the TM/TC interface and CAN bus interface in JMDC to the TTCE (see Fig.25).

Three health-guards are implemented in the JMDC. They are: (1) Overall Tracker Electronics high and low temperature health-guard, which is to provide an overall independent protection of the Tracker Electronics for too high and too low temperatures; (2)Radiator freezing health-guard, which is to warn for freezing of the radiators, which may occur during periods when the PDS does not supply power to the Tracker radiators  heaters(120V) when the Tracker is switched off; (3)JMDC-TTCE communication outage health-guard (in TTCE-Manager), which is to protect the Tracker Electronics for possibly hazardous malfunctioning of the TTCS during a TTCE-JMDC communication outage.

### 5.2.1 TTCE low level controls

There are five low level controls: 1) Pump set point control; 2) Accumulator temperature set

458     point control; 3) Evaporator inlet heating control; 4) Start-up heating control; 5) Cold-orbit heating

459     control; two of them are close-loop controls: 1) accumulator temperature PI control, and 2) cold orbit

460     heater on-off control (see Fig.24).

### 461     5.2.2 Communication

462     The TTCE communicates with JMDC through CAN bus, and the JMDC communicates with ground

463     through the tele-mornitor or tele-control (TM/TC) interface (see Fig.25).

464

## 465   6   In-orbit performance

466     AMS02 was launched to Space on 16[th], May 2011 and transferred from shuttle to International

467     Space Station on 19[th], May 2011.

### 468   6.1   Communication check

469     Communication was checked before  the other tests with the shuttle and the ISS respectively,

470     by sending request of 'READ ALL' to TTCE and getting feedback to verify the entire critical

471     information of temperature, pressure, and so on, all over the cooling system. Ever since the up-and-

472     down data flow was confirmed, which indicating normal communication were established, the

473     following tests were ready to proceed.

### 474   6.2   Functional check

475     Before servicing to AMS Tracker, TTCS conducted a serial of functional tests for each

476     component from May 17[th] to 18[th] while shuttle STS-134 entered the space orbit. All the TTCS

477     components and sensors passed the tests and performed as they did on the ground.

## 6.3 Performance tests

### 6.3.1 Start-up test

The start-up of the TTCS and the Tracker were successfully completed and the process is depicted in Fig.26. To prevent cavitation to the pump, the accumulator temperature was increased to $15\,^{o}C$ in advanced to expel liquid $CO_2$ from the accumulator to the loop, and to liquidize $CO_2$ vapor in the loop. As soon as the sub-cooling, the temperature difference between the accumulator and the pump inlet liquid, was over $5\,^{o}C$, the pump was switched on with rotation speed of 5000RPM. Due to the cooled liquid coming from the condensers, the temperature of the pump inlet dropped; those of the evaporators and the Tracker planes dropped as well. When the Tracker was powered on, the liquid $CO_2$ started to evaporate and smoothly turned to stable two-phase state in half an hour without super-heating. To prevent the Tracker planes' temperatures from rising over the limit of $30\,^{o}C$, the set-point temperature was tuned down to $5\,^{o}C$. The evaporators' temperature decreased with the accumulator's temperature correspondingly. In the end, the TTCS was running at the saturation temperature of $5\,^{o}C$.

### 6.3.2 Nominal operation

Having been tested in space, we found the proper working condition for the normal operation of the TTCS, with the operation temperature of $0\,^{o}C$ and the pump rotation speed of 6000RPM. The temperatures of the evaporators and the Tracker planes stay stable even the radiators temperatures vary from $-35\,^{o}C$ to $-10\,^{o}C$ (see Fig.27). The thermal boundaries of the TTCS, reflected by the two radiator temperatures, are mainly affected by the beta angle in general. If the beta angle stays in the range of $-45$ to 45 degrees, the temperature of the pump inlet is between$-20\,^{o}C$ and $-5.6\,^{o}C$. Thus the TTCS could smoothly run at the saturation temperature of $0\,^{o}C$ and the pump rotation speed of 6000RPM, without any operation interrupt, nor extra heating power consumption. When the beta angle is higher than 45 degrees, or lower than $-45$ degrees, the cold-orbit heater would be activated automatically. We will discuss the cold-orbit heater's performance in the next section.

### 6.3.3 Automatic control for the cold-orbit operation

To maintain the TTCS running automatically, most of the heaters are in auto-control mode, monitored by Pt1000's and controlled by heaters such as the cold-orbit heater (60W) that is used to prevent the $CO_2$ from freezing inside the condensers or being too cold to initiate the $CO_2$ to two-phase in the evaporators. Its switching set-point was set at $-20$ °C, which is adjustable by the ground command. The control scenario is to take the temperature of the pump inlet as the reference, once it goes below $-20$ °C, the cold-orbit heater would automatically switch on; otherwise, it stays off. As shown in Fig.28, every time when the pump-inlet temperature drops to $-20$ °C during the ISS orbit variation, the cold-orbit heater functions automatically as designed. Only 1 °C of temperature oscillation on the evaporator is seen and it comes back to the previous state (above $-20$ °C) within a few minutes, there is no visible temperature variation to the Tracker temperature at all (neither that of Plane_4 nor that of Plane_9).

### 6.3.4  Hot switching from the primary loop to the secondary loop

The centrifugal pumps are the only moving components of the TTCS. To sustain its proper performance, the pumps are used alternatively: each pump operations for three months on average. Thus, the hot switching, without powering down the Tracker, from a running loop to a stand-by loop is scheduled every three months. To verify the temperature stability of the Tracker during the hot switching, a loop was switched from the primary to the secondary loop with operation temperature of $-5$ °C, pump speed of 5000RPM (see Fig.29). When CO2 sub-cooling of over 5°C was guaranteed in the secondary loop, one of the pumps of the secondary loop was switched on. As a result, the temperature of the pump-inlet of the secondary loop started to drop from $-5$ °C  to $-20$ °C, and then varied with time in an orbital period. After running the two loops simultaneously for a while, the pump of the primary loop was switched off. The temperature of the pump-inlet of the primary loop went up steeply because there was no more sub-cooling liquid flowing through. The Tracker plane4's temperature increased by 2 °C, which is explained by the thermal resistance between the evaporator of

527 the primary and the secondary, leading to the difference of resistances of the two loop to the Tracker

528 hybrids. As shown in the Fig.13, the evaporator of the secondary is above that of the primary, the

529 thermal resistance of tubes and liquid would have an impact on the heat transfer efficiency from the

530 Tracker to the TTCS evaporators. As a result, the Tracker plane 4's temperature increased.

### 6.3.5   Performance during the Soyuz undocking.

532 Working on the ISS, the TTCS is indirectly impact by the shuttle's docking and undocking,

533 because the space station must adjust its attitudes. For example, during the Soyuz (42P) undocking on

534 29th, October, 2011, from 07:45 to 09:32, the ISS attitude shifted from (+356.000, +355.500, +0.700)

535 to (0.000, +90.000, 0.000), and back afterward, where the (Yaw, Roll, Pitch) are the three Euler

536 angles that define the ISS attitude. This means that the ISS rotated by $90^o$ in the horizontal. The

537 Tracker's temperature was hardly affected. However, the sub-cooling was not sufficient anymore,

538 because the temperature of the wake radiator increased. The cavitation-prevention alarm was triggered.

539 This alarm was a health-guard in the control electronics to prevent the pump from cavitation. The

540 operation temperature increased automatically by heating up the accumulator until enough sub-

541 cooling was reached. In the case shown in Fig.30, the sub-cooling is set at 5 ℃. The evaporator

542 temperatures show a ripple, while those of the Tracker planes are hardly affected. Implicitly it is

543 shown that the health-guard was working properly.

## 7   Conclusion

545 The AMS-TTCS has been developed for the thermal control of the AMS Tracker. It has been

546 operating reliably since May 19, 2011. The temperature stability of the Tracker layers can be

547 controlled within 1℃ for each loop for various complex thermal environments, including spacecraft

548 dockings and un-dockings; and within 3 ℃  for loop switching, which meets the Tracker operation

549 requirement. It shows promising applications of mechanically pumped two-phase loops for space

550 thermal controls, manned space flights and distant planets exploration.

# Acknowledgements

This work was supported by the Prophase Research of National Basic Research Program of China under Grant No. 2006CB708613, the Science and Technology Program of People's Government of Guangdong Province, China.

Supports to the thermal vacuum test of the TTCS are acknowledged, in particular from Dr. X.D. Cai (MIT), Prof. Bruna, Bertucci, (U. of Perugia). The authors would like to thank AIDC (Taiwan) for the manufacture of the Tracker radiators, the condensers and the heat exchangers, and integration of the TTCB; CSIST (Taiwan), H. Jinchi, Y.J. Fanchiang, and S.H. Wang for the manufacture of the electronics; CAST (China) for the development and manufacture of the accumulators; NLR for the efforts for condenser & HX design, detailed box design and TV test;, and NIKHEF for the design and manufacture of the evaporators. Supports from NASA during the development of the TTCS are appreciated. Special thanks go to ESA, ESTEC Laboratory (Noordwijk), for providing thermal vacuum test facilities.

**Figure captions**

601 Fig.1 The AMS-02 working on the International Space Station. The two red arrows point to the two

602 TTCS radiators, respectively.

603 Fig. 2 The 3D model graph of the two redundant TTCS loops with all the other AMS parts hidden.

604 Fig. 3 The schematic diagram of the upgraded TTCS design, with the only change of bottom

605 evaporator, which differs from the original design in heat load (-19.5W), outer ring's shape (square),

606 and position.

607 Fig. 4 The schematic SINDA/FLUINT model (fluid part) of the TTCS.

608 Fig. 5 The variation of the inlet and the outlet temperatures of the heat exchanger with the that of

609 orbital heat flux; (a) Cold case (-75,+0,+0,-15)，FR=2 g/s, Tset=-15 $^{o}$C; (b) hot case (+75,-15,+0,-15),

610 FR=3 g/s, Tset=15 $^{o}$C. Label TL stands for the temperature of the fluid at the inlet and the outlet of

611 the heat exchanger; Label Rad stands for the average temperature of the radiator out plate.

612 Fig. 6 The variation of the temperatures of the primary loop in the case (+75-15-20-15), 4g/s, and at

613 20 $^{o}$C.

614 Fig. 7 The variation of temperatures of the condenser, indicating that the CO2 should freeze after 0.8

615 hour at Tset=258 K in a cold case (-75+0+0-15). Labels TL#### stand for the temperature of the

616 condenser lumps in the SINDA model, and the numbers following TL are the lump numbers. The

617 frozen point of CO2 is -56 $^{o}$C.

618 Fig. 8  The modal shape of the TTCS accumulator.

619 Fig. 9  (a) The radiator design drawing; (b) the radiator condenser interface.

620 Fig. 10  A Tracker radiator, before and after the top plate was covered.

621 Fig. 11  3D design drawing of the condenser (a). The cross-section of the condenser plate (b).

622 Fig. 12 A condenser under construction.

623 Fig. 13  The 3D design drawing of the top evaporator. It consists of two redundant evaporators (blue

624 lines) connected with the Tracker front-end electronics. The hybrids were thermally connected to

625   the evaporator outer rings by cooper braids. G, E and F are temperature sensors along the

626   evaporator.

627   Fig. 14  The top evaporator inner rings that have been integrated to the AMS-Tracker. The copper

628   soldered on the evaporator tubes (the fine metal tubes) thermally connects with the front-end

629   electronics hybrids (vertical blocks in the photo) through carbon fiber bars (not shown in the photo).

630   Fig. 15  The 3D design drawing of the bottom evaporator, the only upgraded part of the TTCS.

631   Fig. 16  The TTCB assembly drawing.

632   Fig. 17 The integrated TTCB QM. ① Accumulator; ② Cold-orbit heaters.

633   Fig. 18 The centrifugal pump produced by PDT.

634   Fig. 19 3D design drawing of the accumulator; ①accumulator vessel, ②braces, ③subcooled $CO_2$ tube,

635   ④Peltier element set,  liquid pipe (that connects the accumulator to the loop, ⑥ accumulator heater

636   heat pipe.

637   Fig. 20 A flight model accumulator.

638   Fig. 21 The flight model heat exchanger.

639   Fig. 22  The TTCE block diagram

640   Fig. 23  The dual redundant block structure inside TTCE crate.

641   Fig. 24 The schematic diagram of the TTCS control loops. The liquid flow control, the start-up heater

642   control, and the defreezing control are impletemented by the ground commends.

643   Fig. 25 The communication interfaces of the TTCE.

644   Fig. 26 TTCS start-up for operation temperature of 5 $^o$C, at pump rotation speed of 5000RPM. The

645   noise of the DPS is clearly associated with the appearance of the two-phase flow after the Tracker

646   switched on.

647   Fig. 27 Routine performance of the TTCS at the operation temperature of 0 $^o$C and pump rotation

648   speed of 6000 RPM.

649   Fig. 28 Cold-orbit heater auto-function performance at the pump rotation speed of 6000RPM and at

650   the saturation temperature of 0 $^o$C.

651    Fig. 29 TTCS performance during hot switching from the primary to the secondary loop at the

652    saturation temperature of $-5\,^{\circ}\text{C}$, and the pump rotation speed of 5000RPM.

653    Fig. 30 TTCS performance during the undocking of the Soyuz (42P) with the attitude change from

654    (+356.000, +355.500, +0.700) to (0.000, +90.000, 0.000), and a duration of 107 minutes. The sub-

655    cooling health guard functions as designed.

656    Figures

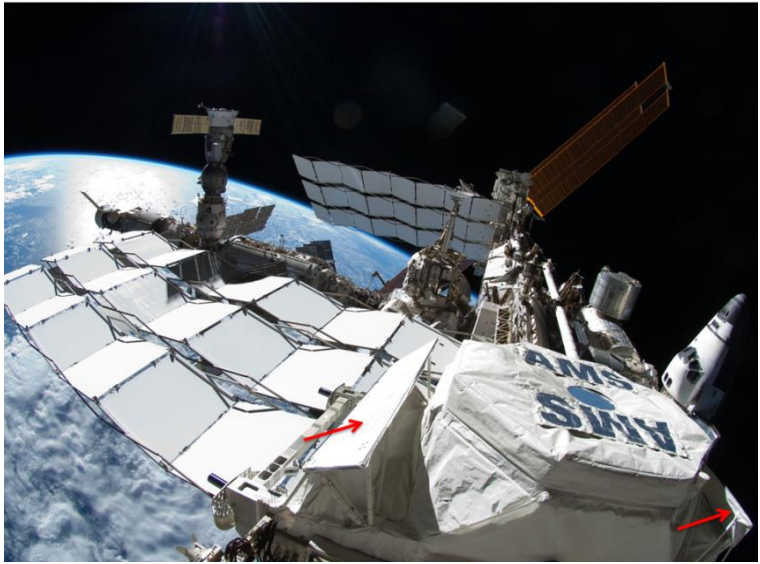

657

658    Fig.1 The AMS-02 working on the International Space Station. The two red arrows point to the two

659    TTCS radiators, respectively.

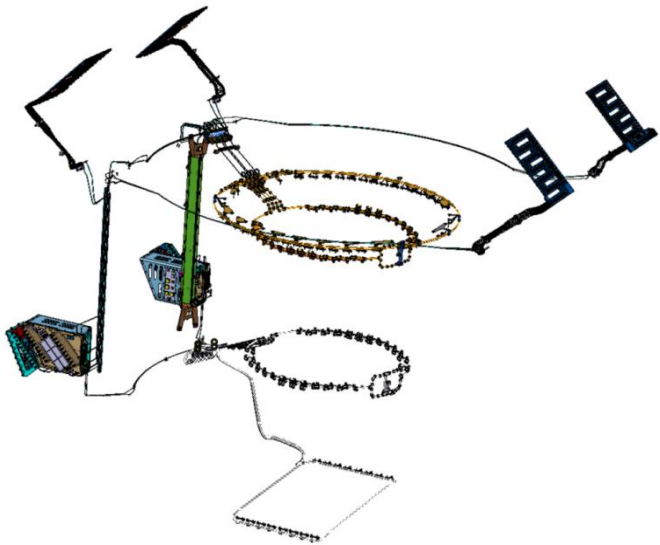

660

661    Fig. 2 The 3D model graph of the two redundant TTCS loops with all the other AMS parts hidden.

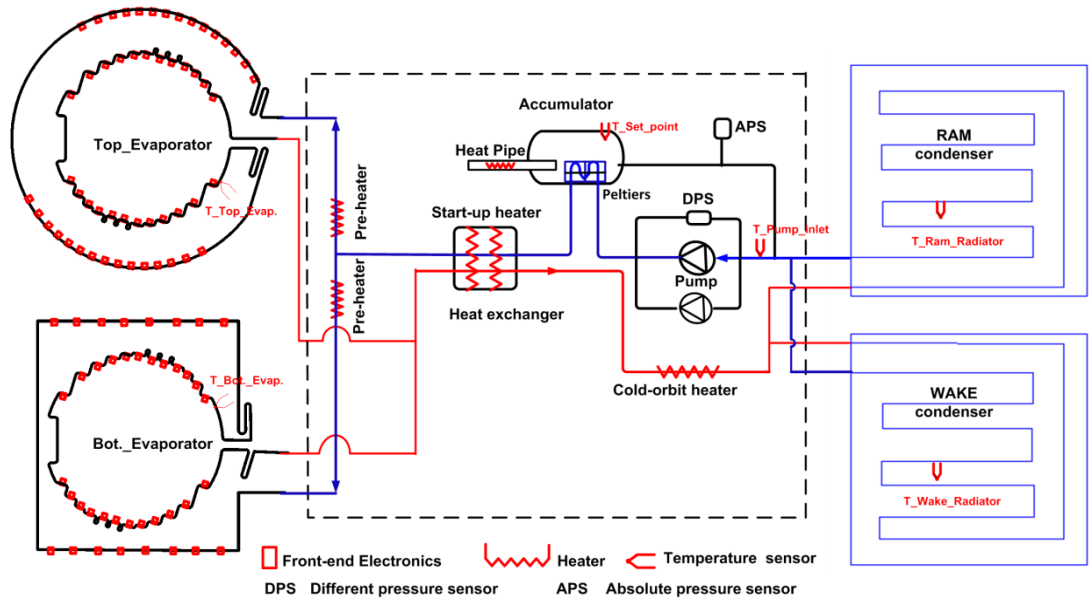

662

663  Fig. 3 The schematic diagram of the upgraded TTCS design, with the only change of bottom

664  evaporator, which differs from the original design in heat load (-19.5W), outer ring's shape (square),

665  and position.

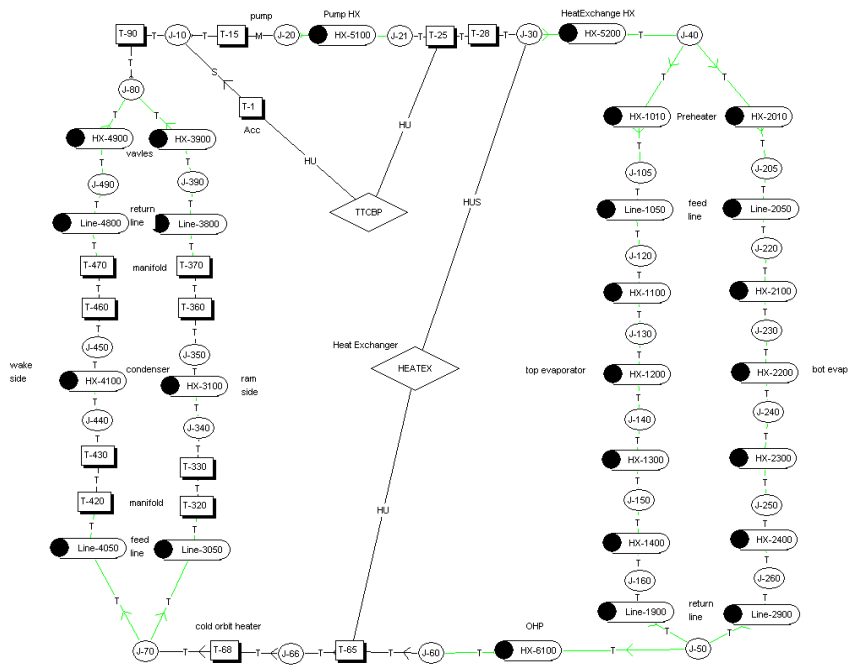

666

667  Fig. 4 The schematic SINDA/FLUINT model (fluid part) of the TTCS.

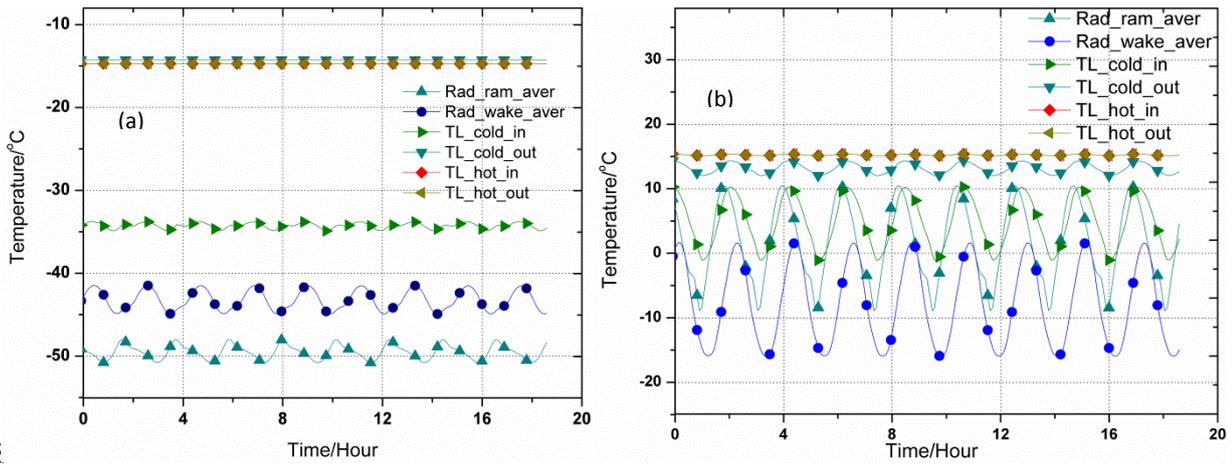

668

669 Fig. 5 The variation of the inlet and the outlet temperatures of the heat exchanger with the that of

670 orbital heat flux; (a) Cold case (-75,+0,+0,-15) , FR=2 g/s, Tset=-15 $^{\circ}$C; (b) hot case (+75,-15,+0,-15),

671 FR=3 g/s, Tset=15 $^{\circ}$C. Label TL stands for the temperature of the fluid at the inlet and the outlet of

672 the heat exchanger; Label Rad stands for the average temperature of the radiator out plate.

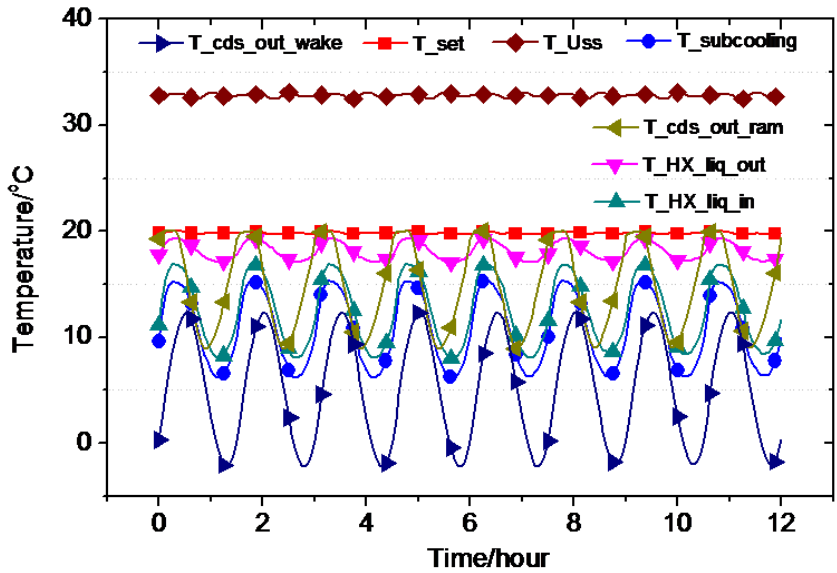

673

674 Fig. 6 The variation of the temperatures of the primary loop in the case (+75-15-20-15), 4g/s, and at

675 20 $^{\circ}$C.

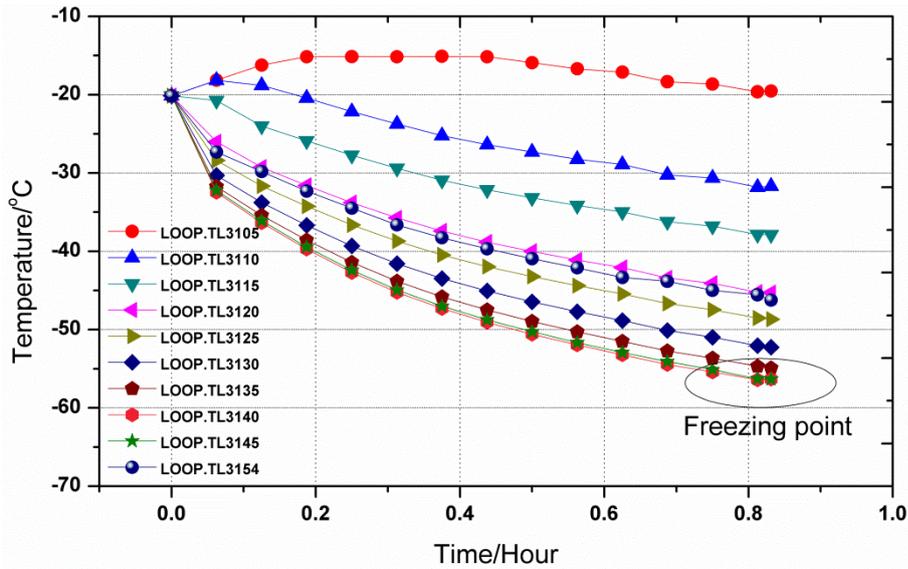

676

677 Fig. 7 The variation of temperatures of the condenser, indicating that the CO2 should freeze after 0.8

678 hour at Tset=258 K in a cold case (-75+0+0-15). Labels TL#### stand for the temperature of the

679 condenser lumps in the SINDA model, and the numbers following TL are the lump numbers. The

680 frozen point of CO2 is -56 $^{\circ}$C.

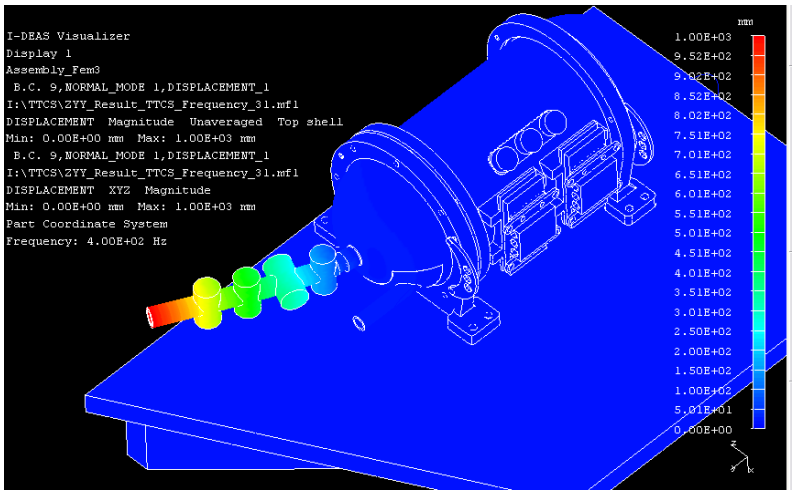

681

682 Fig. 8  The modal shape of the TTCS accumulator.

(a)

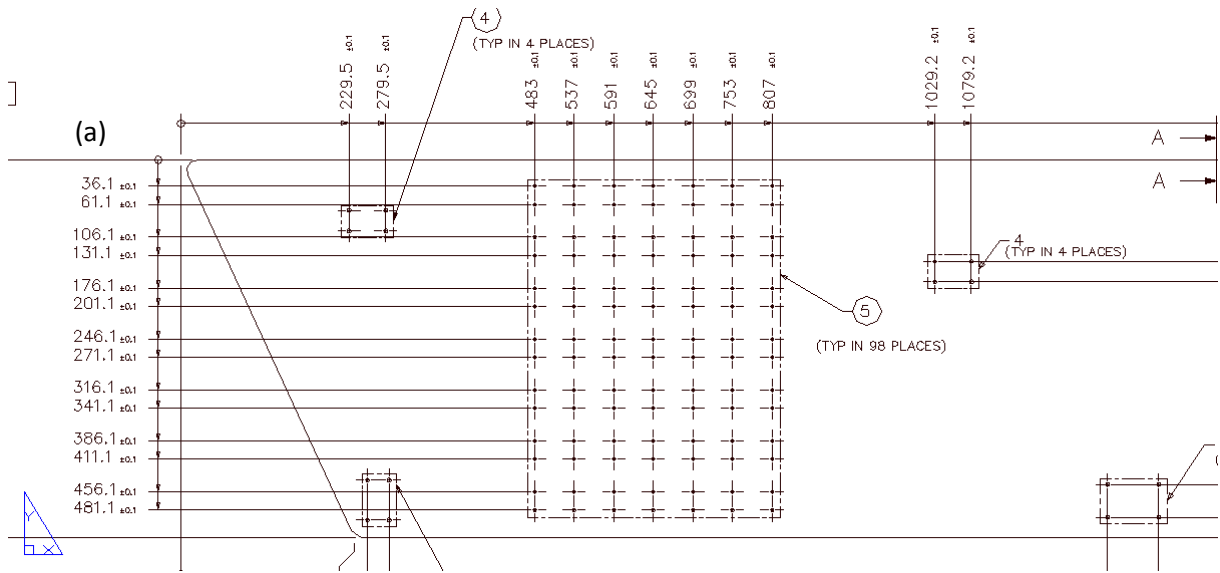

683

(b)

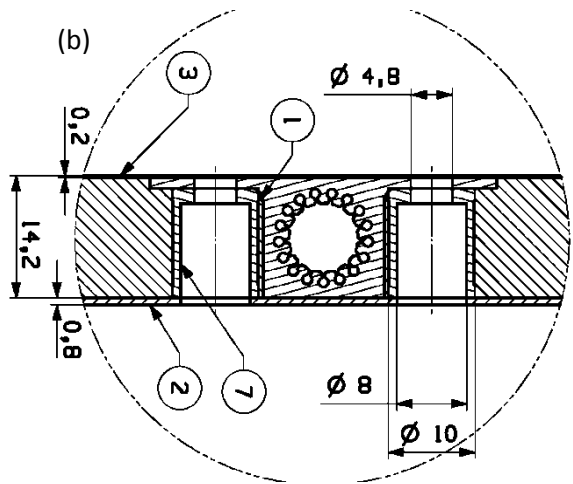

684

685    Fig. 9  (a) The radiator design drawing; (b) the radiator condenser interface.

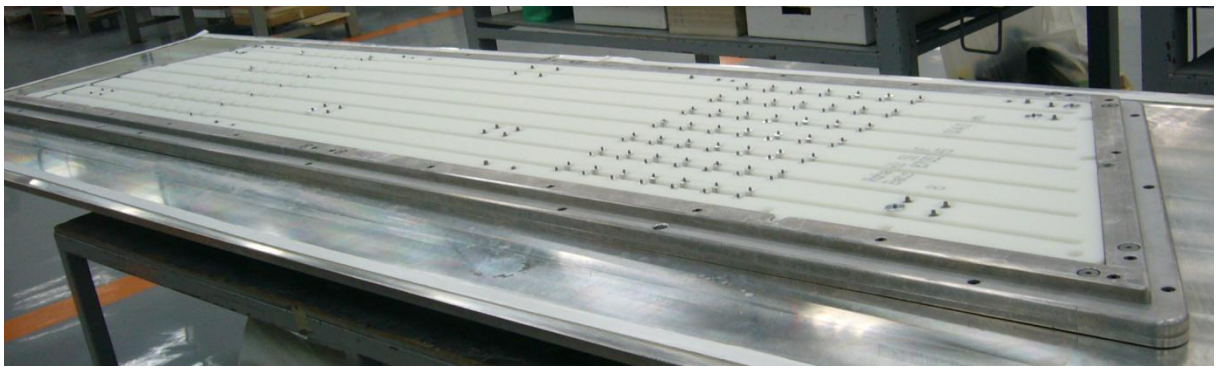

686

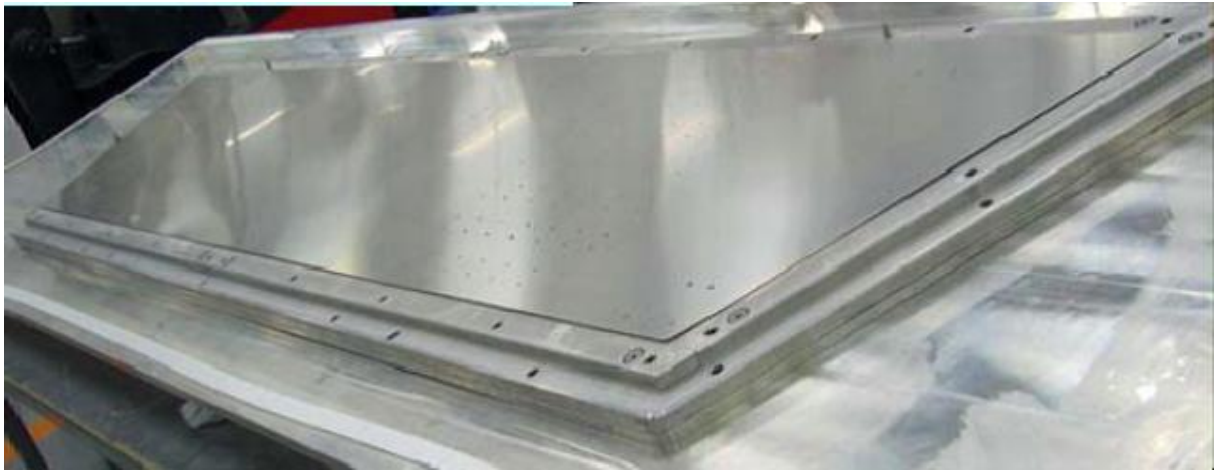

687

688    Fig. 10  A Tracker radiator, before and after the top plate was covered.

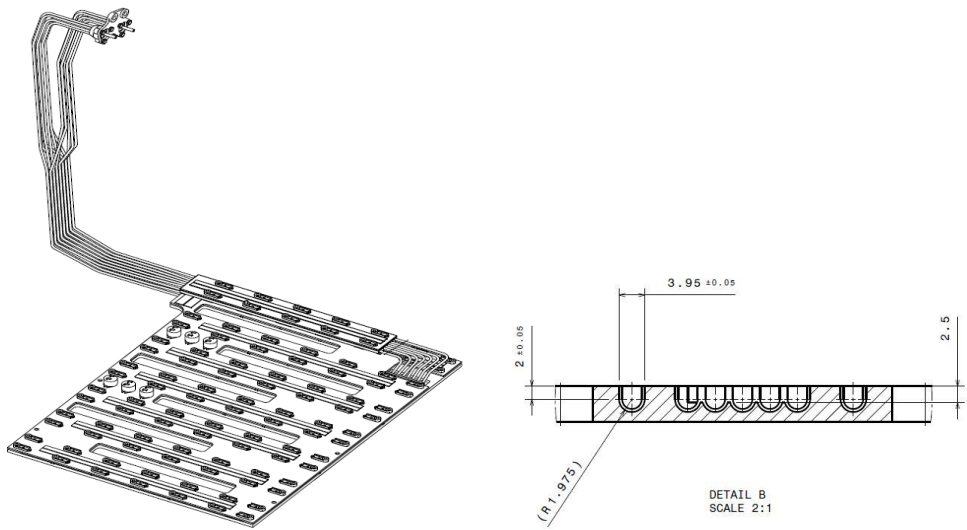

689

690    Fig. 11  3D design drawing of the condenser (a). The cross-section of the condenser plate (b).

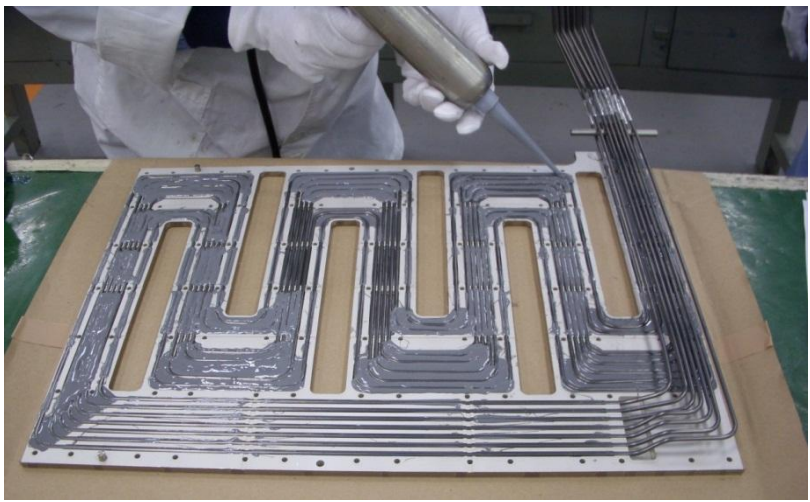

691

692    Fig. 12 A condenser under construction.

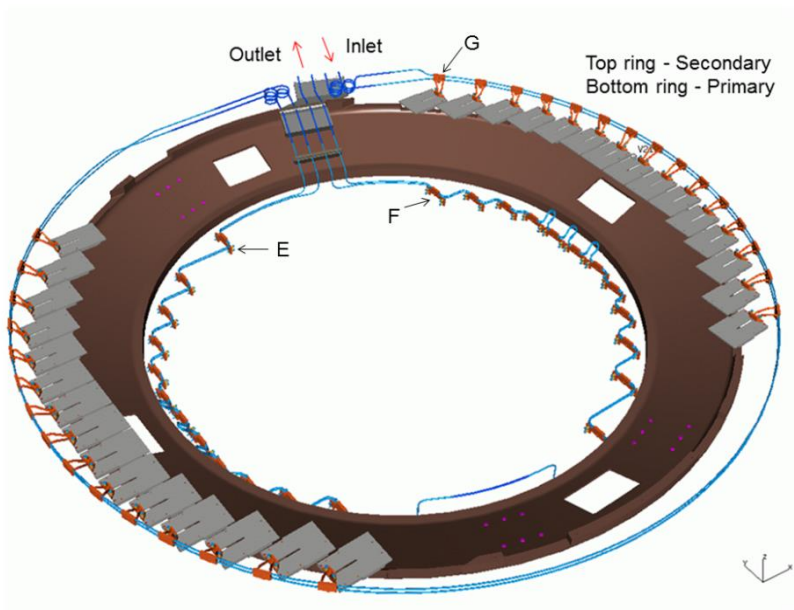

693

694    Fig. 13  The 3D design drawing of the top evaporator. It consists of two redundant evaporators (blue

695    lines) connected with the Tracker front-end electronics. The hybrids were thermally connected to

696    the evaporator outer rings by cooper braids. G, E and F are temperature sensors along the

697    evaporator.

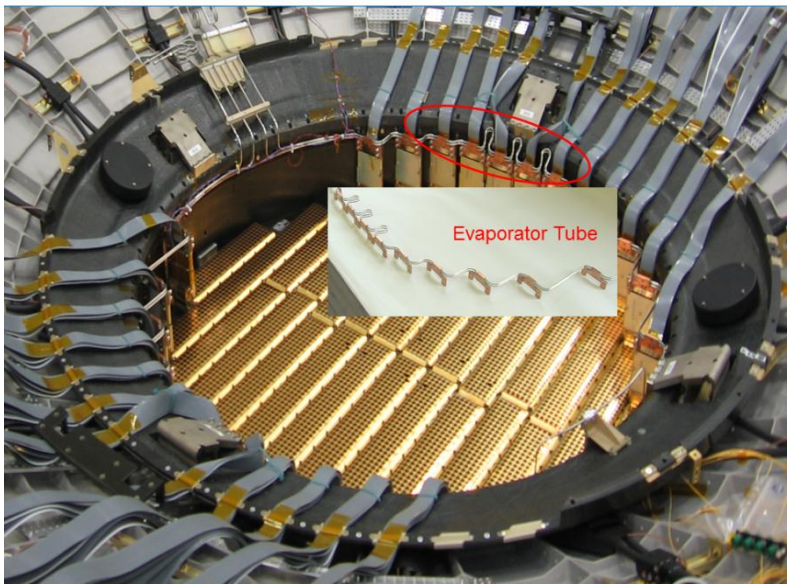

698

699    Fig. 14  The top evaporator inner rings that have been integrated to the AMS-Tracker. The copper

700    soldered on the evaporator tubes (the fine metal tubes) thermally connects with the front-end

701    electronics hybrids (vertical blocks in the photo) through carbon fiber bars (not shown in the photo).

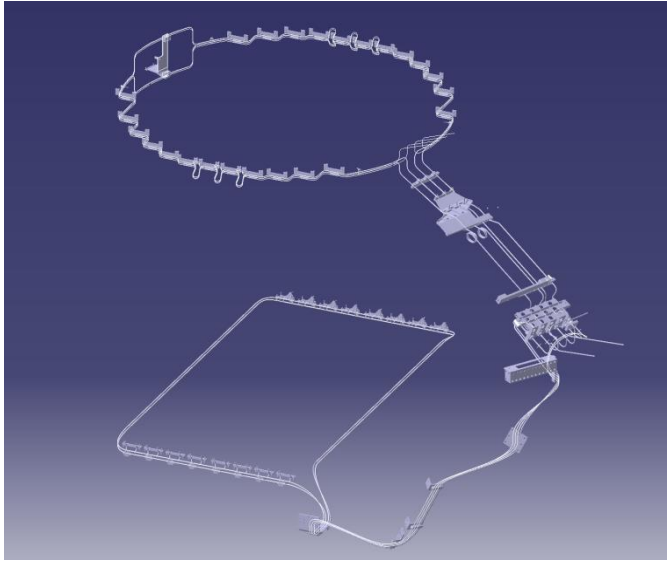

702

703    Fig. 15  The 3D design drawing of the bottom evaporator, the only upgraded part of the TTCS.

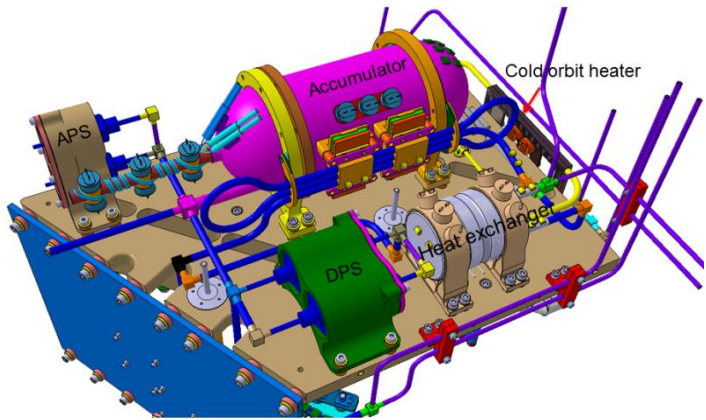

704

705    Fig. 16  The TTCB assembly drawing.

706 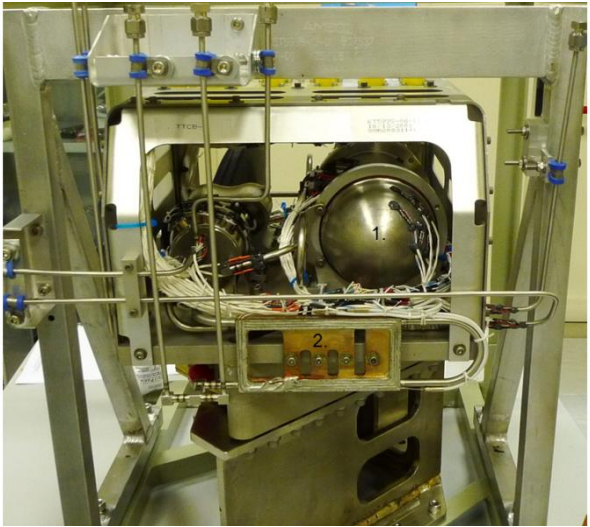

707 Fig. 17 The integrated TTCB QM. ① Accumulator; ② Cold-orbit heaters.

708 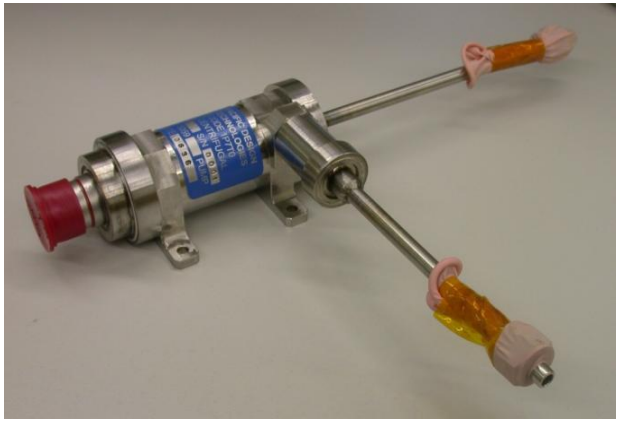

709 Fig. 18 The centrifugal pump produced by PDT.

710 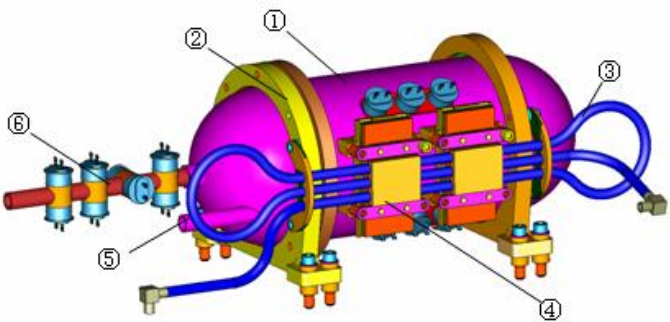

711 Fig. 19 3D design drawing of the accumulator; ①accumulator vessel, ②braces, ③subcooled $CO_2$ tube,

712 ④Peltier element set,  liquid pipe (that connects the accumulator to the loop, ⑥ accumulator heater

713 heat pipe.

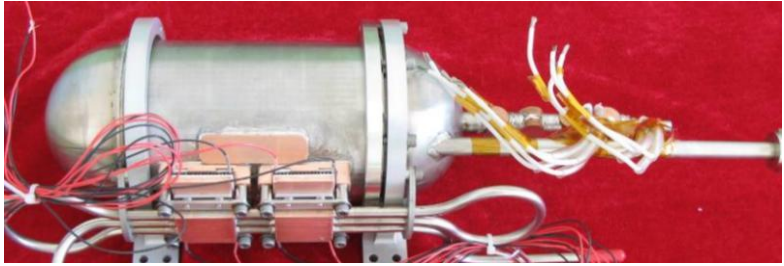

714

715    Fig. 20 A flight model accumulator.

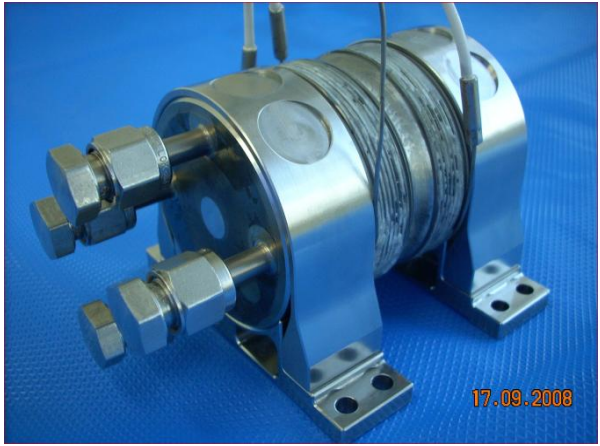

716

717    Fig. 21 The flight model heat exchanger.

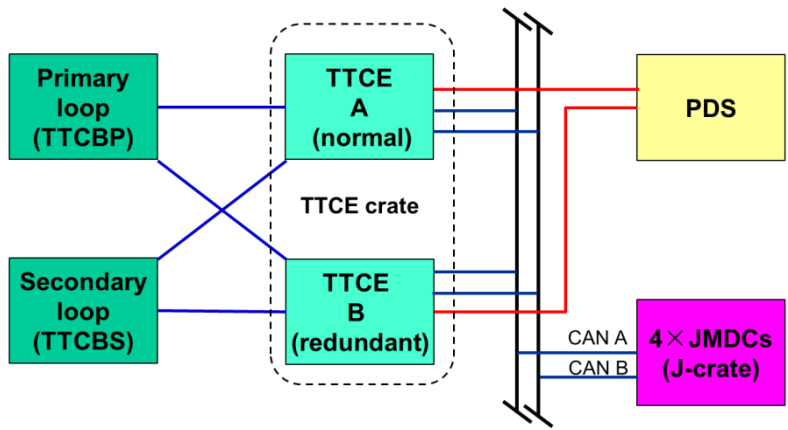

718

719    Fig. 22  The TTCE block diagram

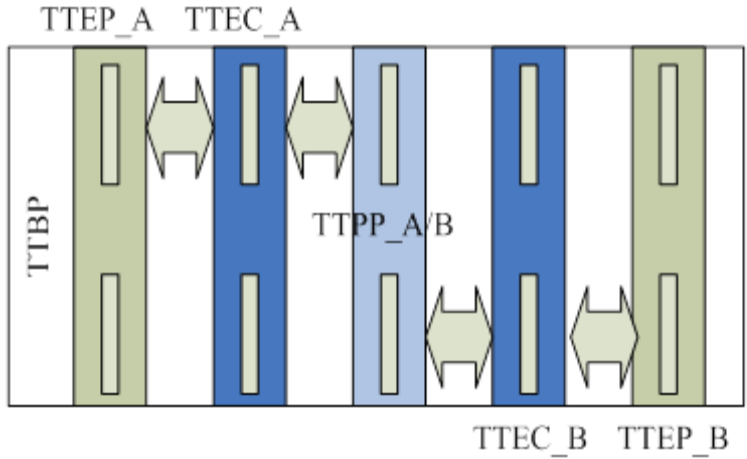

720

721     Fig. 23  The dual redundant block structure inside TTCE crate.

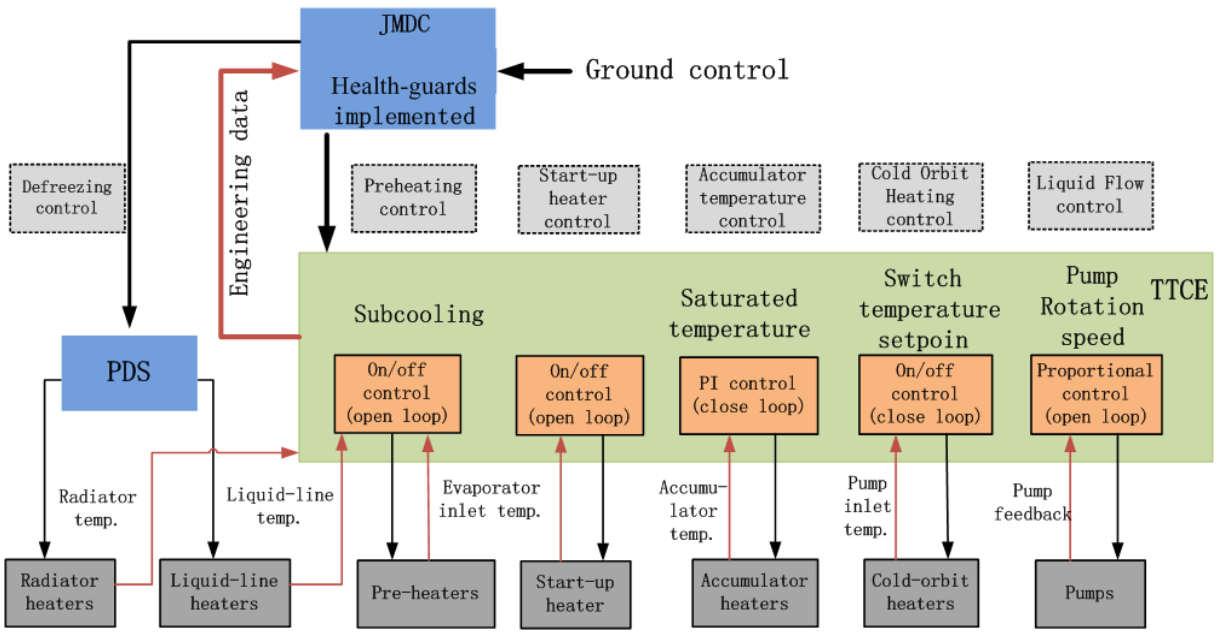

722

723     Fig. 24 The schematic diagram of the TTCS control loops. The liquid flow control, the start-up heater

724     control, and the defreezing control are implemented by the ground commends.

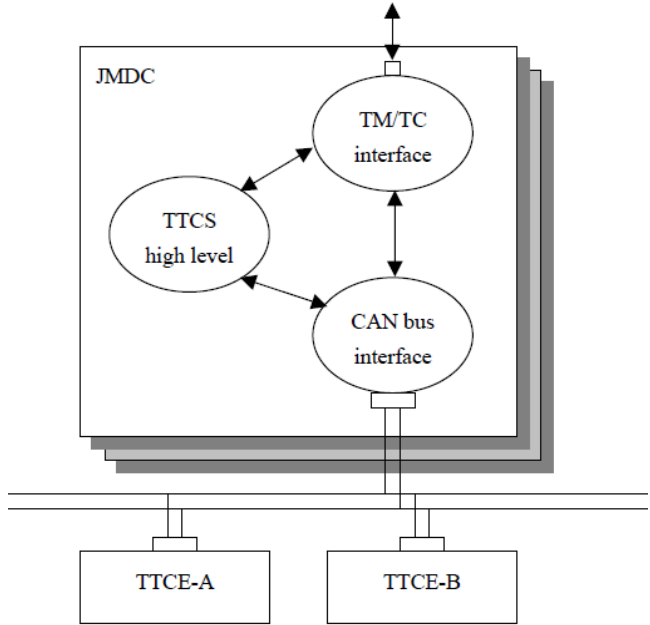

725

Fig. 25 The communication interfaces of the TTCE.

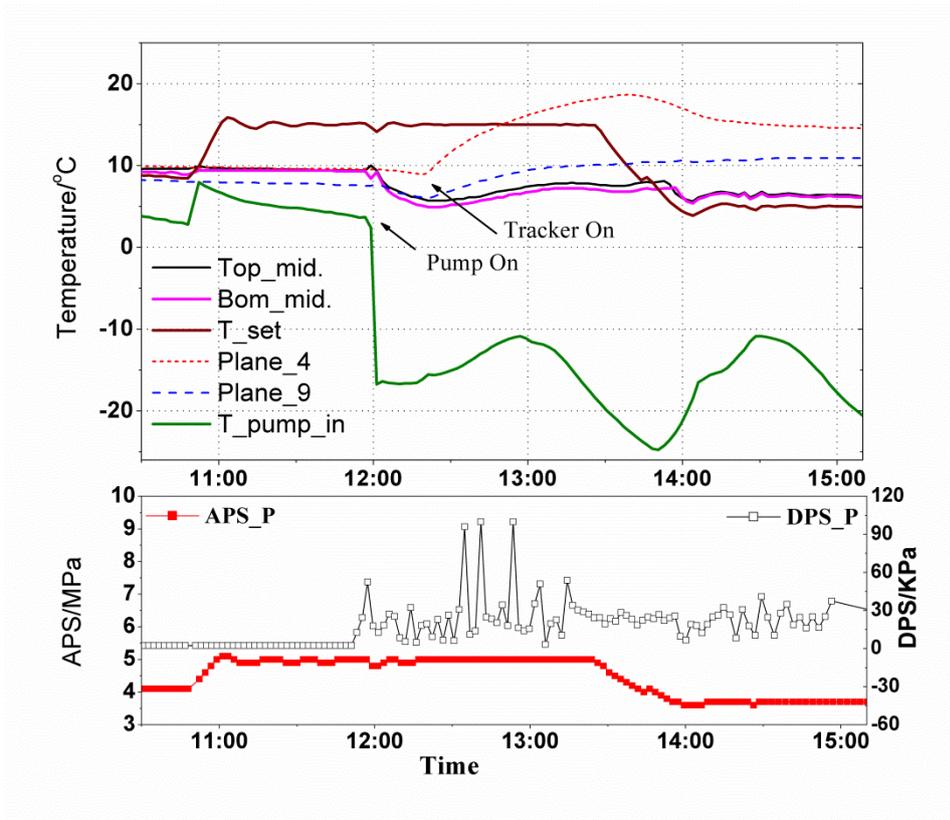

727

Fig. 26 TTCS start-up for operation temperature of 5 $^{\circ}$C, at pump rotation speed of 5000RPM. The noise of the DPS is clearly associated with the appearance of the two-phase flow after the Tracker switched on.

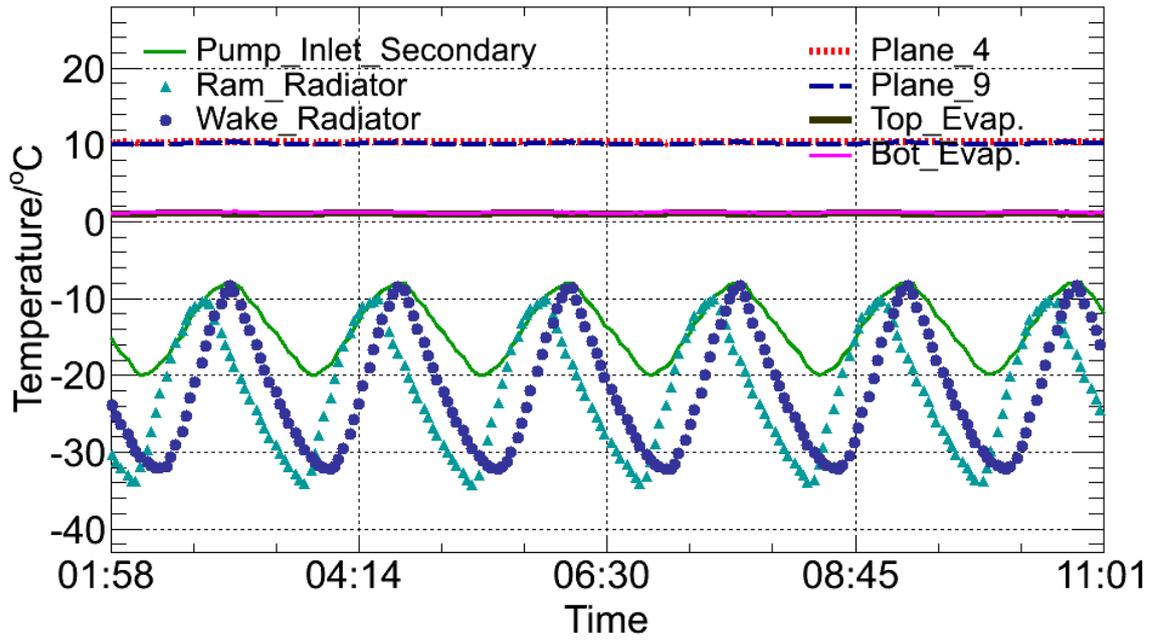

731

732 Fig. 27 Routine performance of the TTCS at the operation temperature of 0 °C and pump rotation

733 speed of 6000 RPM.

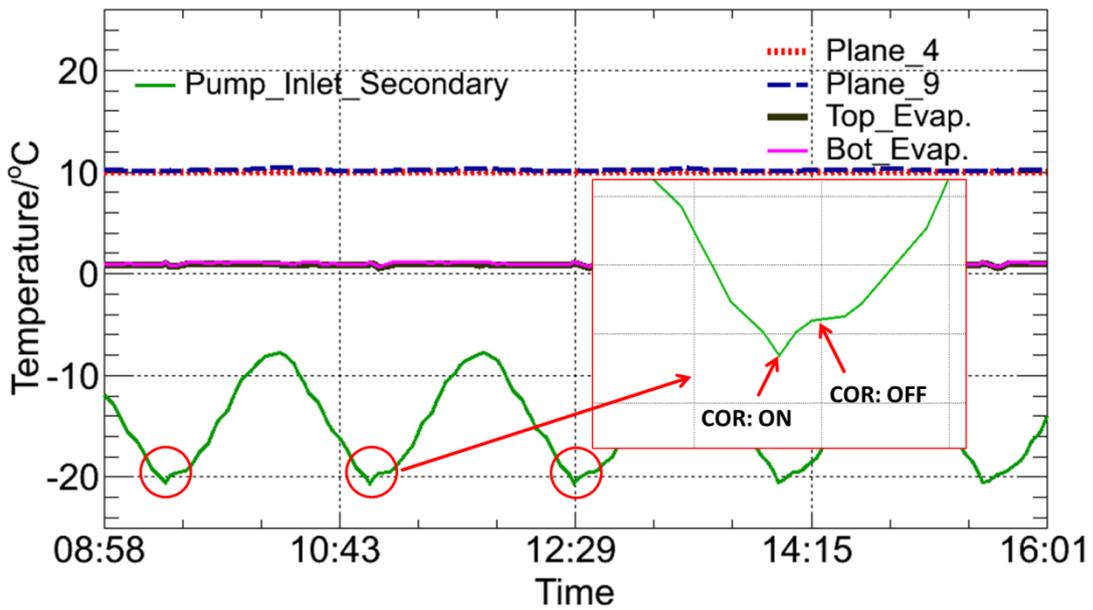

734

735 Fig. 28 Cold-orbit heater auto-function performance at the pump rotation speed of 6000RPM and at

736 the saturation temperature of 0 °C.

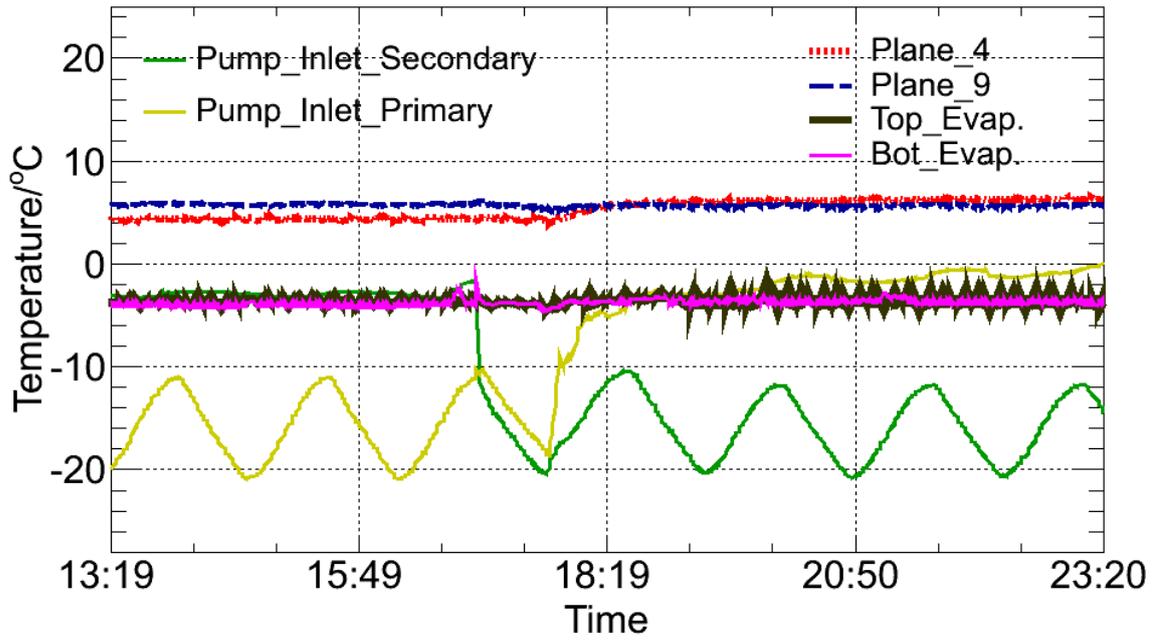

737

Fig. 29 TTCS performance during hot switching from the primary to the secondary loop at the

739   saturation temperature of −5 °C, and the pump rotation speed of 5000 RPM.

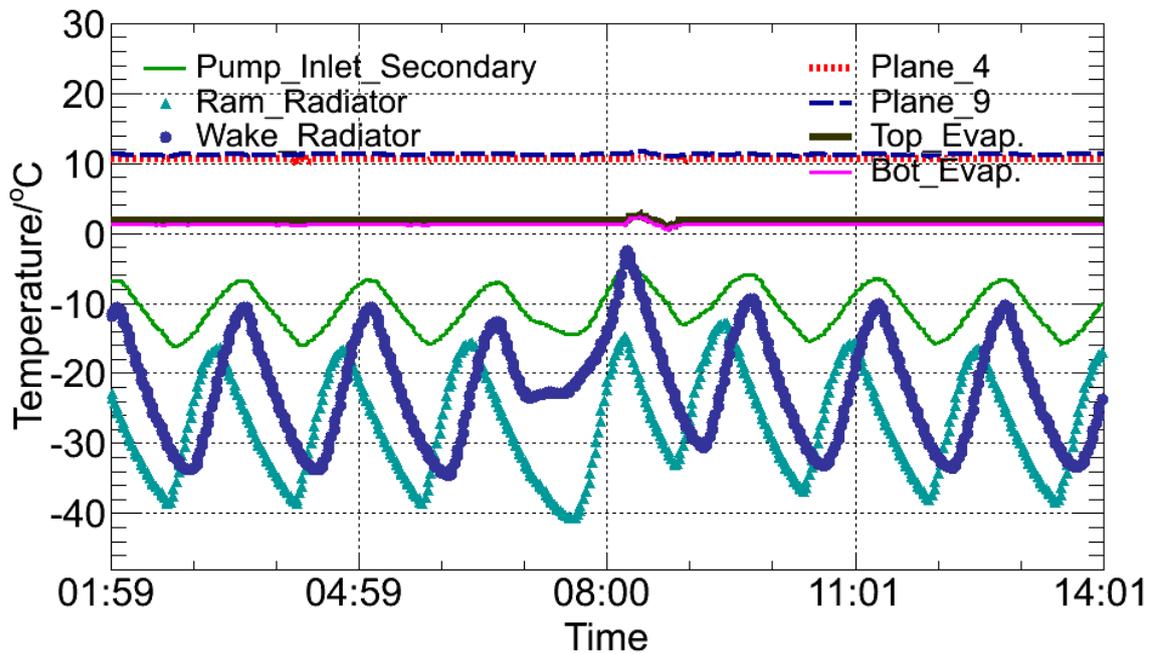

740

Fig. 30 TTCS performance during the undocking of the Soyuz (42P) with the attitude change from

742   (+356.000, +355.500, +0.700) to (0.000, +90.000, 0.000), and a duration of 107 minutes. The sub-

743   cooling health guard functions as designed.

Table 1 Functions of the TTCS main components.

| Components | Functions |
| --- | --- |
| Evaporator | To collect and transport the heat from the Tracker to the $CO_2$ loop |
| Condenser | To conduct the heat from the loop to the radiators that dissipate the heat to the space |
| Accumulator | To compensate the mass of $CO_2$ in the loop and to control the operation temperature of the system |
| Pump | To provide driving power for the loop |
| Heat-exchanger | To exchange heat between the saturated $CO_2$ and the sub-cooled liquid $CO_2$ |
| Pre-heater | To heat the sub-cooled liquid $CO_2$ into saturated state before flowing into the evaporator |
| Start-up heater | To avoid sub-cooled liquid $CO_2$ flow into the Tracker to damage the electronics at extremely cold environmental condition |
| Cold orbit heater | To prevent the $CO_2$ from being cooled to the freezing point by effectively heating the fluid |
| De-freezing heater | To defreeze the solid $CO_2$ inside the tubes from the manifolds to the condensers |
| Radiator heaters | To defreeze the solid $CO_2$ inside the condensers mounted on the radiators |
| Component box | To assemble the components except for the evaporators and the condensers |

Table 2 The Beta and Euler angles range of the ISS.

| Angle | Variation Range |
| --- | --- |
| Beta angle | −75° to +75° |
| Yaw (Z) attitude angle | −15° to +15° |
| Roll (X) attitude angle | −15° to +15° |
| Pitch(Y) attitude angle | 0° to +25° with docked STS |
| | −20° to +15° with undocked STS |

Table 3  Summary of the minimum MofS for all the load cases.

| Components | Static:MofS | | Thermal:MofSs | |
|---|---|---|---|---|
| | Yield | Ultimate | Yield | Ultimate |
| Accumulator | 0.03 | 0.11 | 0.02 | 0.02 |
| Fixed Bracket & Clamp Collar&Wedge | 1.10 | 2.11 | 1.00 | 1.82 |
| Sliding Bracket | 4.01 | 3.56 | 2.76 | 2.95 |
| Heat Pipe | 1.05 | 2.46 | 0.92 | 1.14 |
| Peltier Fixed | 2.07 | 2.26 | 0.51 | 1.20 |
| TS & Peltier Heat Exchanger & Peltier heat exchanger press &Spring Support | 0.61 | 1.13 | 0.68 | 1.22 |
| Joints | Static: bolt MofS | | Thermal: bolt MofS | |
| Accu.Bracket Clamp&Collar Bolt | 0.12 | | 0.09 | |
| PipeFix&Clamp Bolt | 0.123 | | 0.122 | |
| Press&Saddle Bolt | 0.107 | | 0.106 | |

Table 4  Summary of the minmum MofS for all load cases (Fail-safe)

| Components | Static: MofS | | Thermal: MofS | |
| --- | --- | --- | --- | --- |
| | Yield | Ultimate | Yield | Ultimate |
| Accumulator | 0.54 | 1.80 | 0.53 | 1.55 |
| Fixed Bracket & Clamp Collar&Wedge | 1.61 | 5.17 | 1.49 | 4.66 |
| Sliding Bracket | 6.91 | 8.85 | 3.68 | 6.92 |
| Heat Pipe | 1.07 | 9.19 | 1.87 | 7.31 |
| Peltier Fixed | 4.94 | 16.41 | 0.76 | 3.17 |
| TS & Peltier Heat Exchanger & Peltier heat exchanger press & Spring Support | 0.98 | 3.20 | 0.24 | 2.64 |
| Joints | Static: bolt MofS | | Thermal:bolt MofS | |
| Accumulator.Bracket Clamp&Collar Bolt | 0.34 | | 0.30 | |
| PipeFix&Clamp Bolt | 0.50 | | 0.49 | |
| Press&Saddle Bolt | 0.19 | | 0.20 | |

Table 5  Minimum Margins of Safeties and locations

| Structural | Yield MofS | Ultimate MofS | Location | Bolt MofS | Location |
|---|---|---|---|---|---|
| Normal | 4.97 | 3.35 | Side-plate | 0.047 | Cover/Base-plate |
| Fail-safe | 5.28 | 6.21 | Side-plate | 0.22 | Pump Bracket/Start up radiator |

Nomenclature

| | |
|---|---|
| AHP | Accumulator heat pipe |
| AMS | Alpha Magnetic Spectrometer |
| APS | Absolute pressure sensor |
| CAN (bus) | Controller Area Network |
| DoF | Degree of Freedom |
| DPS | Differential pressure sensor |
| EM | Engineering Model |
| EMC | Electro Magnetic compactibility |
| ESTEC | European Space Technology Centre |
| FEM | Finite element method |
| FM | Flight Model |
| HX | heat exchanger |
| ISS | International Space Station |
| JMDC | Mission Computers |
| MofS | Margin of Safety |
| PDS | Power Distribution System |
| PWM | Pulse width modulation |
| QM | Qualification Model |
| RICH | Ring-imaging Cherenkov (detector) |
| RPM | Revolutions Per Minute |
| STS | Space Shuttle Mission |
| TC | Tele-control |
| TM | Tele-monitor |
| TS | Thermostat |
| TTBP | Tracker Thermal Back Plane |
| TTCB | Tracker Thermal Component Box |
| TTCE | Tracker Thermal Control Electronics |
| TTCS | Tracker Thermal Control System |
| TTEC | Tracker Thermal Electronic Control Board |
| TTEP | Tracker Thermal Electronic Power Board |
| TTPP | Tracker Thermal Pump & Pressure sensors Board |
| TVT | Thermal Vacuum Test |
| USS | Universal Support Structure |

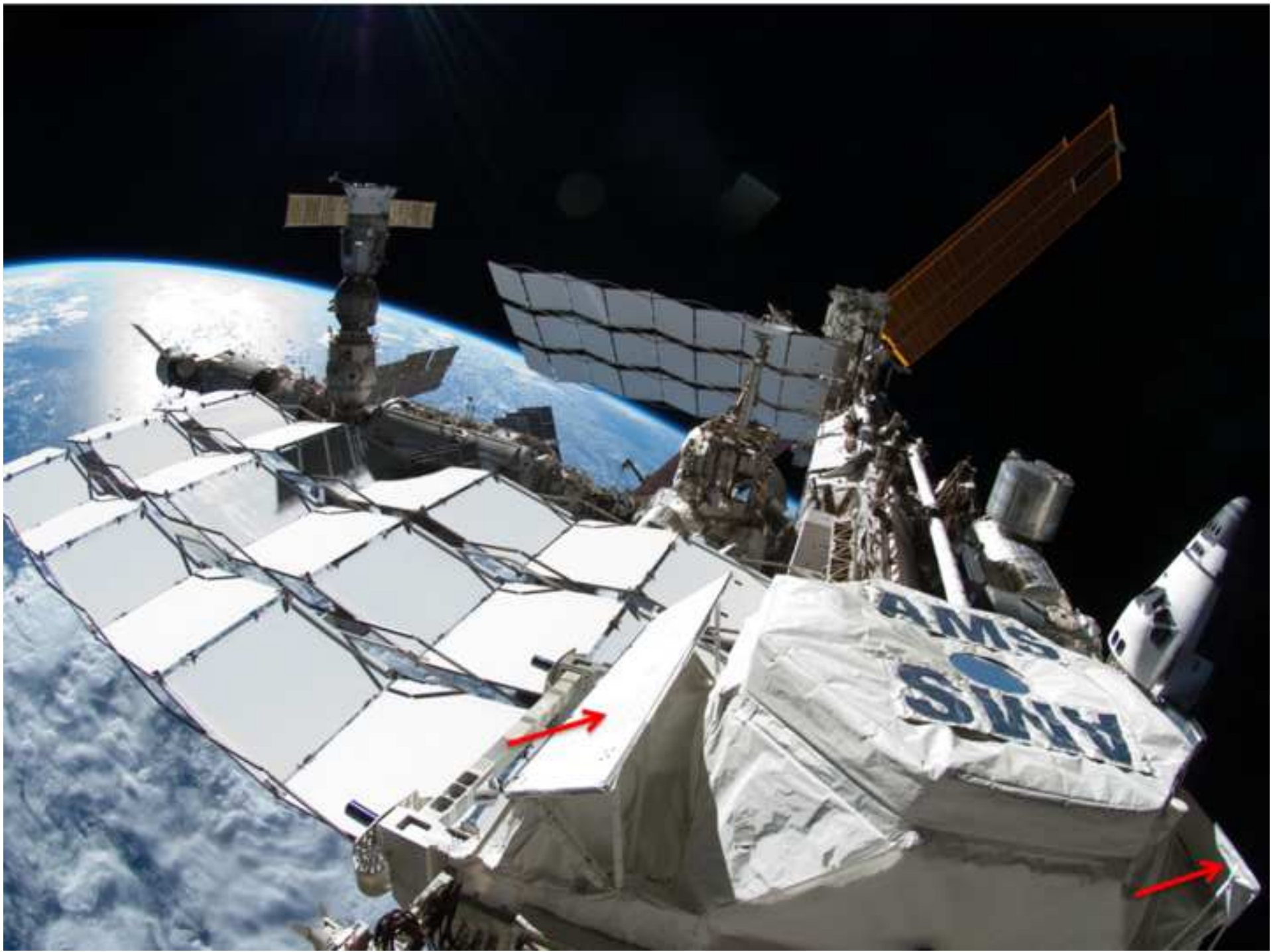

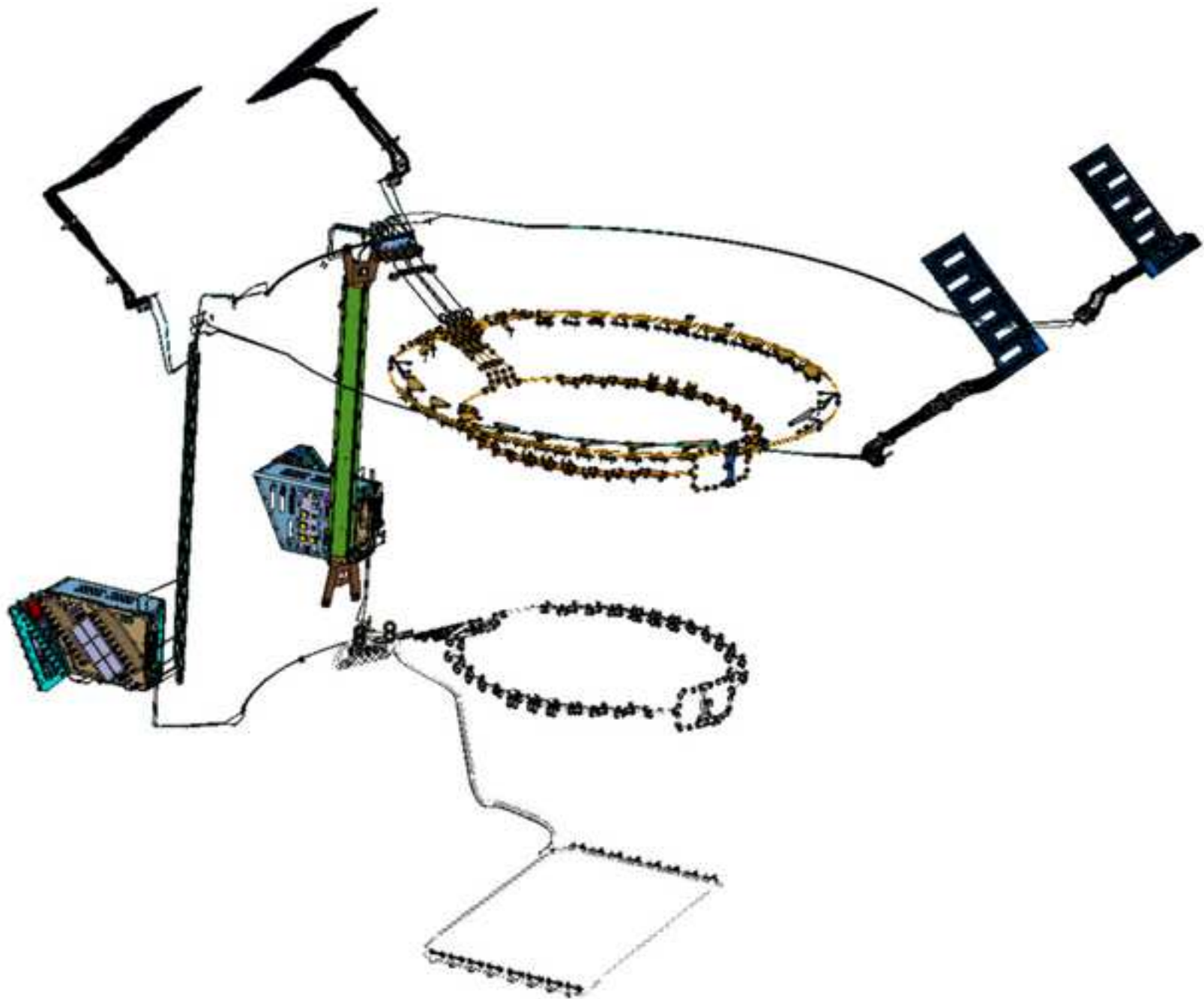

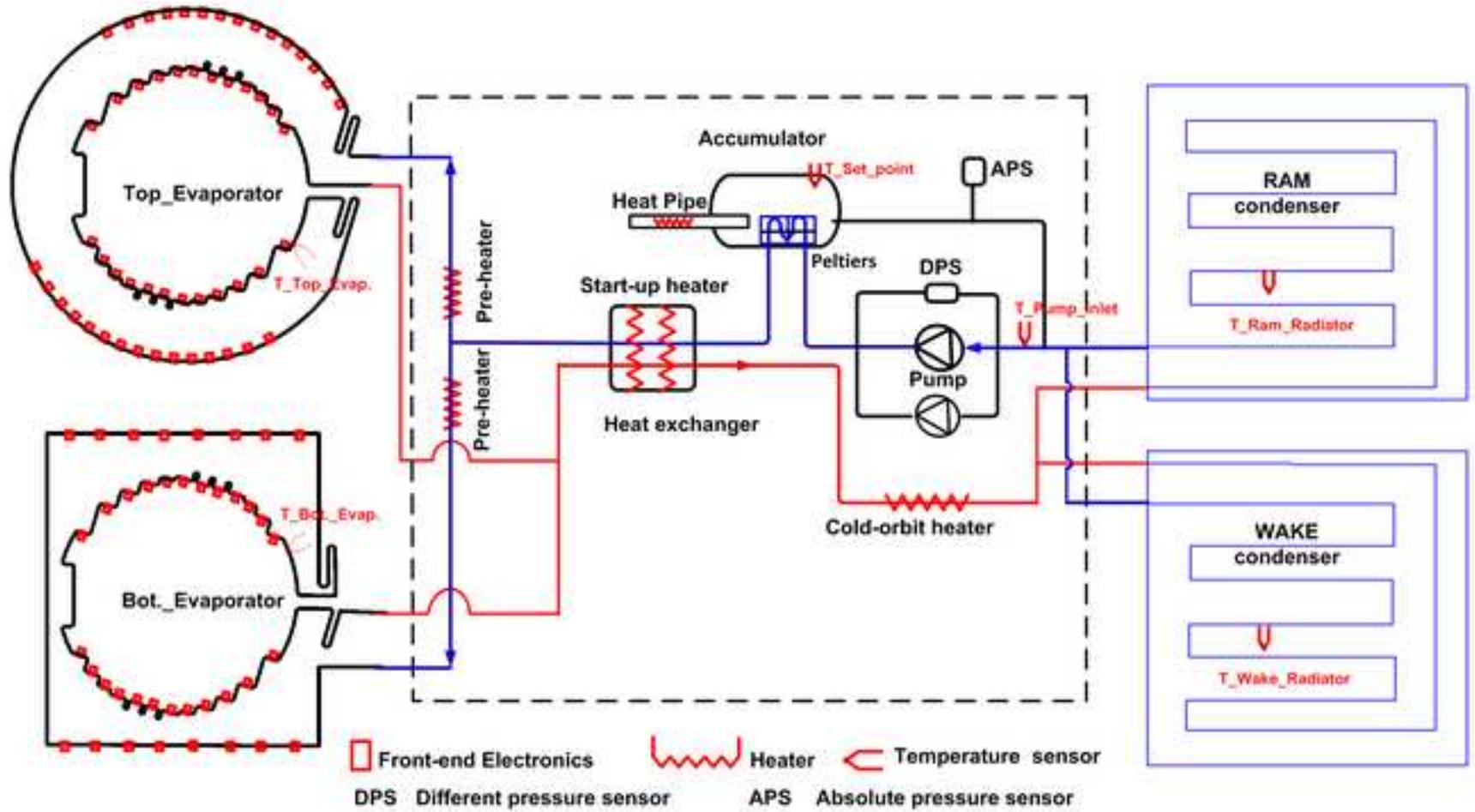

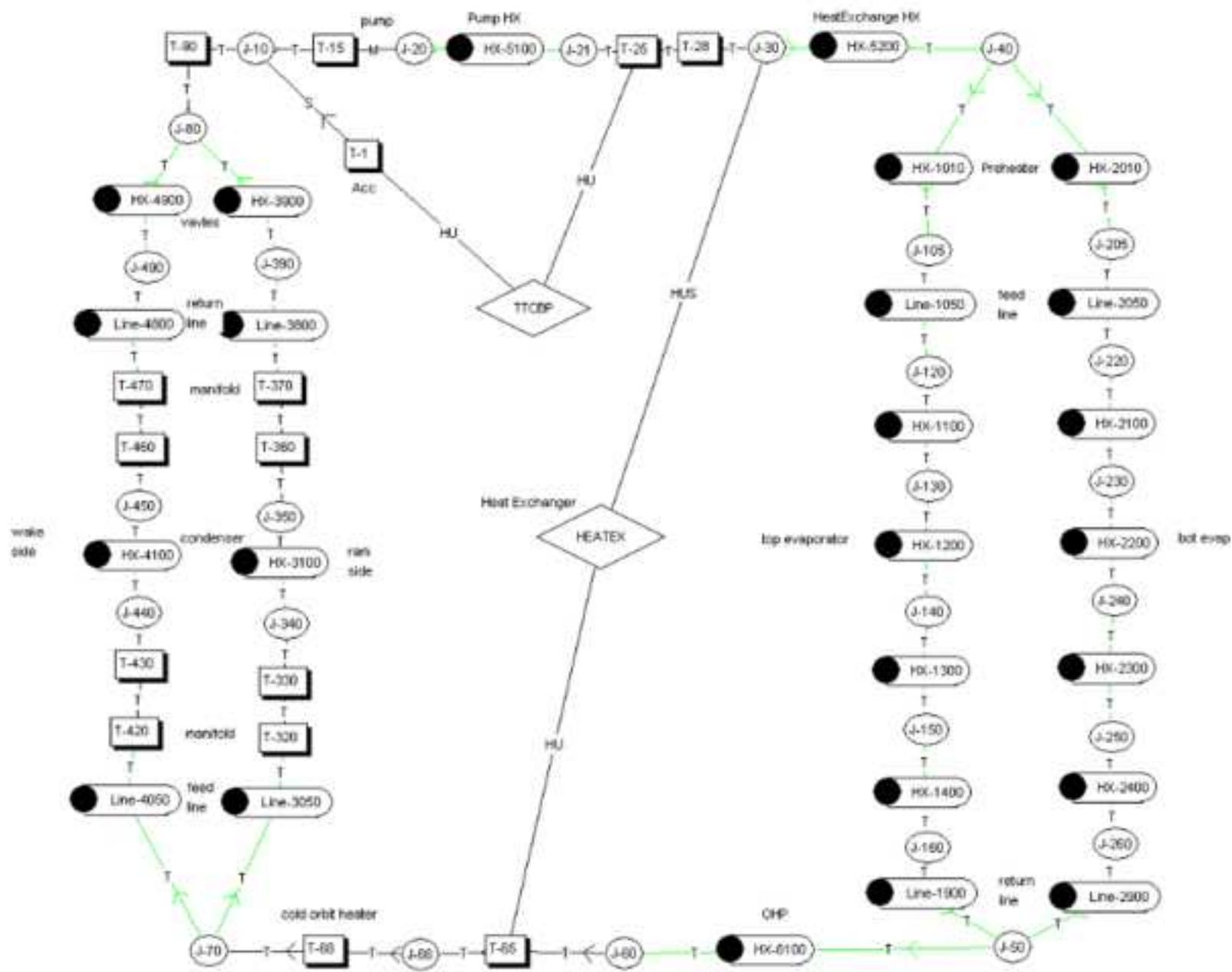

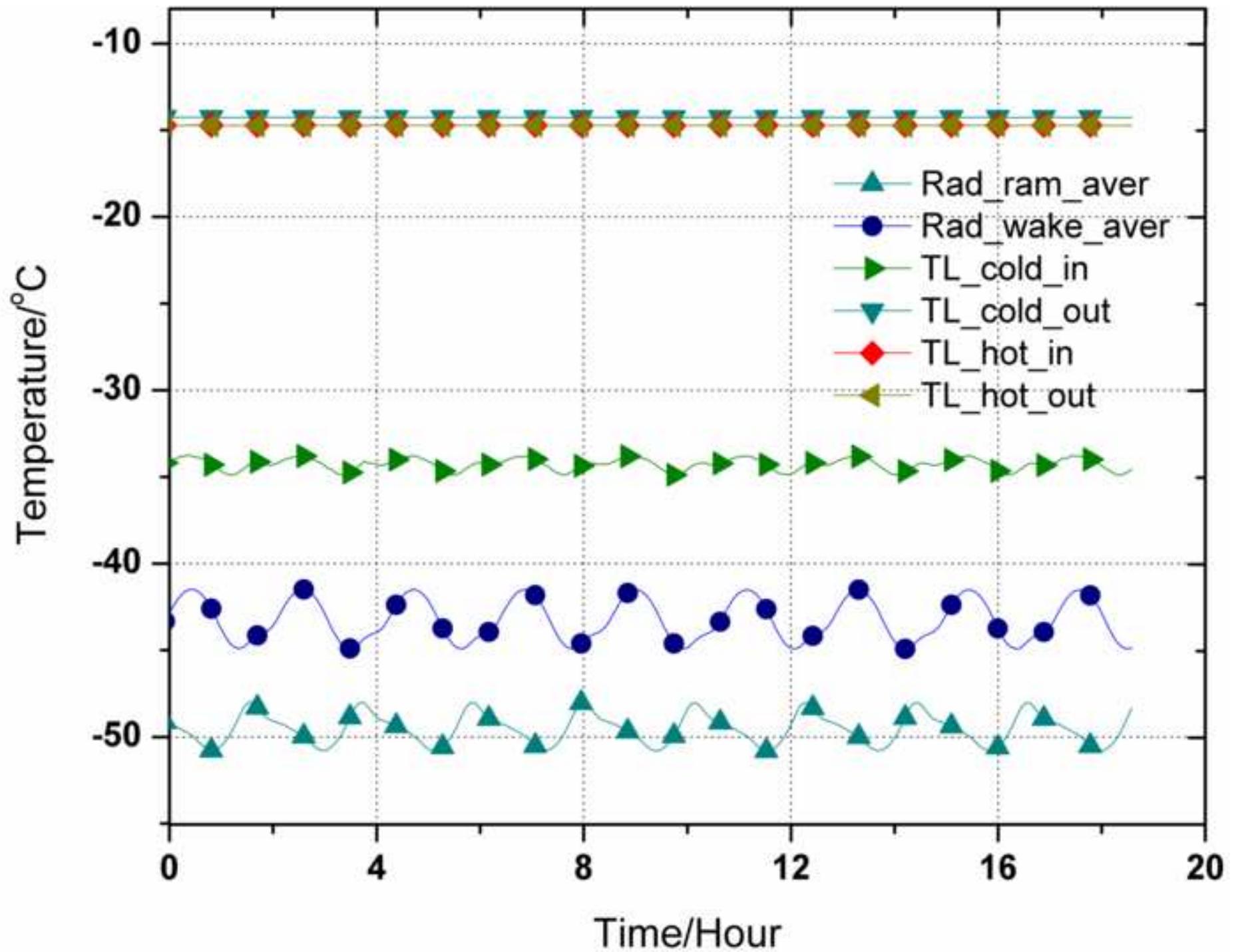

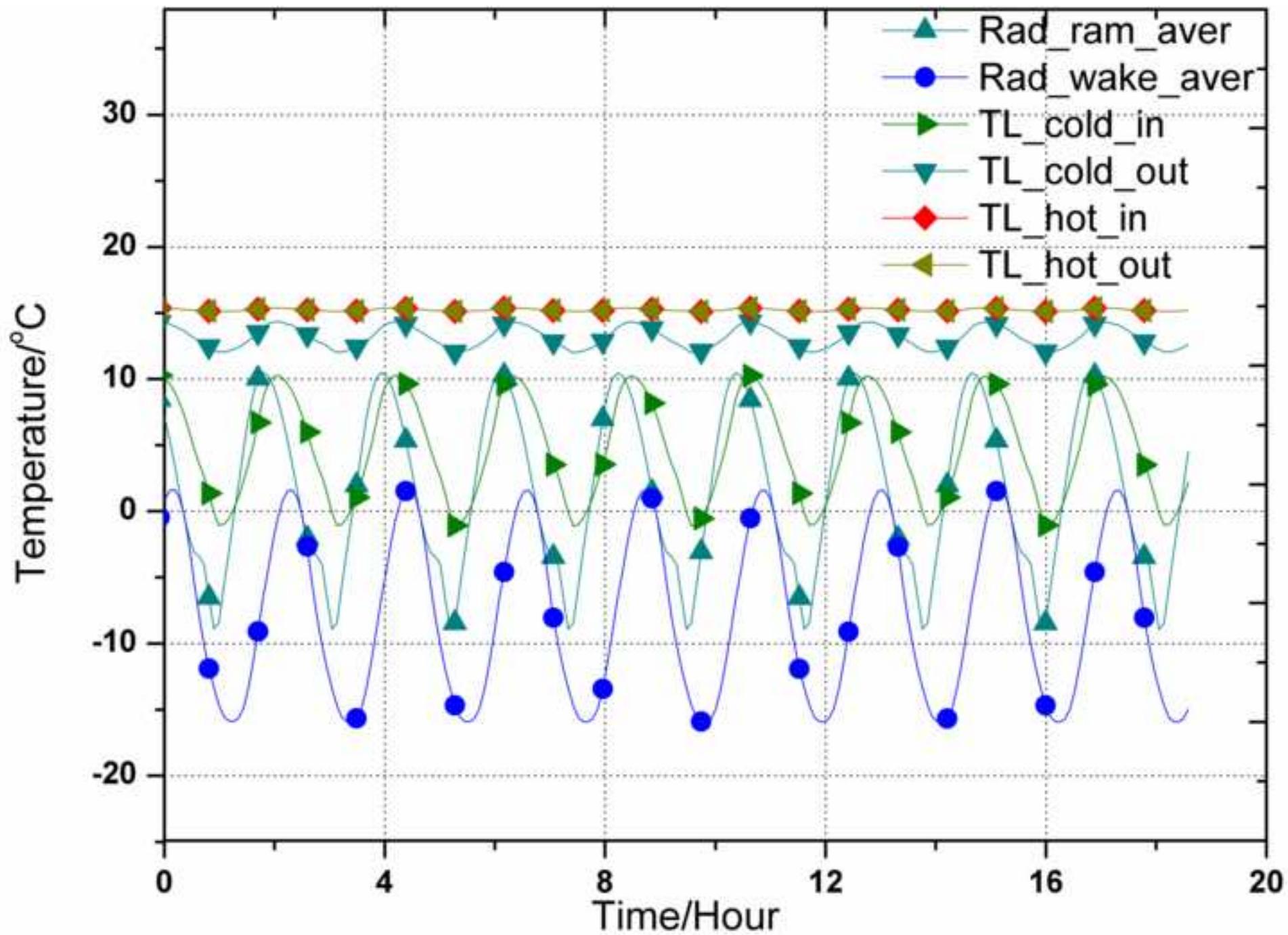

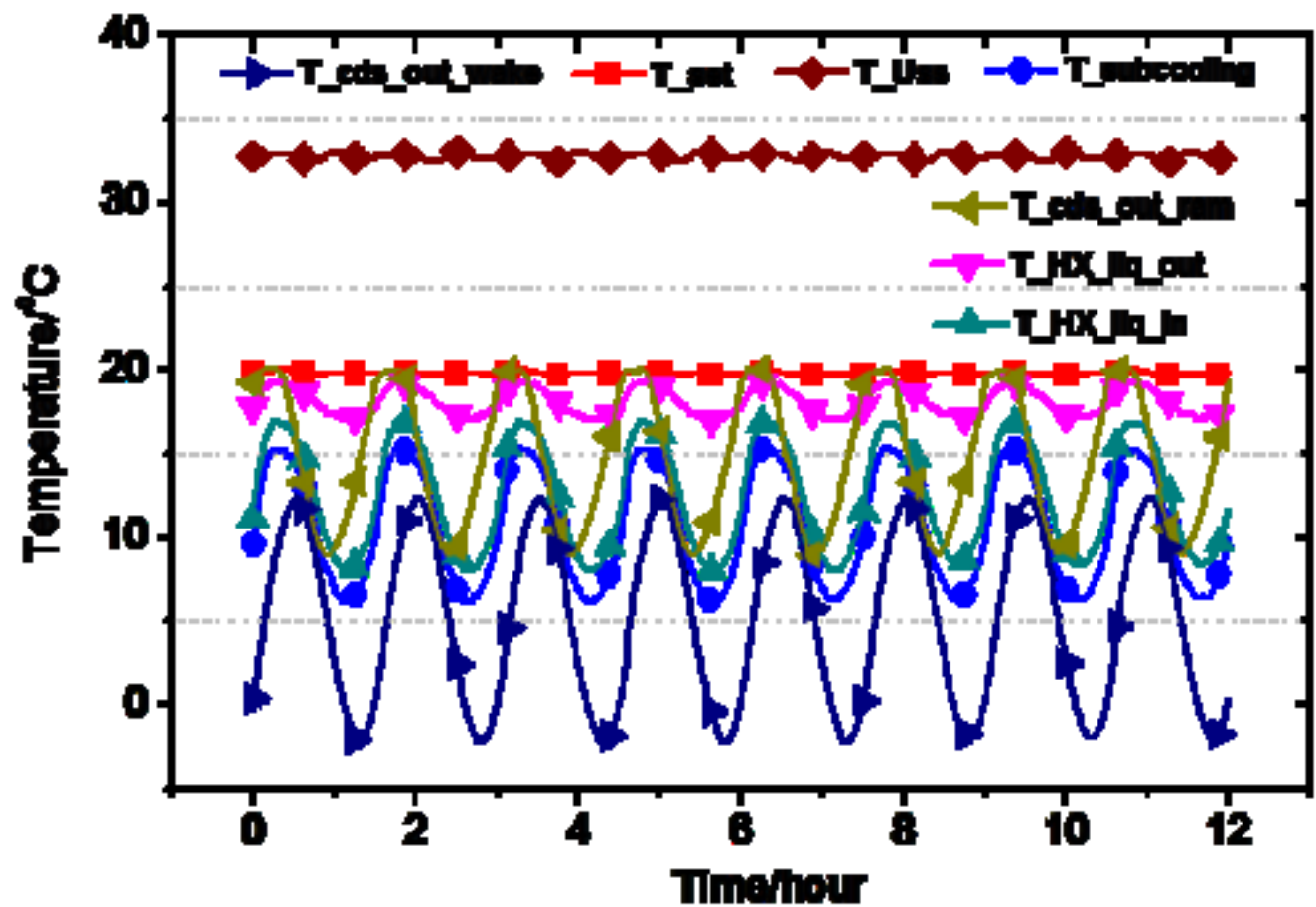

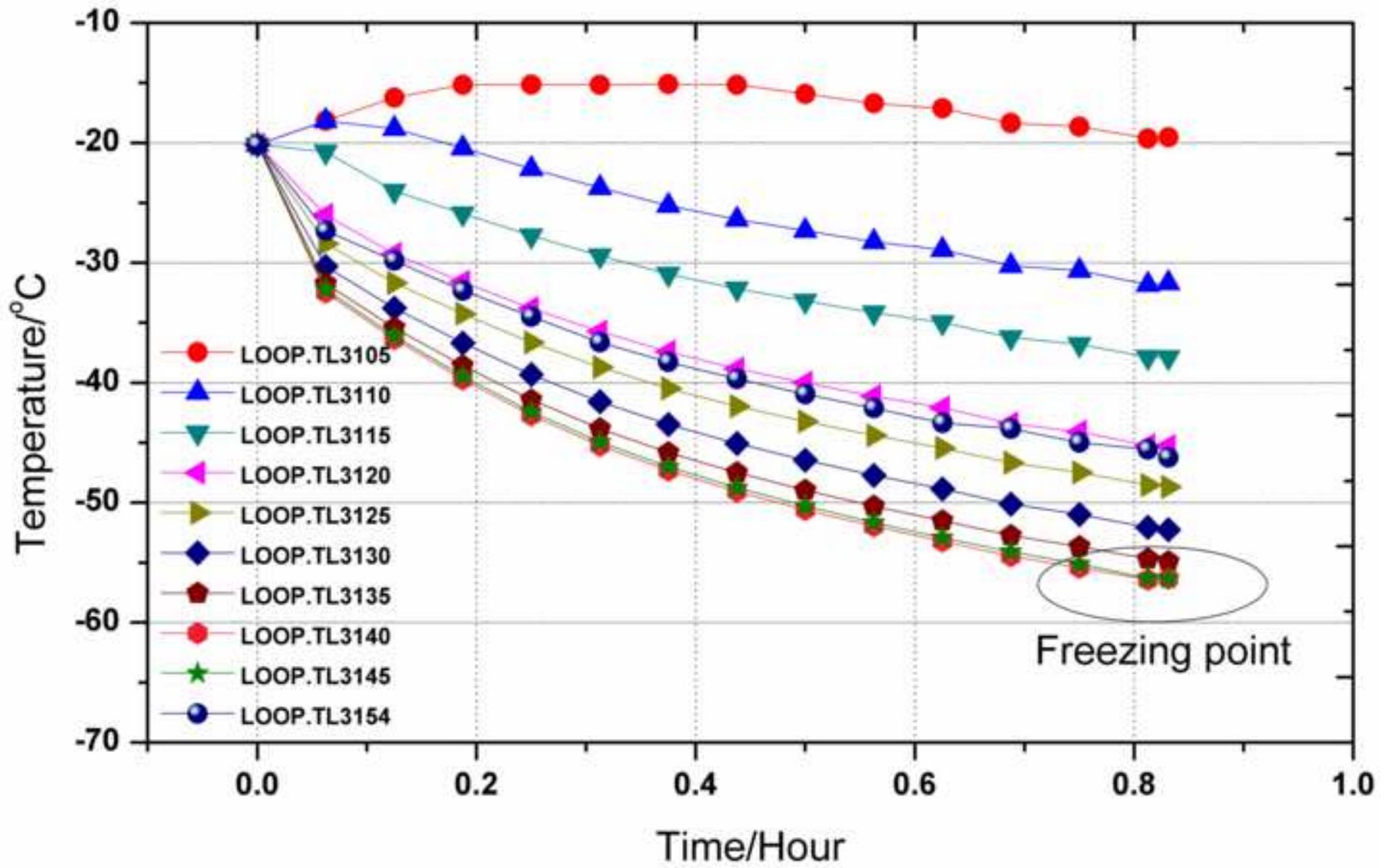

**Figure(s)**
**Click here to download high resolution image**

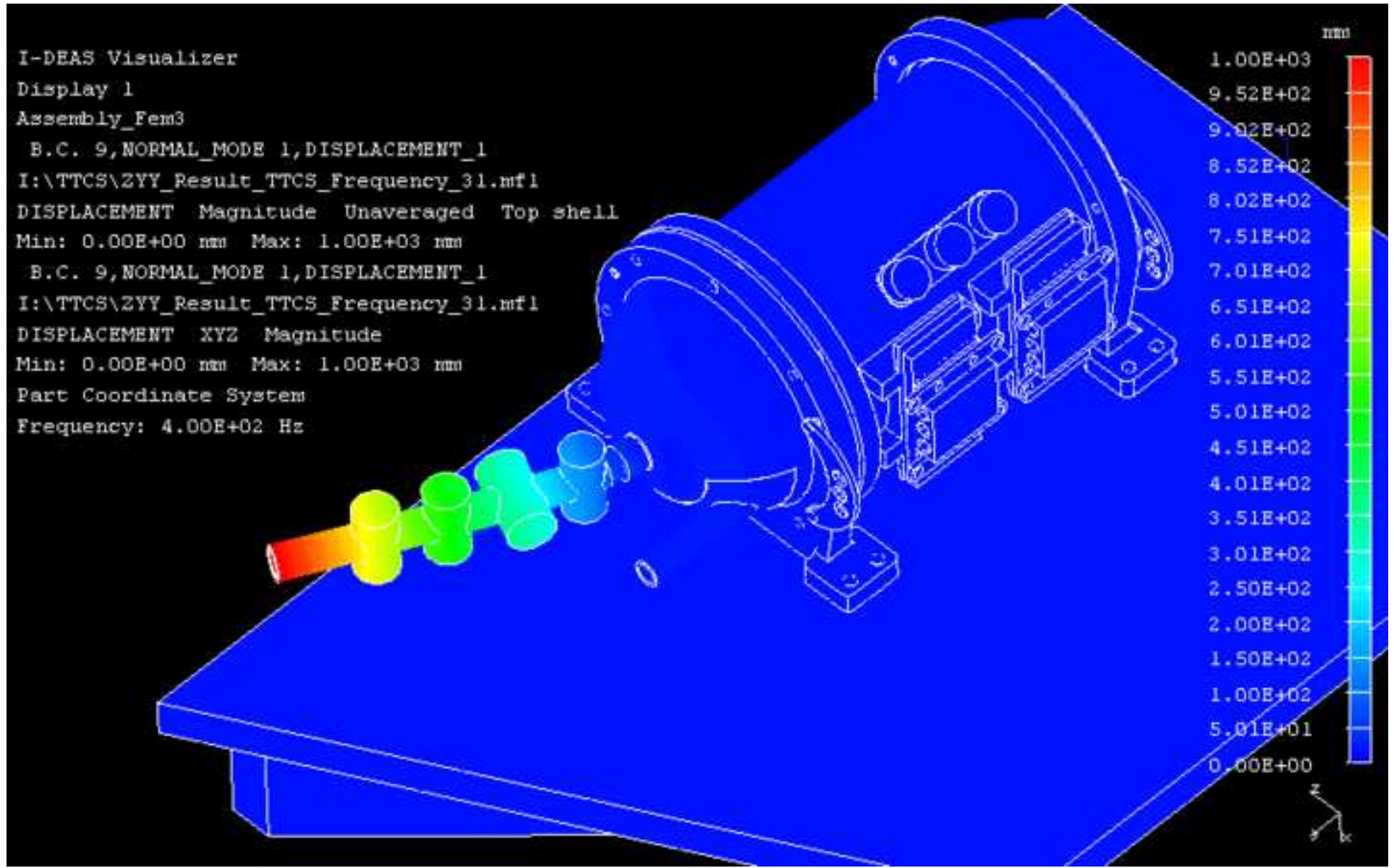

**Figure(s)**
Click here to download high resolution image

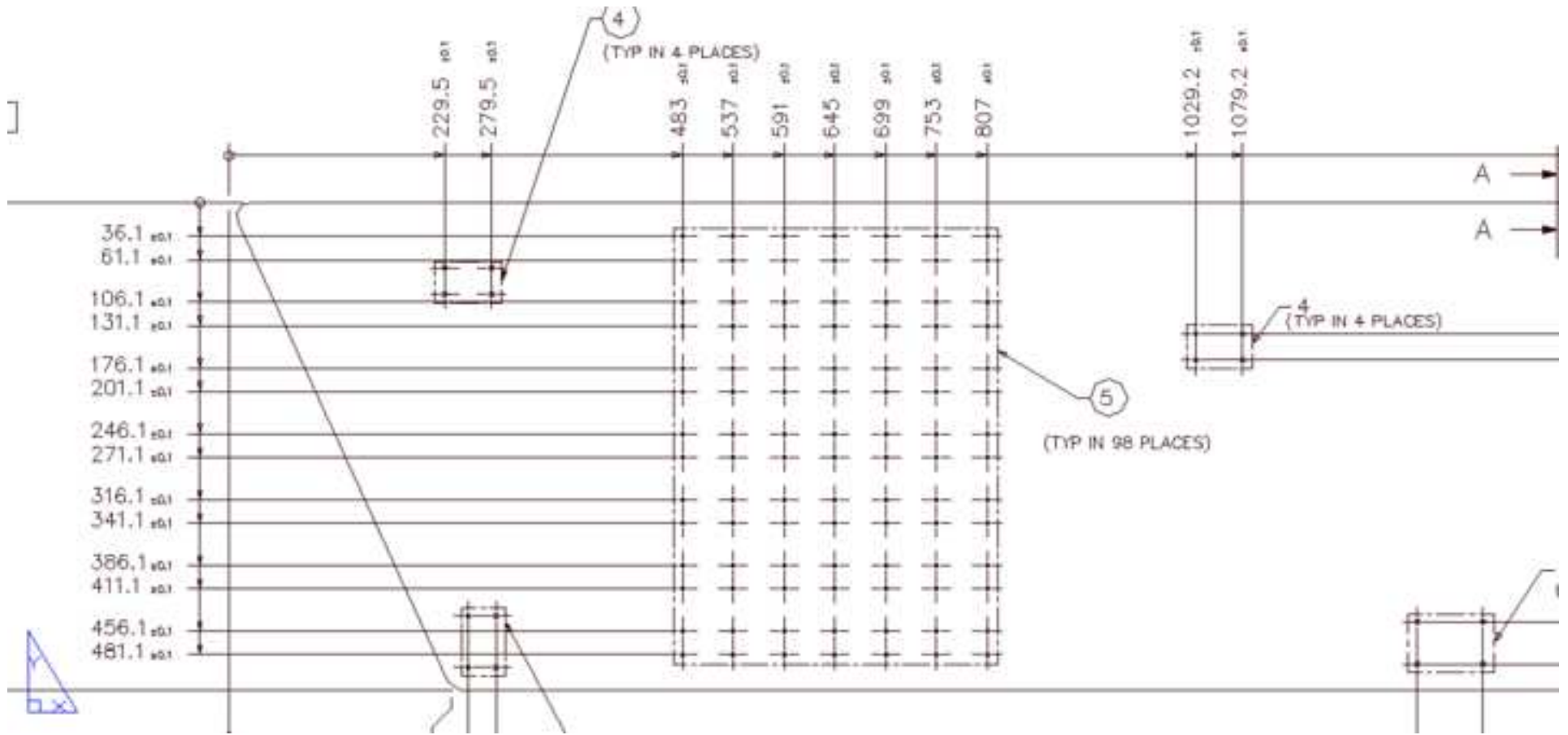

**Figure(s)**
Click here to download high resolution image

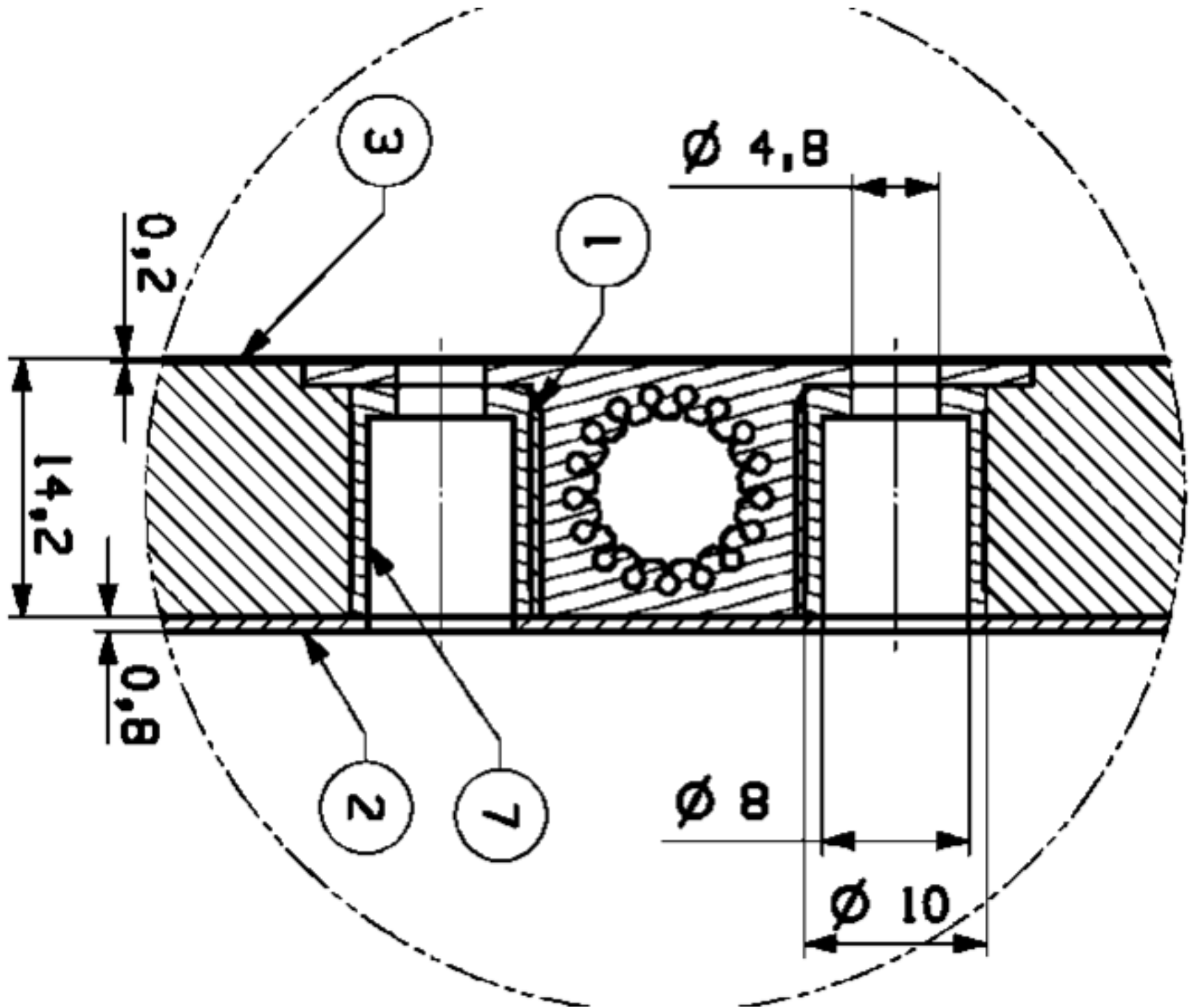

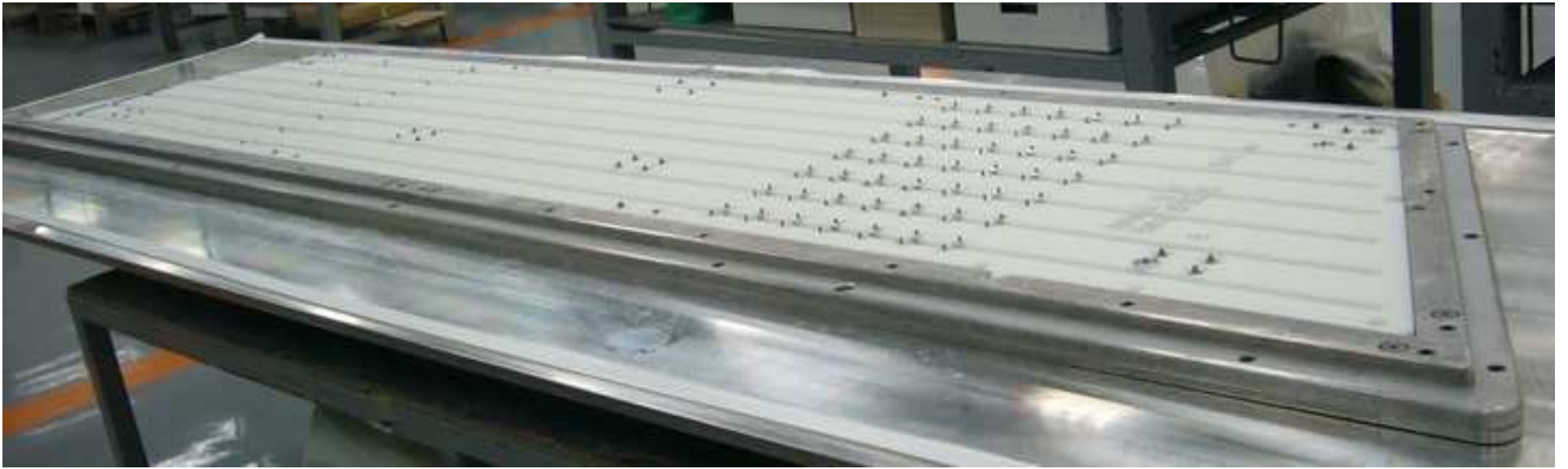

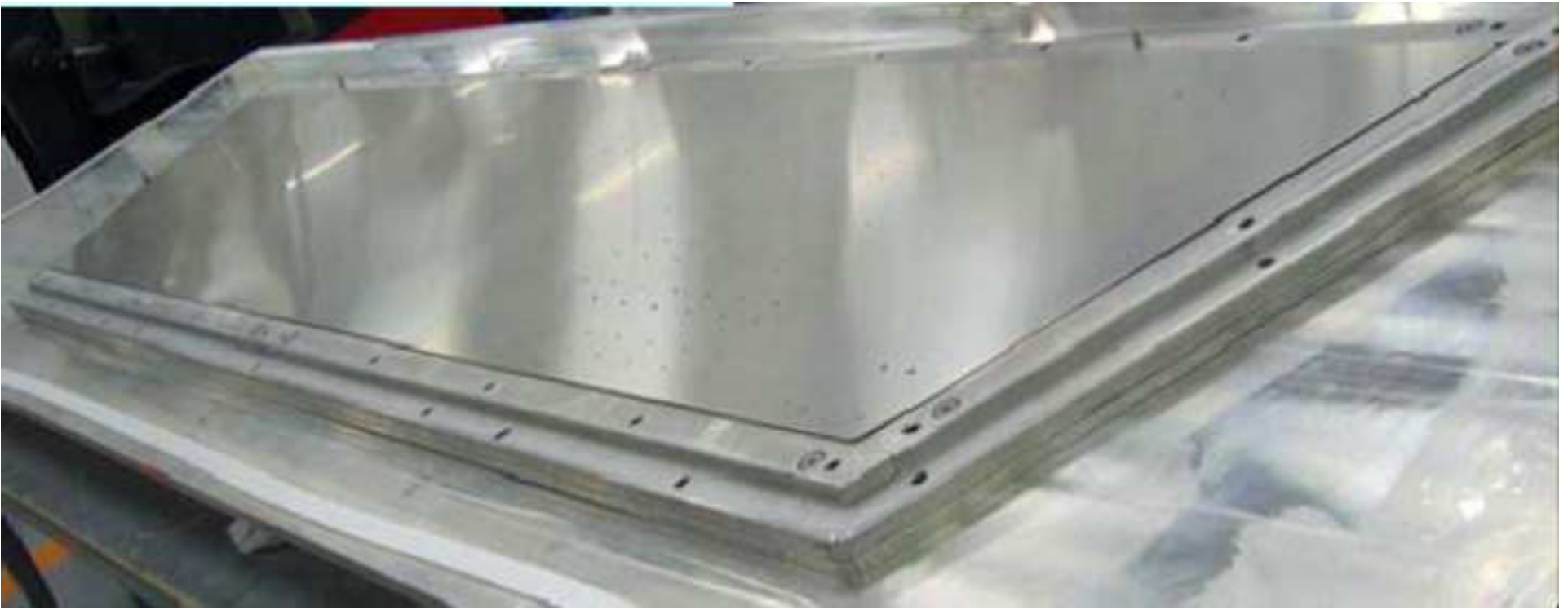

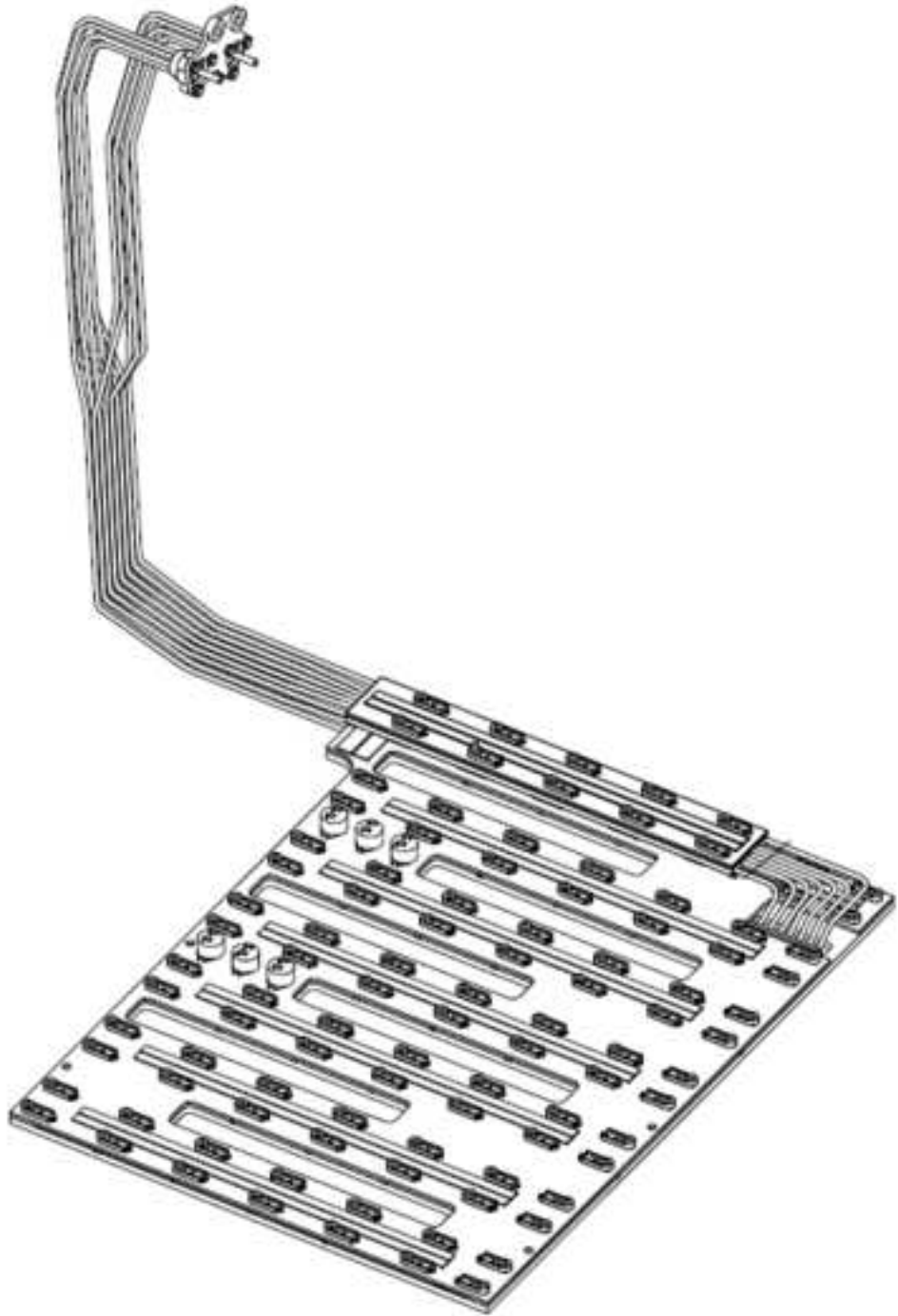

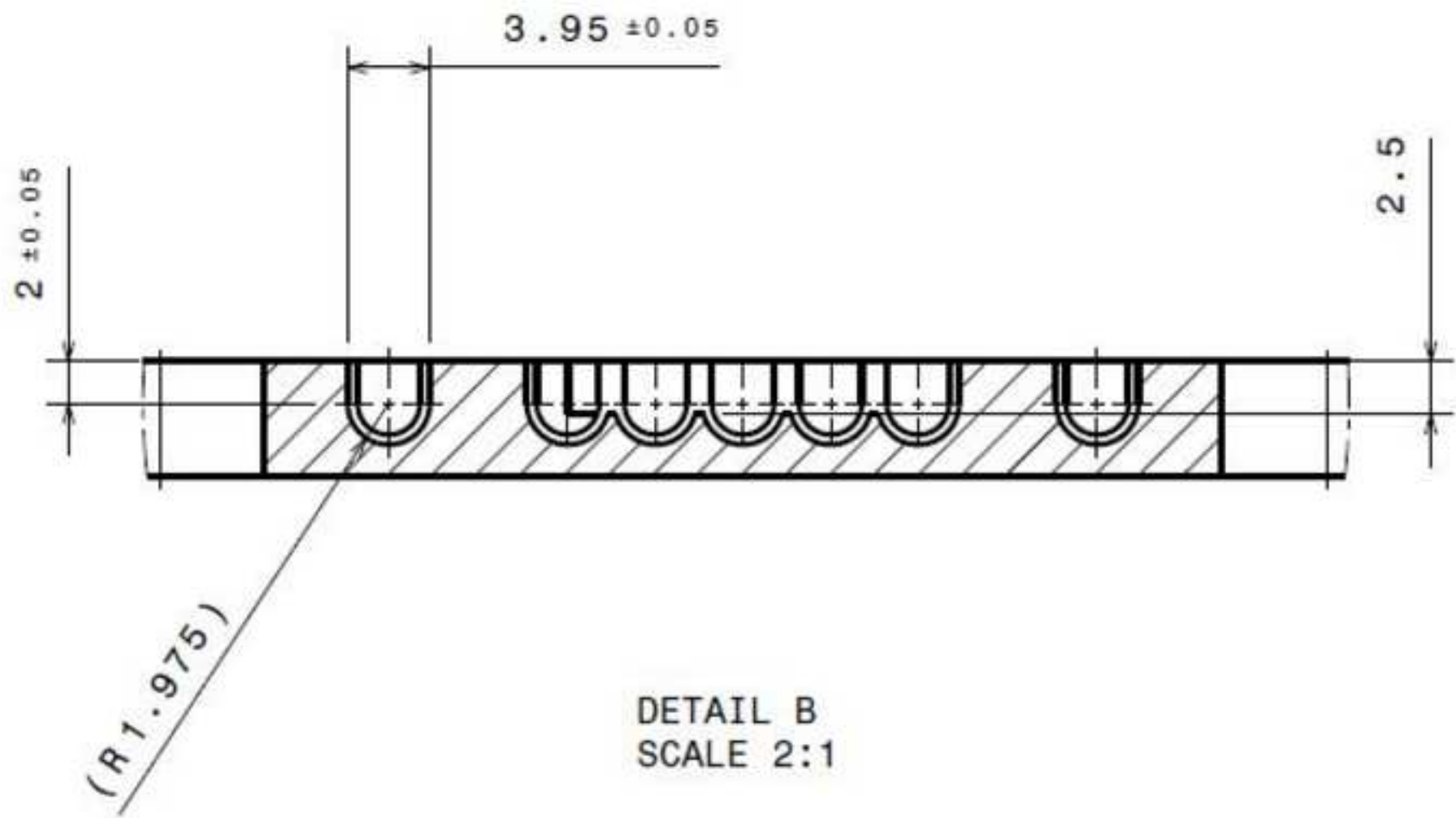

DETAIL B
SCALE 2:1

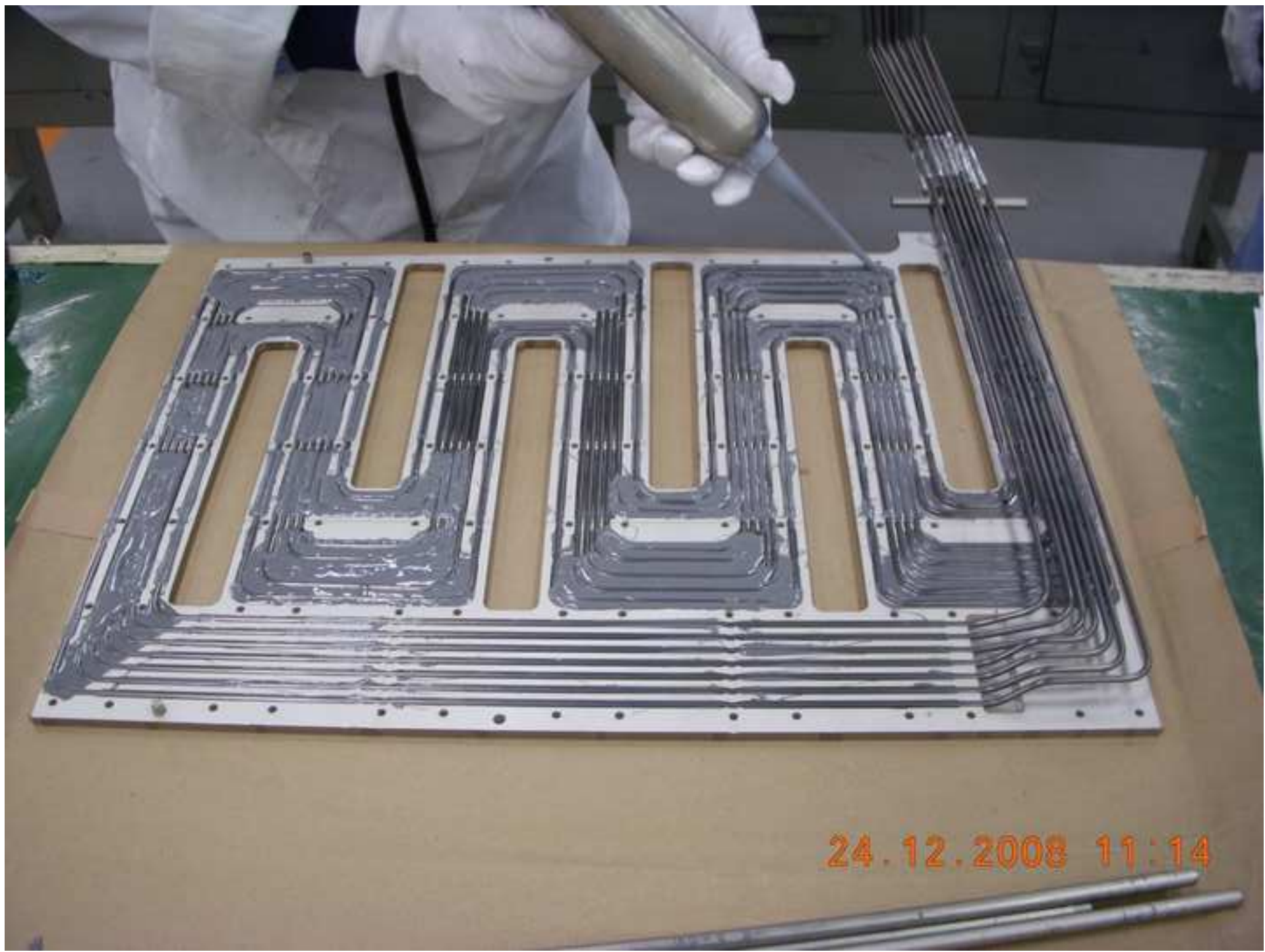

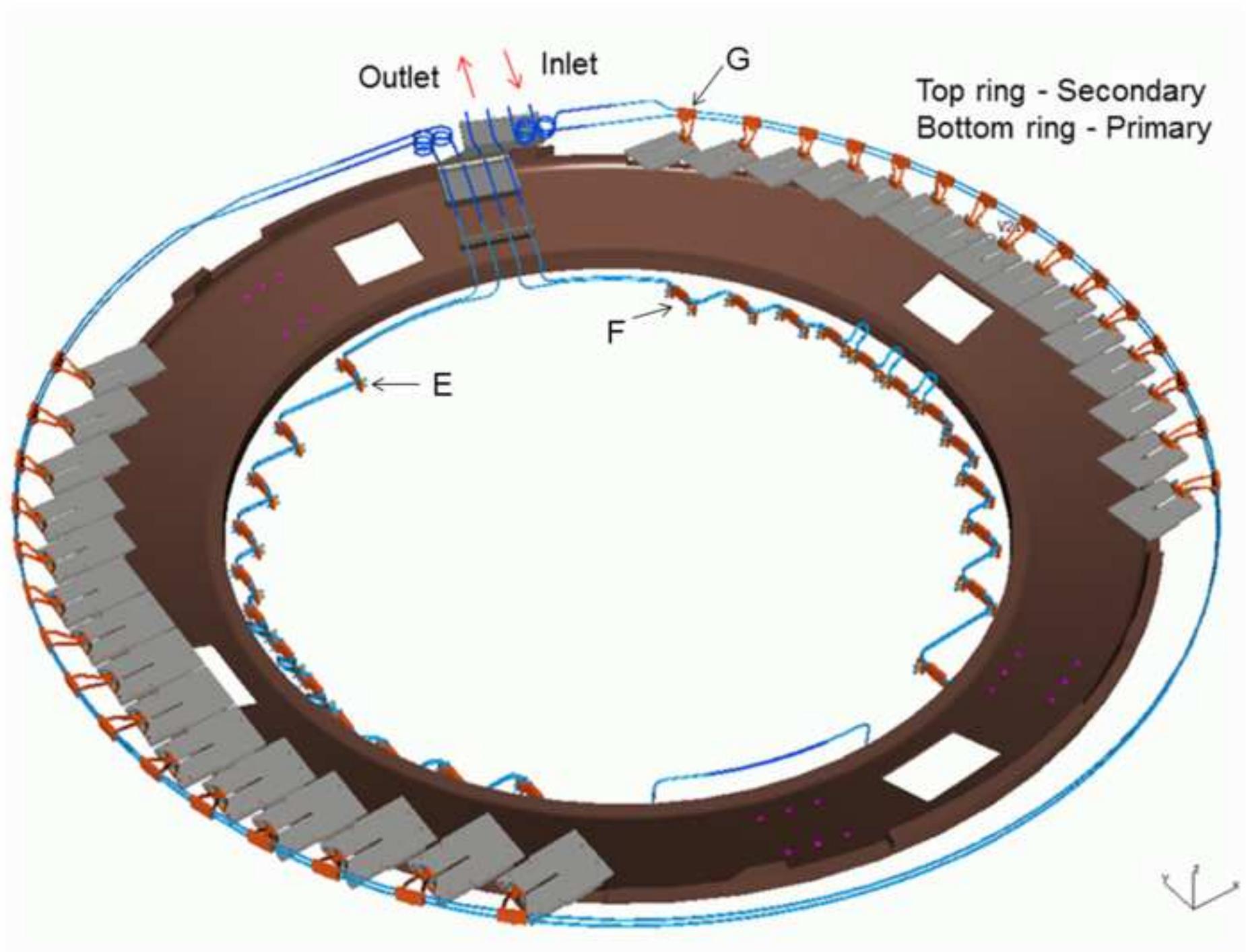

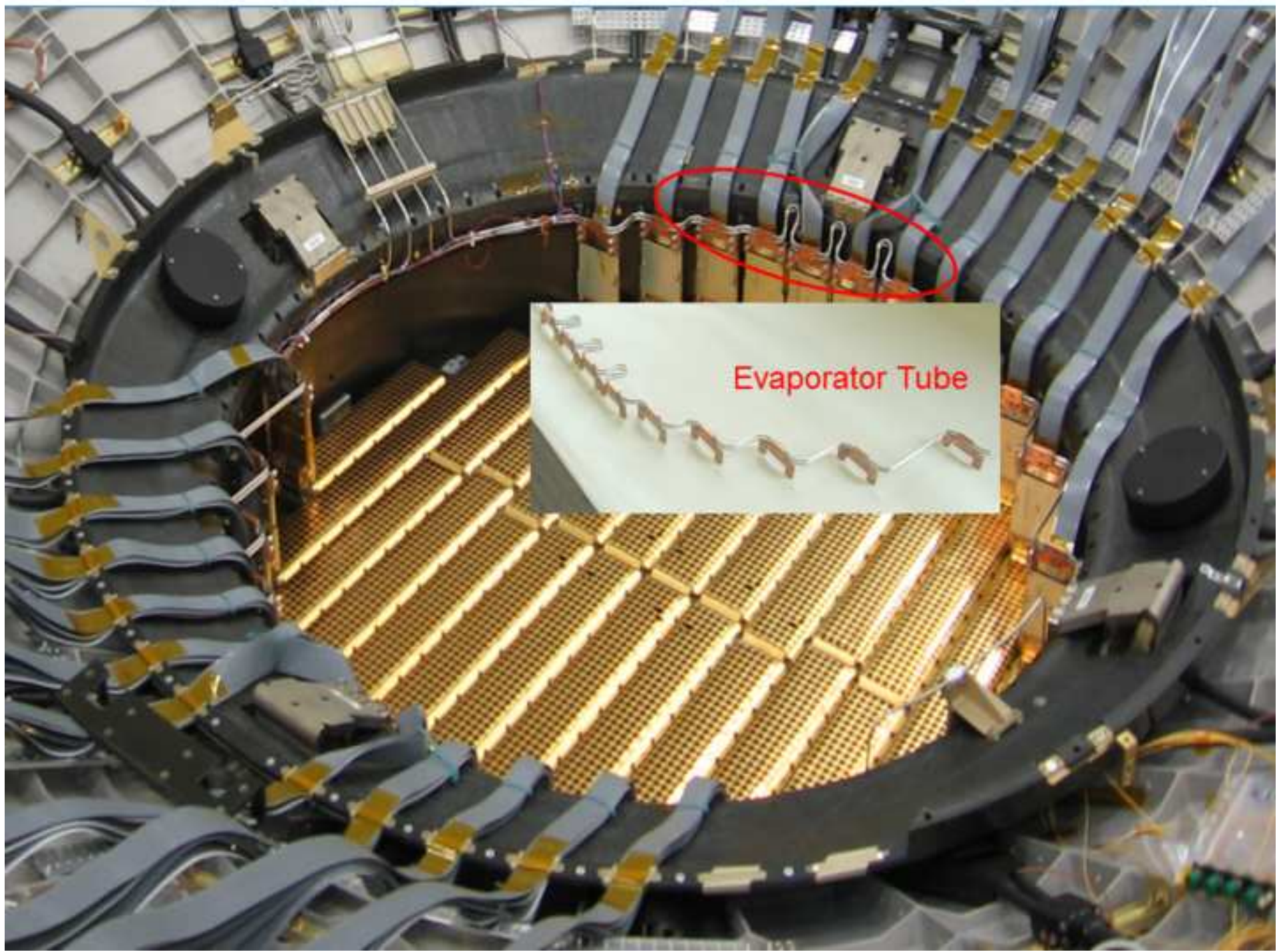

Evaporator Tube

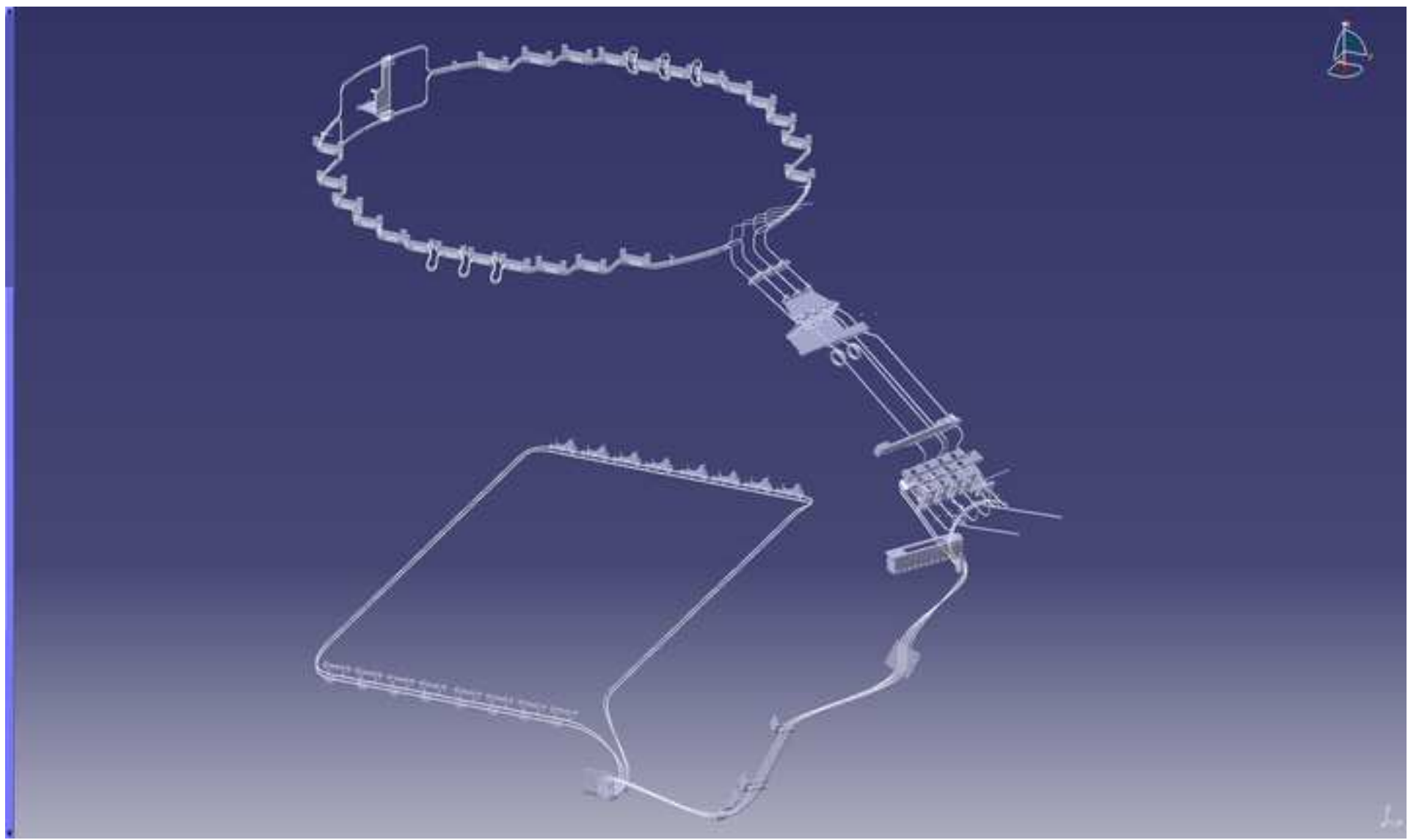

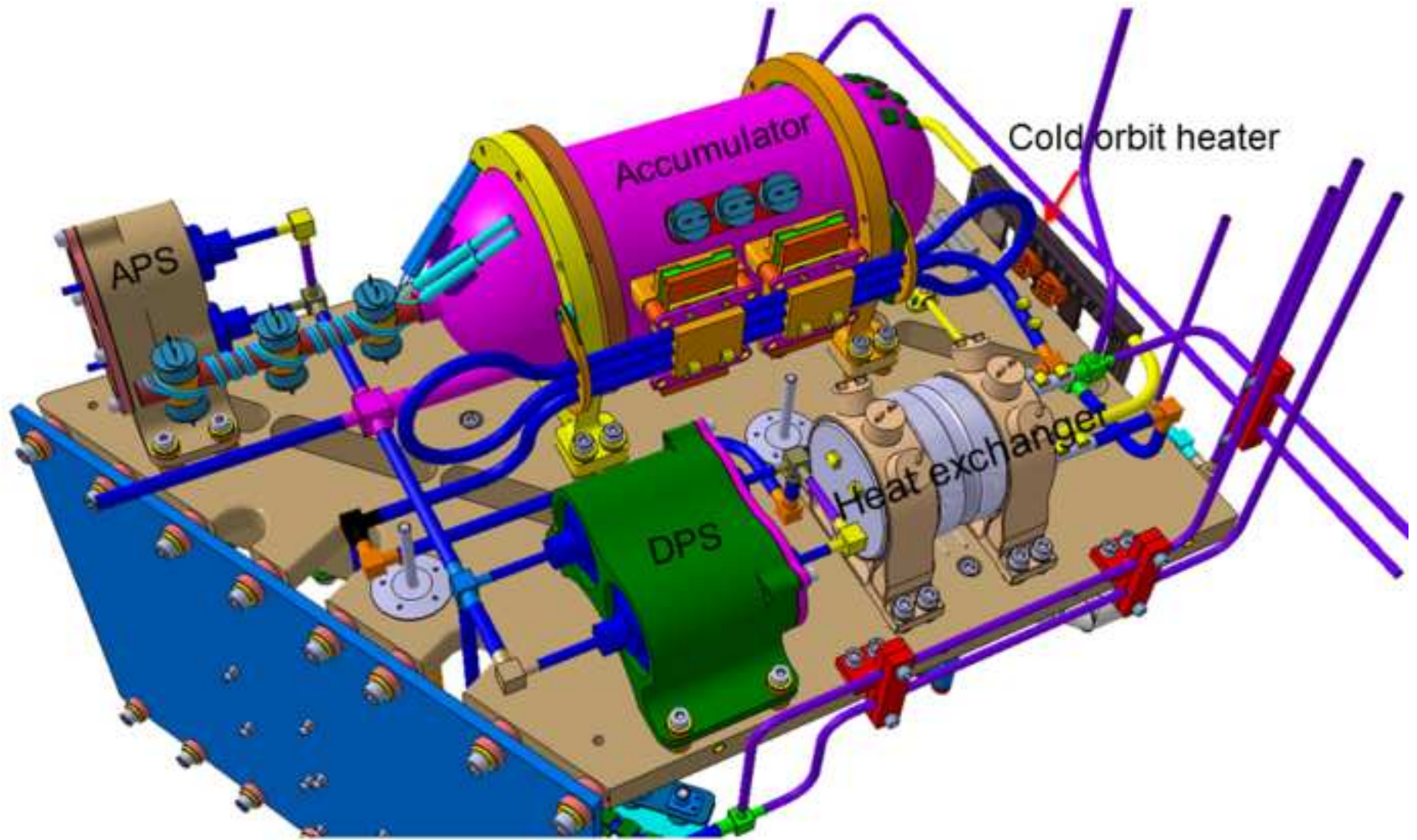

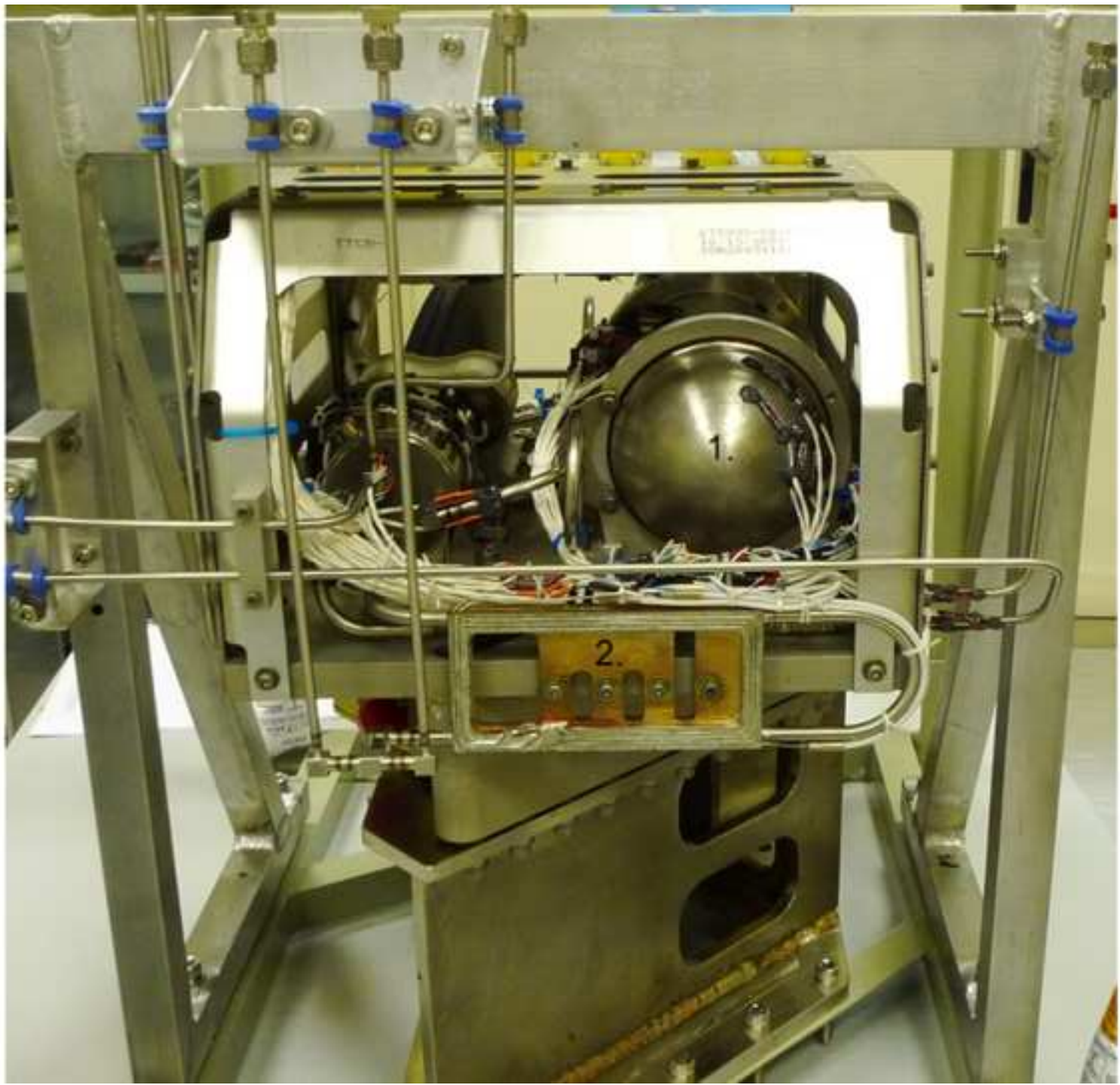

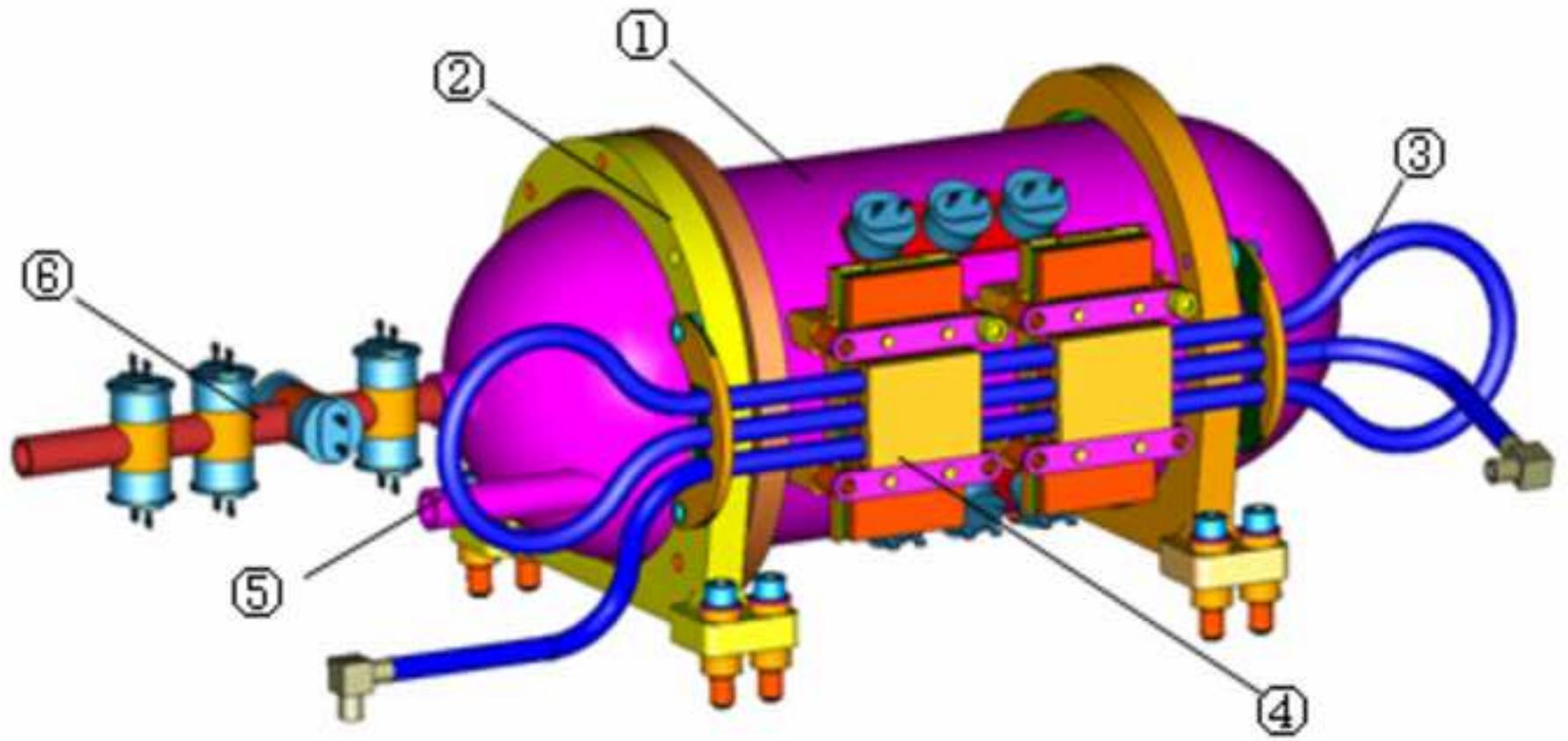

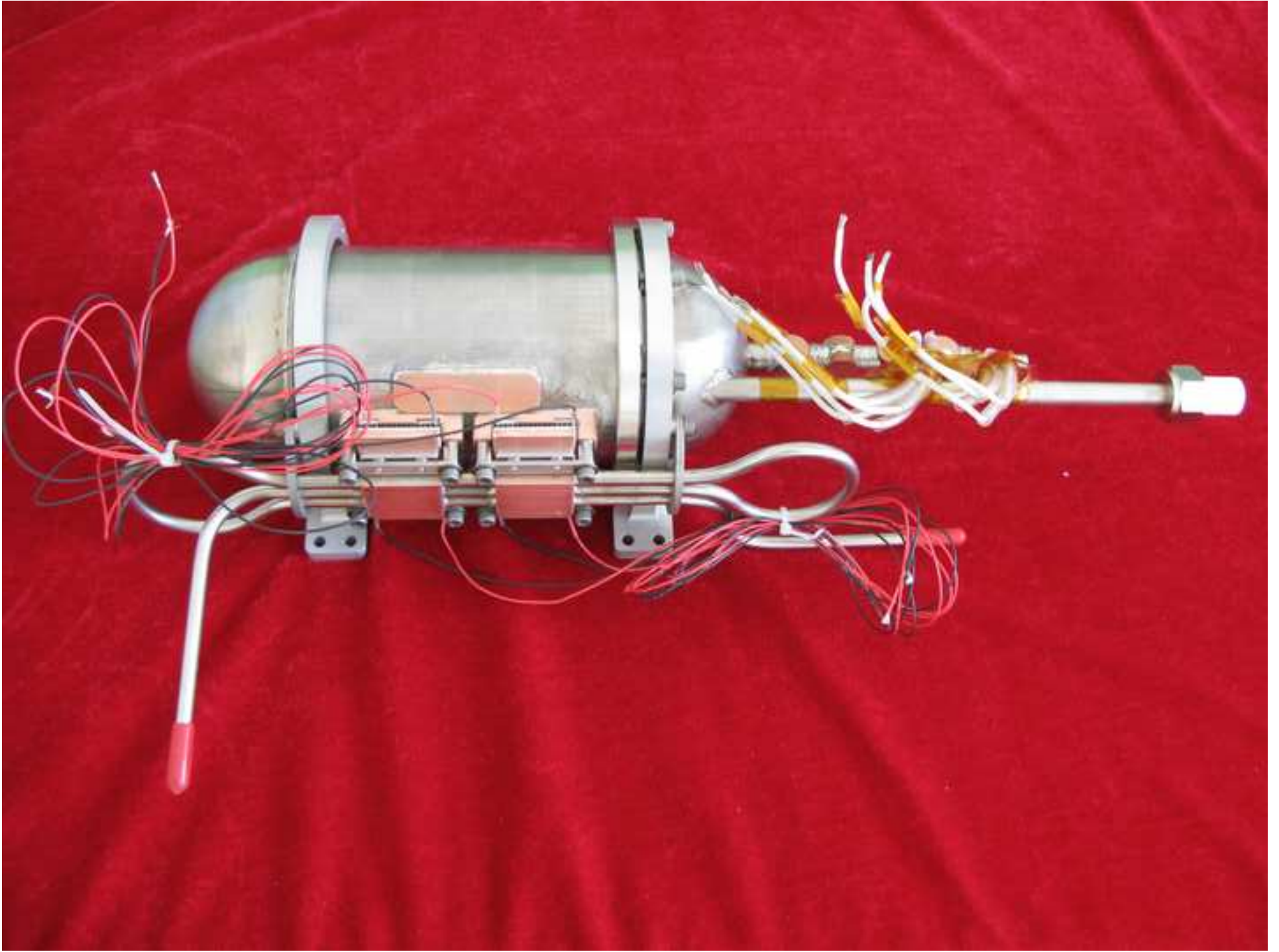

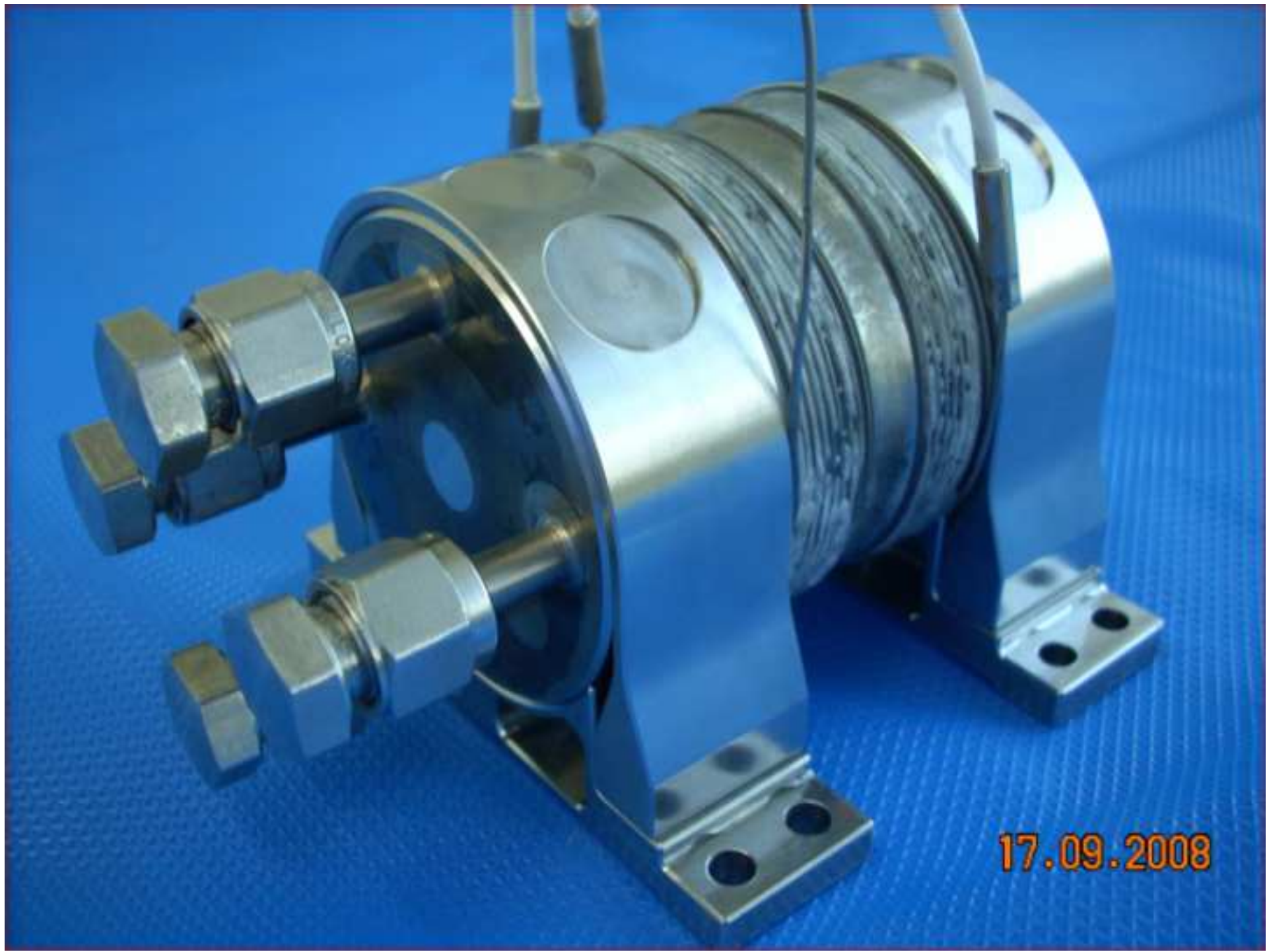

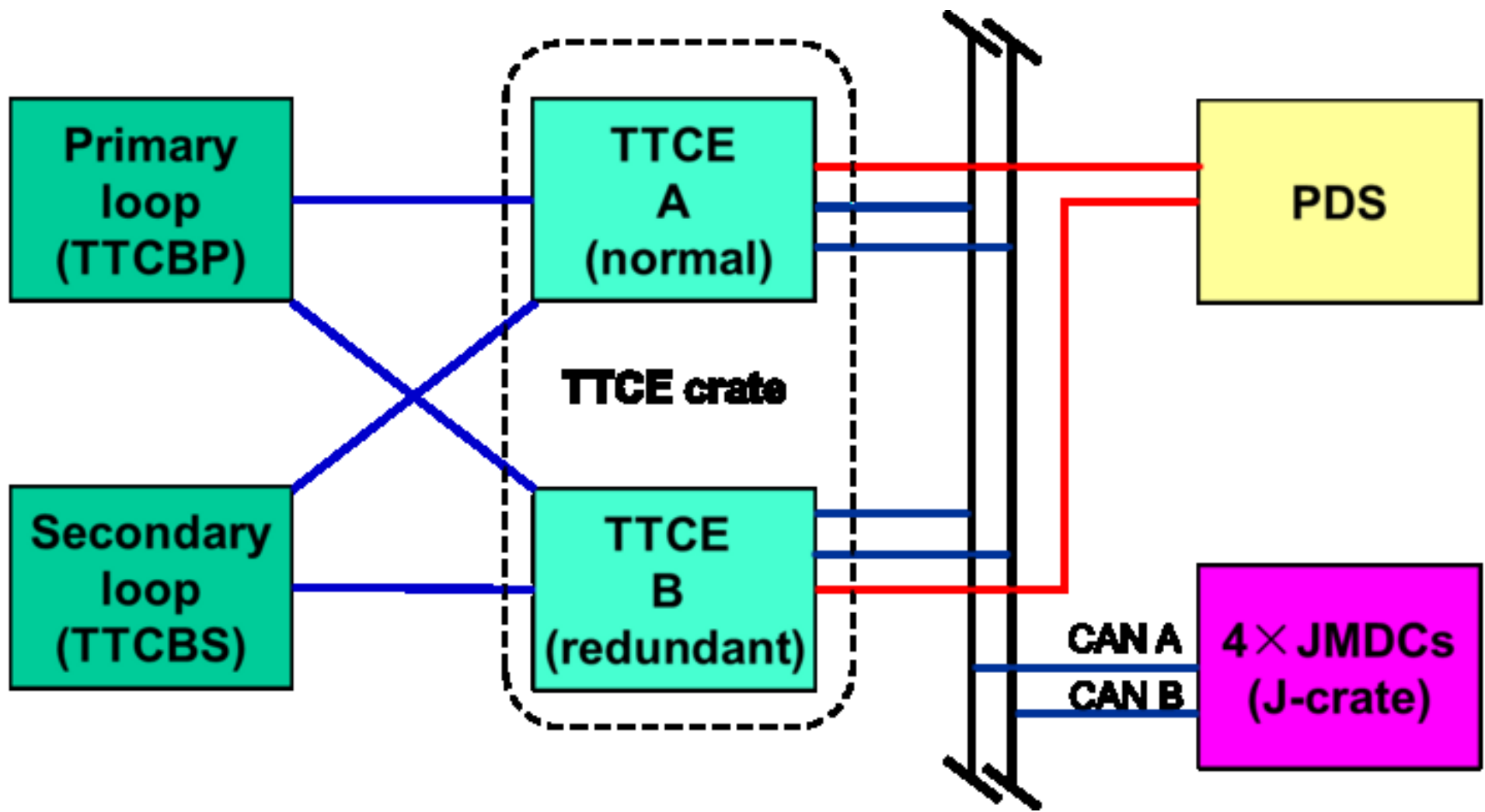

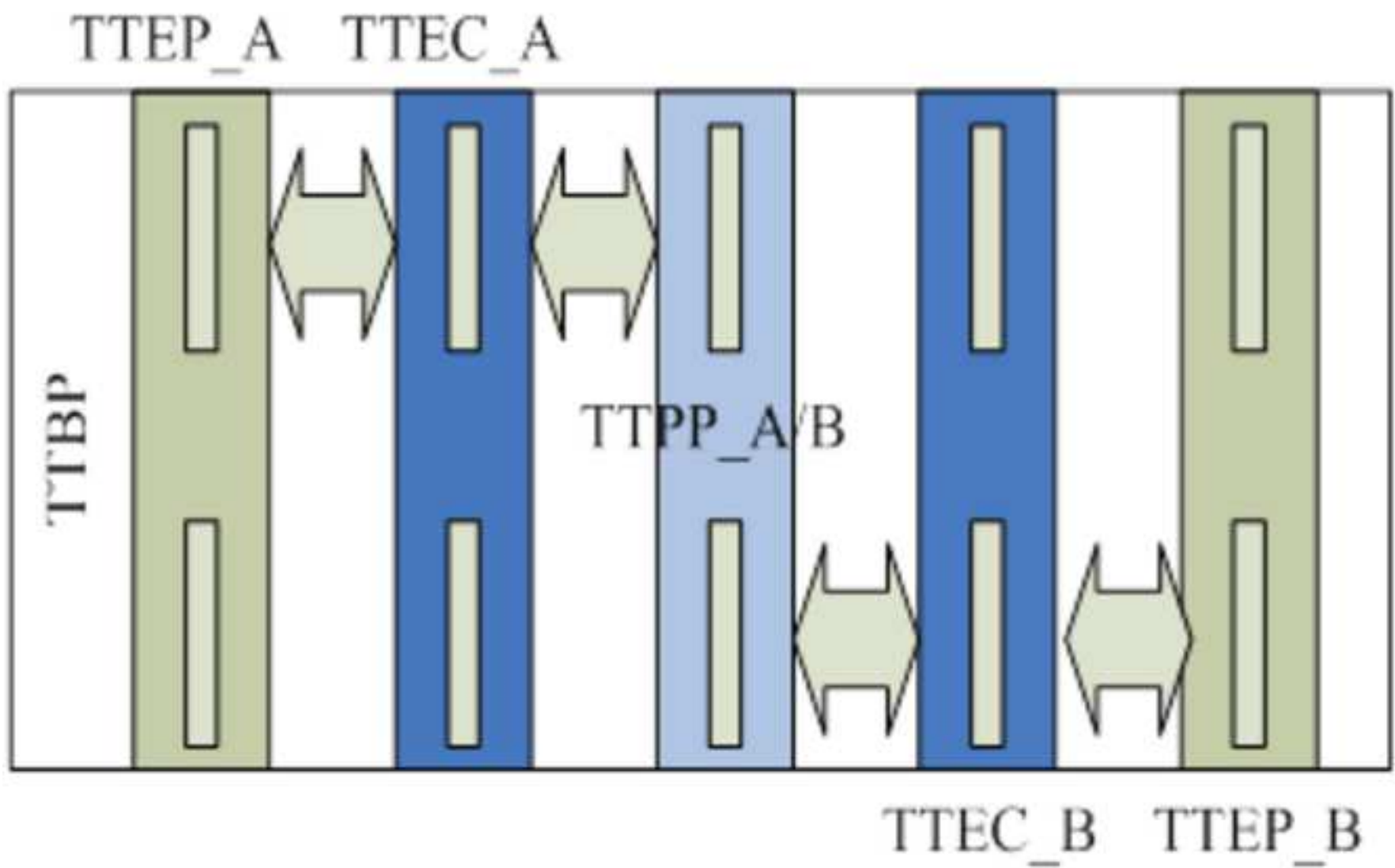

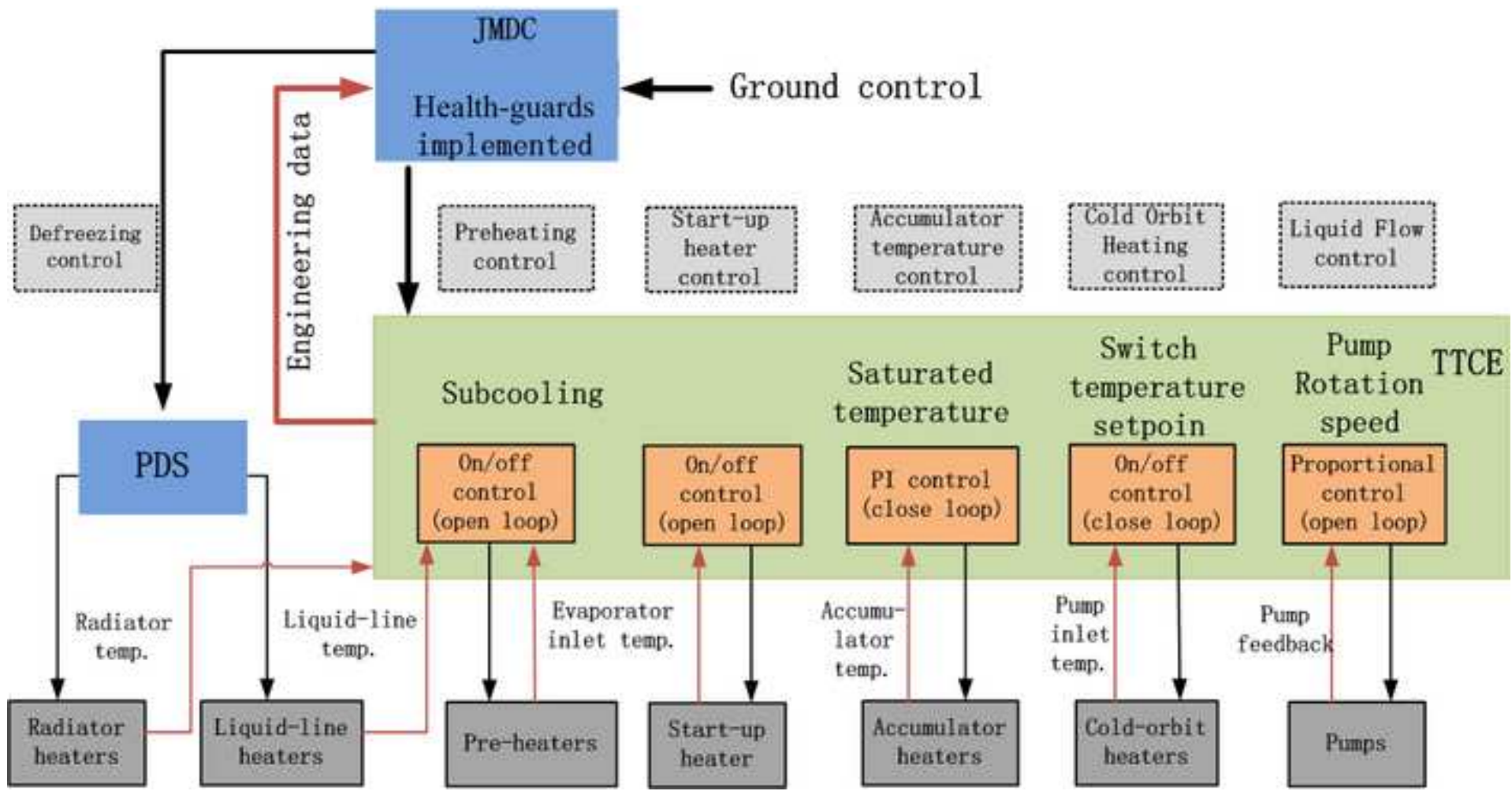

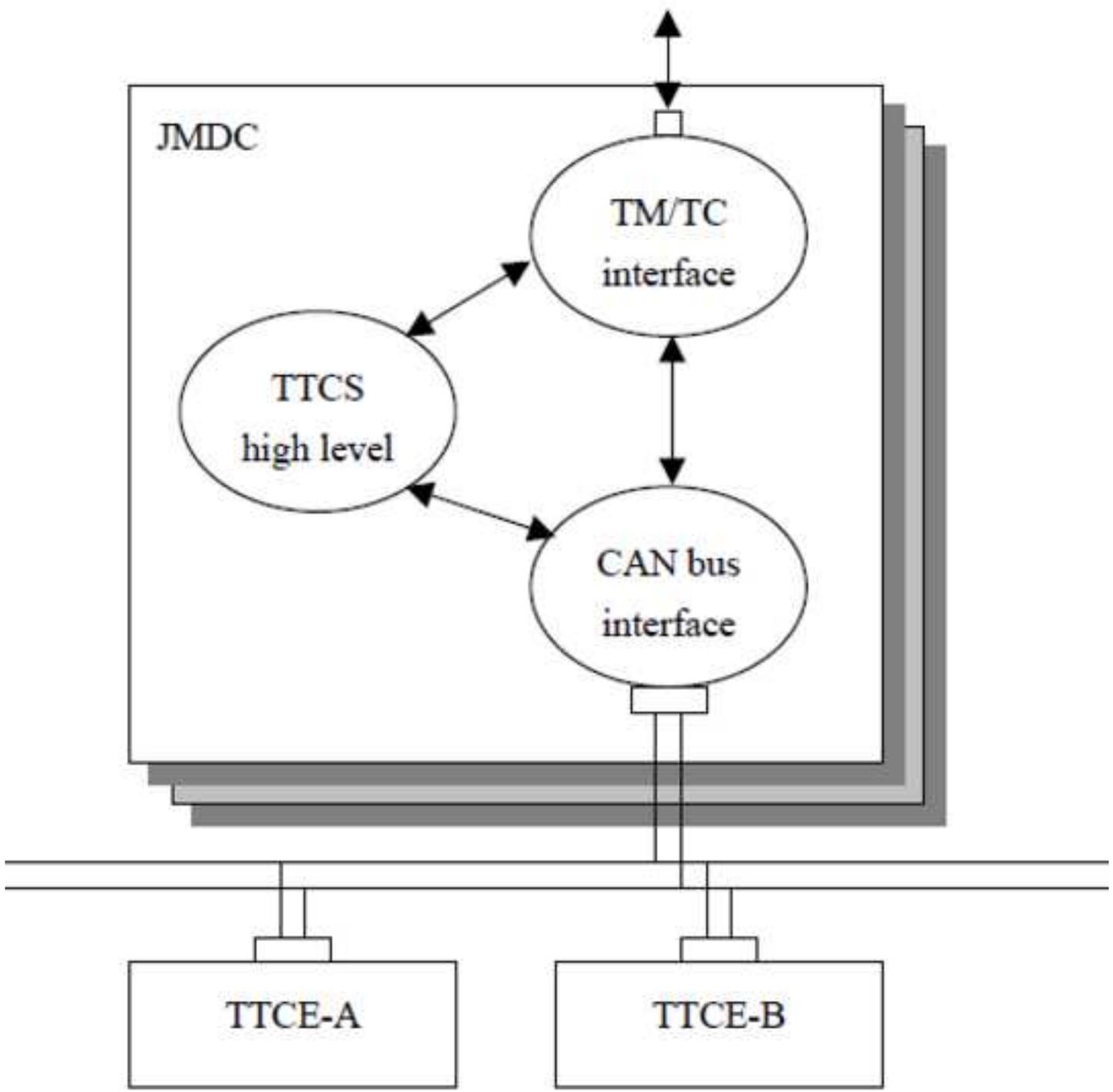

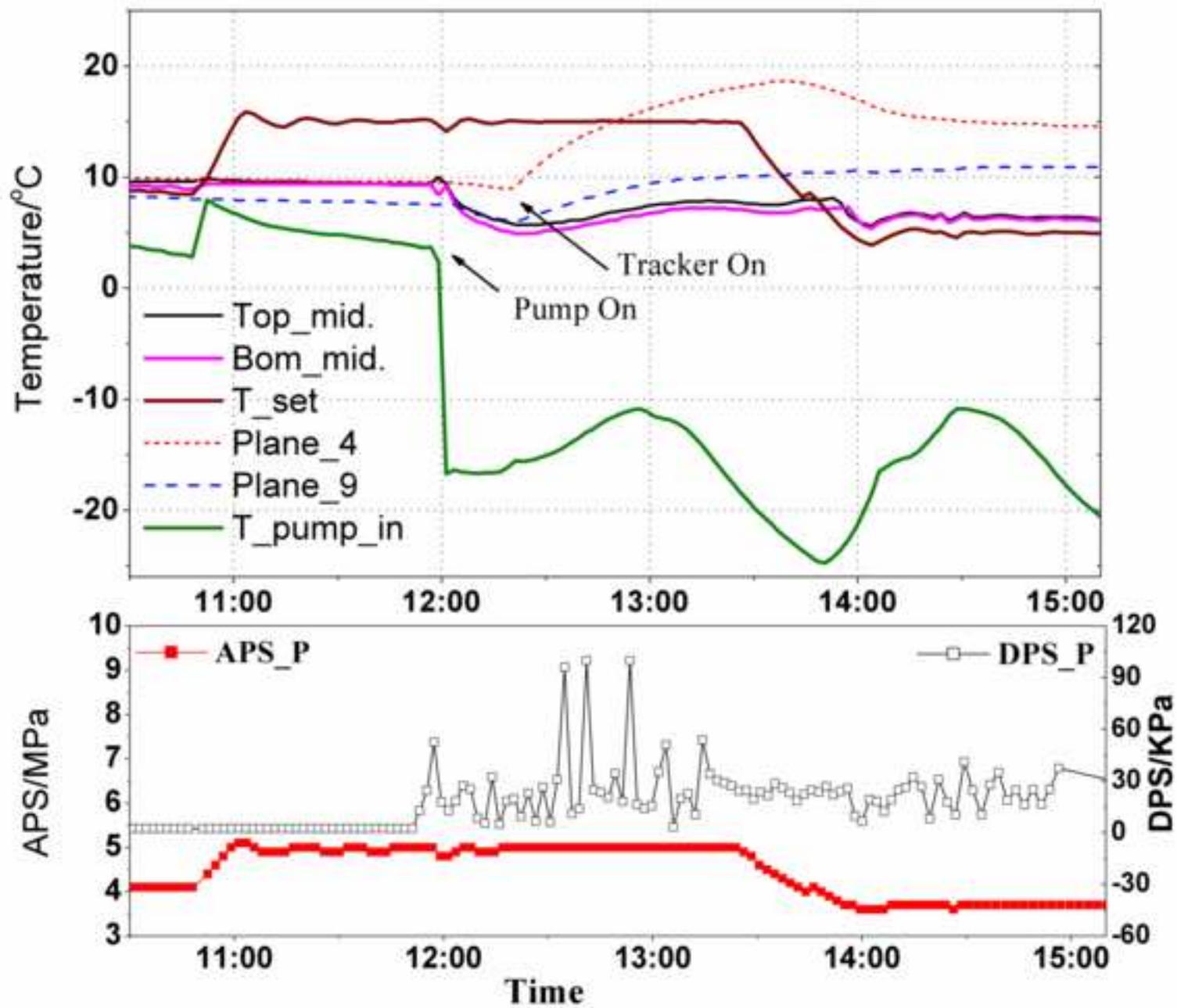

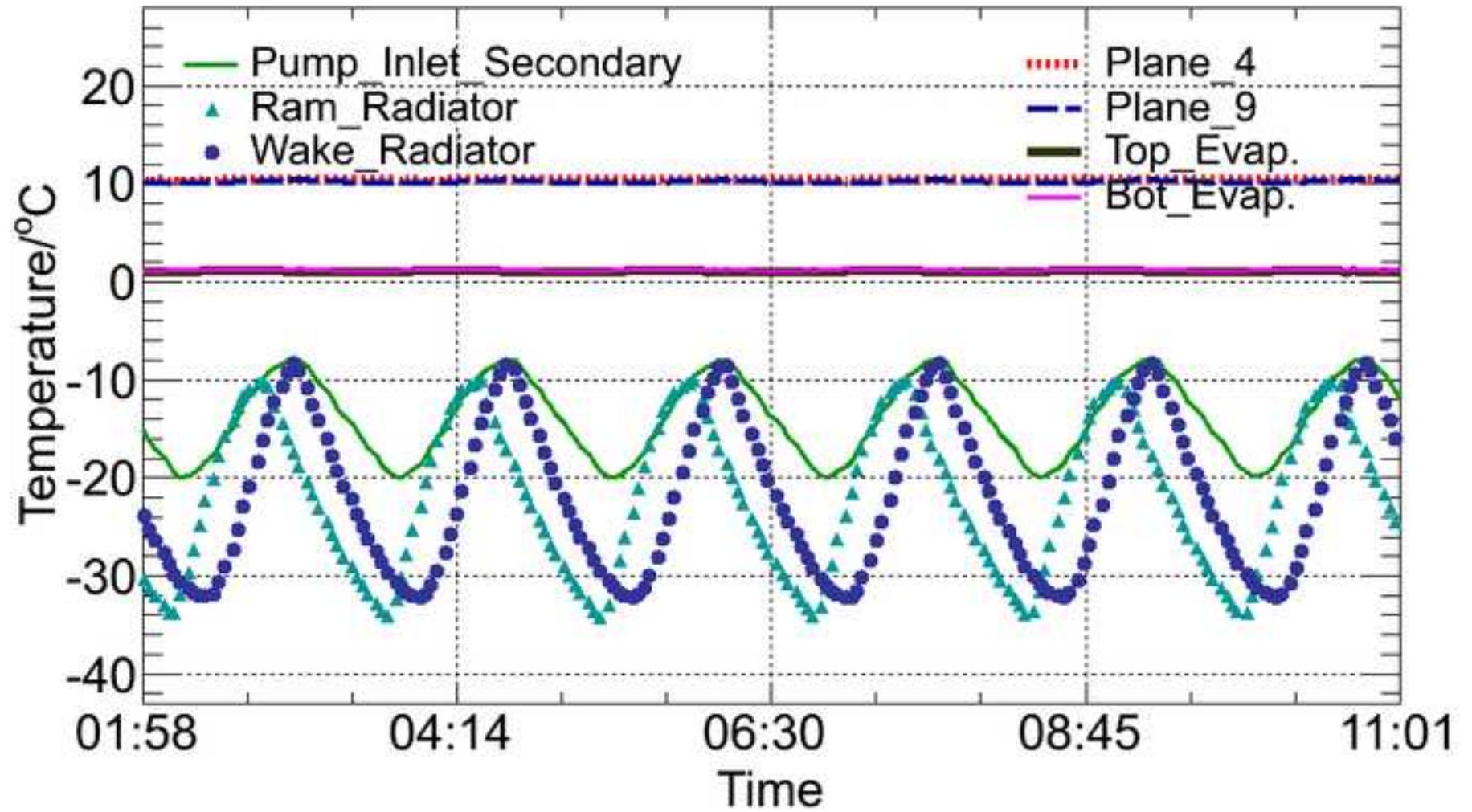

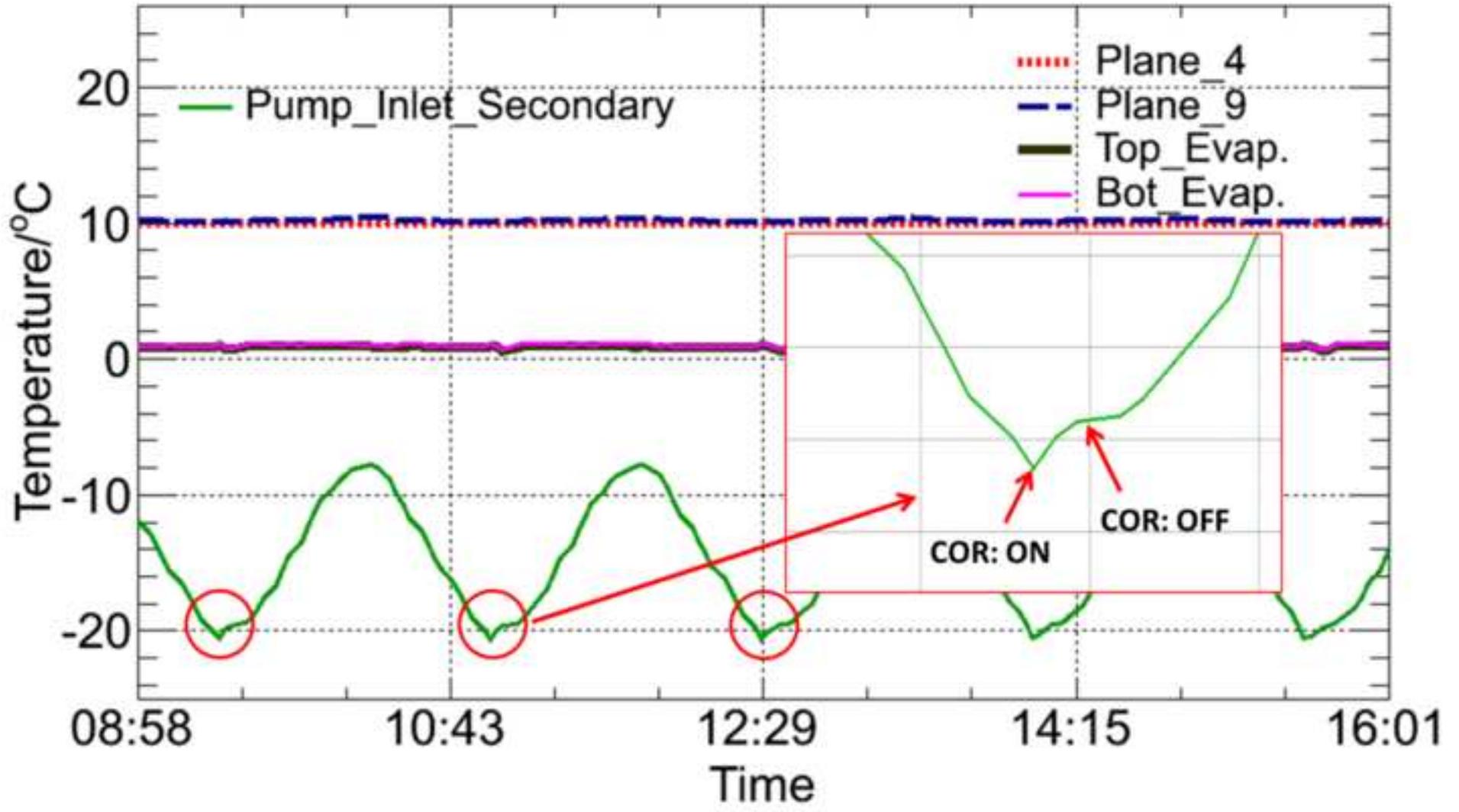

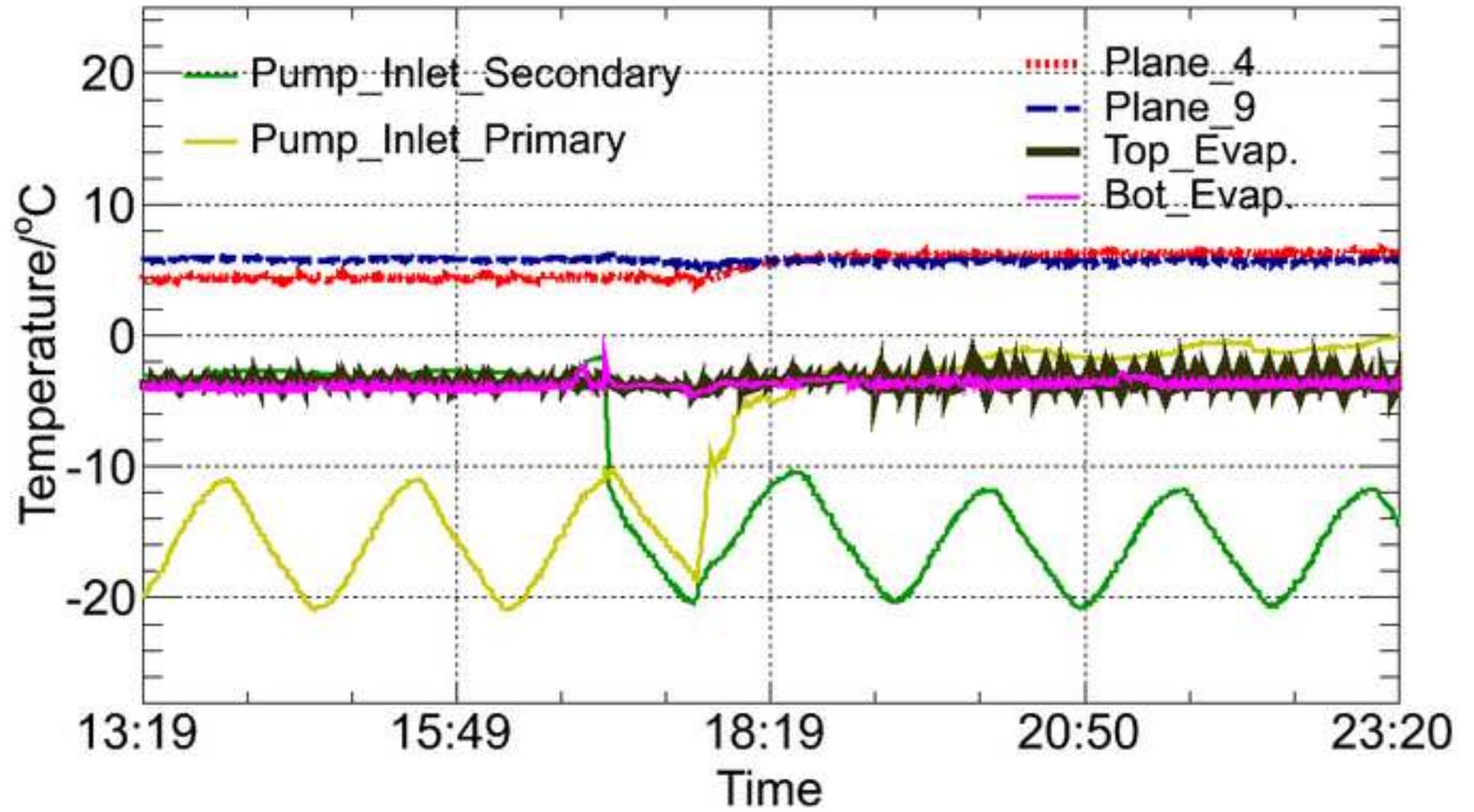

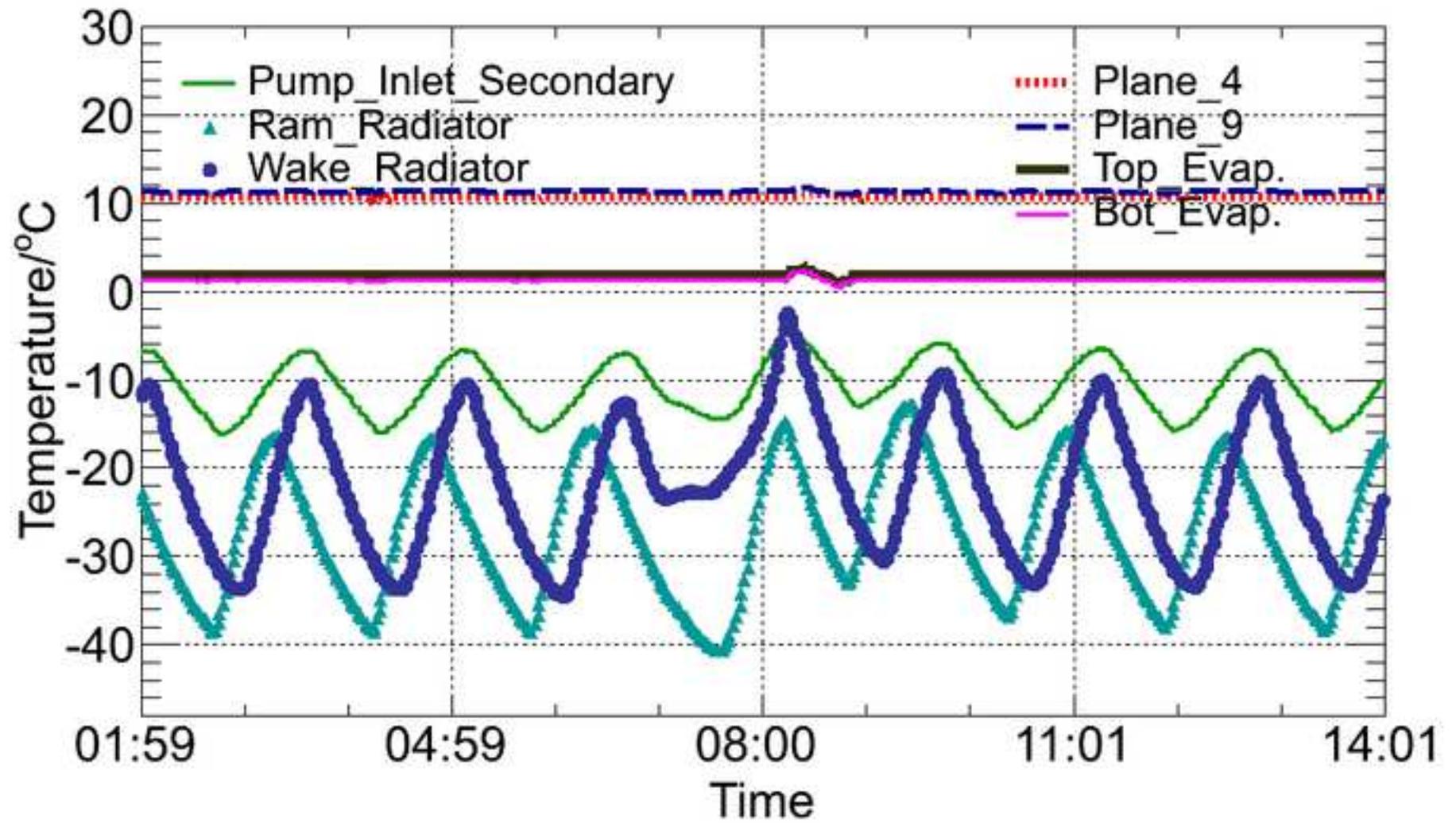

**Table(s)**

Table 5  Minimum Margins of Safeties and locations

| Structural | Yield MofS | Ultimate MofS | Location | Bolt MofS | Location |
|---|---|---|---|---|---|
| Normal | 4.97 | 3.35 | Side-plate | 0.047 | Cover/Base-plate |
| Fail-safe | 5.28 | 6.21 | Side-plate | 0.22 | Pump Bracket/Start up radiator |

| Nomenclature | |
|---|---|
| AHP | Accumulator heat pipe |
| AMS | Alpha Magnetic Spectrometer |
| APS | Absolute pressure sensor |
| CAN (bus) | Controller Area Network |
| DoF | Degree of Freedom |
| DPS | Differential pressure sensor |
| EM | Engineering Model |
| EMC | Electro Magnetic compactibility |
| ESTEC | European Space Technology Centre |
| FEM | Finite element method |
| FM | Flight Model |
| HX | heat exchanger |
| ISS | International Space Station |
| JMDC | Mission Computers |
| MofS | Margin of Safety |
| PDS | Power Distribution System |
| PWM | Pulse width modulation |
| QM | Qualification Model |
| RICH | Ring-imaging Cherenkov (detector) |
| RPM | Revolutions Per Minute |
| STS | Space Shuttle Mission |
| TC | Tele-control |
| TM | Tele-monitor |
| TS | Thermostat |
| TTBP | Tracker Thermal Back Plane |
| TTCB | Tracker Thermal Component Box |
| TTCE | Tracker Thermal Control Electronics |
| TTCS | Tracker Thermal Control System |
| TTEC | Tracker Thermal Electronic Control Board |
| TTEP | Tracker Thermal Electronic Power Board |
| TTPP | Tracker Thermal Pump & Pressure sensors Board |
| TVT | Thermal Vacuum Test |
| USS | Universal Support Structure |

Table 1 Functions of the TTCS main components.

| Components | Functions |
| --- | --- |
| Evaporator | To collect and transport the heat from the Tracker to the $CO_2$ loop |
| Condenser | To conduct the heat from the loop to the radiators that dissipate the heat to the space |
| Accumulator | To compensate the mass of $CO_2$ in the loop and to control the operation temperature of the system |
| Pump | To provide driving power for the loop |
| Heat-exchanger | To exchange heat between the saturated $CO_2$ and the sub-cooled liquid $CO_2$ |
| Pre-heater | To heat the sub-cooled liquid $CO_2$ into saturated state before flowing into the evaporator |
| Start-up heater | To avoid sub-cooled liquid $CO_2$ flow into the Tracker to damage the electronics at extremely cold environmental condition |
| Cold orbit heater | To prevent the $CO_2$ from being cooled to the freezing point by effectively heating the fluid |
| De-freezing heater | To defreeze the solid $CO_2$ inside the tubes from the manifolds to the condensers |
| Radiator heaters | To defreeze the solid $CO_2$ inside the condensers mounted on the radiators |
| Component box | To assemble the components except for the evaporators and the condensers |

Table 2 The Beta and Euler angles range of the ISS.

| Angle | Variation Range |
|---|---|
| Beta angle | −75° to +75° |
| Yaw (Z) attitude angle | −15° to +15° |
| Roll (X) attitude angle | −15° to +15° |
| Pitch(Y) attitude angle | 0° to +25° with docked STS |
| | −20° to +15° with undocked STS |

Table 3  Summary of the minimum MofS for all the load cases.

| Components | Static:MofS | | Thermal:MofSs | |
|---|---|---|---|---|
| | Yield | Ultimate | Yield | Ultimate |
| Accumulator | 0.03 | 0.11 | 0.02 | 0.02 |
| Fixed Bracket & Clamp Collar&Wedge | 1.10 | 2.11 | 1.00 | 1.82 |
| Sliding Bracket | 4.01 | 3.56 | 2.76 | 2.95 |
| Heat Pipe | 1.05 | 2.46 | 0.92 | 1.14 |
| Peltier Fixed | 2.07 | 2.26 | 0.51 | 1.20 |
| TS & Peltier Heat Exchanger & Peltier heat exchanger press &Spring Support | 0.61 | 1.13 | 0.68 | 1.22 |
| Joints | Static: bolt MofS | | Thermal: bolt MofS | |
| Accu.Bracket Clamp&Collar Bolt | 0.12 | | 0.09 | |
| PipeFix&Clamp Bolt | 0.123 | | 0.122 | |
| Press&Saddle Bolt | 0.107 | | 0.106 | |

Table 4  Summary of the minmum MofS for all load cases (Fail-safe)

| Component Name | Static: MofS | | Thermal: MofS | |
|---|---|---|---|---|
| | Yield | Ultimate | Yield | Ultimate |
| Accumulator | 0.54 | 1.80 | 0.53 | 1.55 |
| Fixed Bracket & Clamp Collar&Wedge | 1.61 | 5.17 | 1.49 | 4.66 |
| Sliding Bracket | 6.91 | 8.85 | 3.68 | 6.92 |
| Heat Pipe | 1.07 | 9.19 | 1.87 | 7.31 |
| Peltier Fixed | 4.94 | 16.41 | 0.76 | 3.17 |
| TS & Peltier Heat Exchanger & Peltier heat exchanger press & Spring Support | 0.98 | 3.20 | 0.24 | 2.64 |
| Joints | Static: bolt MofS | | Thermal:bolt MofS | |
| Accumulator.Bracket Clamp&Collar Bolt | 0.34 | | 0.30 | |
| PipeFix&Clamp Bolt | 0.50 | | 0.49 | |
| Press&Saddle Bolt | 0.19 | | 0.20 | |